\documentclass{MITcsail}

\usepackage[utf8]{inputenc}

\usepackage{bbm}
\usepackage{graphicx}%
\usepackage{subcaption}
\usepackage{booktabs} 
\usepackage{multirow}%
\usepackage{amsmath,amssymb,amsfonts}%
\usepackage{amsthm}%
\usepackage{mathrsfs}%
\usepackage[title]{appendix}%
\usepackage{xcolor}%
\usepackage{textcomp}%
\usepackage{manyfoot}%
\usepackage{booktabs}%
\usepackage{algorithm}%
\usepackage{algorithmicx}%
\usepackage{algpseudocode}%
\usepackage{listings}%
\usepackage{color}
\usepackage{soul}

\usepackage{graphicx}
\usepackage{tabularx}
\usepackage{array}
\usepackage{booktabs} 
\usepackage[most]{tcolorbox}

\usepackage{tcolorbox}
\tcbuselibrary{listings,skins,breakable}
\usepackage{upquote}

\usepackage{hyperref} 
\usepackage[backend=biber]{biblatex} 
\renewcommand{\textcolor}[2]{#2}

\newtcblisting{promptblock}[1][]{%
  breakable,
  enhanced,
  colback=gray!4,
  colframe=gray!55,
  boxrule=0.6pt,
  arc=1mm,
  left=2mm,right=2mm,top=1.2mm,bottom=1.2mm,
  listing engine=listings,
  listing only,
  title={#1},
  fonttitle=\bfseries\small,
  attach boxed title to top left={yshift=-2mm,xshift=2mm},
  boxed title style={colback=white,colframe=gray!55},
  listing options={
    basicstyle=\ttfamily\small,
    columns=fullflexible,
    keepspaces=true,
    breaklines=true,
    breakatwhitespace=false,
    breakindent=0pt, 
    prebreak={}, 
    postbreak={},
    showstringspaces=false,
    tabsize=2,
    literate={"}{{"}}1 
  }
}

\tcbset{colback=gray!5!white, colframe=gray!80!black, boxrule=0.5pt, arc=2mm, left=2pt, right=2pt, top=2pt, bottom=2pt}

\title{TacEleven: generative tactic discovery for football open play}

\renewcommand{\thefootnote}{\fnsymbol{footnote}}
\newcommand{\equalcontrib}{\footnotemark[2]}   
\newcommand{\correspondence}{\footnotemark[1]} 
\newcommand{\projectlead}{\footnotemark[3]}    

\hypersetup{colorlinks=true, linkcolor=black, urlcolor=black, citecolor=black}
\usepackage{fontawesome5}
\newcommand{\emailicon}[1]{\textsuperscript{\href{mailto:#1}{\scriptsize\faEnvelope[regular]}}}

\author{
Siyao Zhao~\emailicon{zhaosiyao2023@ia.ac.cn}~$^{1,2}$\equalcontrib,
Hao Ma~\emailicon{mahao2021@ia.ac.cn}~$^{1,3}$\equalcontrib\projectlead,
Zhiqiang Pu~\emailicon{zhiqiang.pu@ia.ac.cn}~$^{1,2,3}$\correspondence,
Jingjing Huang~\emailicon{huangjingjing2024@ia.ac.cn}~$^{1,3}$,
Yi Pan~\emailicon{yi.pan@ia.ac.cn}~$^{1}$,\\
Shijie Wang~\emailicon{shijie.wang2022@outlook.com}~$^{4}$,
Zhi Ming~\emailicon{z.ming@aja.fr}~$^{5}$
\\
\vspace{1em}
\normalfont{\small $^{1}$The Key Laboratory of Cognition and Decision Intelligence for Complex Systems, Institute of Automation, Chinese Academy of Sciences}\\
\normalfont{\small $^{2}$School of Advanced Interdisciplinary Sciences, University of Chinese Academy of Sciences}\\
\normalfont{\small $^{3}$School of Artificial Intelligence, University of Chinese Academy of Sciences}\\
\normalfont{\small $^{4}$Shanghai AI Laboratory}\\
\normalfont{\small $^{5}$Association de la Jeunesse Auxerroise}\\
\vspace{2em}
}

\begin{document}
\maketitle

\footnotetext[2]{\ These authors contributed equally to this work.}
\footnotetext[3]{\ Project leader.}
\footnotetext[1]{\ Correspondence: zhiqiang.pu@ia.ac.cn}

\renewcommand{\thefootnote}{\arabic{footnote}}
\hypersetup{colorlinks=true, citecolor=brown, linkcolor=red, urlcolor=magenta}

\begin{abstract}

Creating offensive advantages during open play is fundamental to football success. However, due to the highly dynamic and long-sequence nature of open play, the potential tactic space grows exponentially as the sequence progresses, making automated tactic discovery extremely challenging. To address this, we propose TacEleven, a generative framework for football open-play tactic discovery developed in close collaboration with domain experts from AJ Auxerre, designed to assist coaches and analysts in tactical decision-making. TacEleven consists of two core components: a language-controlled tactical generator that produces diverse tactical proposals, and a multimodal large language model-based tactical critic that selects the optimal proposal aligned with a high-level stylistic tactical instruction. The two components enables rapid exploration of tactical proposals and discovery of alternative open-play offensive tactics. We evaluate TacEleven across three tasks with progressive tactical complexity: counterfactual exploration, single-step discovery, and multi-step discovery, through both quantitative metrics and a questionnaire-based qualitative assessment. The results show that the TacEleven-discovered tactics exhibit strong realism and tactical creativity, with 52.50 \% of the multi-step tactical alternatives rated adoptable in real-world elite football scenarios, highlighting the framework’s ability to rapidly generate numerous high-quality tactics for complex long-sequence open-play situations. TacEleven demonstrates the potential of creatively leveraging domain data and generative models to advance tactical analysis in sports.

\end{abstract}

\section{Introduction}

Offensive tactics are crucial to success in elite football, where the margin between victory and defeat is often determined by fleeting moments of attacking advantage. Data-driven analysis \cite{garcia2021game, jamil2021using, amichay2025characterizing} has been used to understand the subtle dynamics of tactical effectiveness, giving coaches and analysts retrospective insights.
Recent advances in artificial intelligence (AI) demonstrate a crucial shift in sports analytics \cite{kim2022soccercpd, wang2022individual, fernandez2020soccermap}, moving from passive decision-support toward automated tactical generation \cite{Pu2024, tuyls2021game}. A notable milestone in this evolution is DeepMind's TacticAI \cite{TacticAI}, which leverages deep learning \cite{DL} to propose optimized set pieces for corner kicks. \textcolor{teal}{However, existing methods remain constrained to static situations\cite{Goes2020TheTO, van2021would, van2021leaving, tuyls2021game}, leaving open-play scenarios that are pivotal to creating attacking advantages largely unexplored.}

\textcolor{teal}{Building on the progress in automated tactical generation, the next step lies in open-play tactic discovery.} 
Open-play tactics account for the vast majority of attacking sequences \cite{fernandez2018wide}, \textcolor{blue}{such as quick combination play in tight spaces, diagonal runs behind the defense, or switching play to exploit weak flanks.}
However, the dynamic and long-sequence nature of open play creates a vast combinatorial search space that surpasses the practical limits of human exploration, confining coaches and analysts to incremental variations of known tactics. This highlights the need for automated approaches capable of advancing tactical discovery and innovation.

\begin{figure}[h!tbp]
    \centering
    \includegraphics[width=\linewidth]{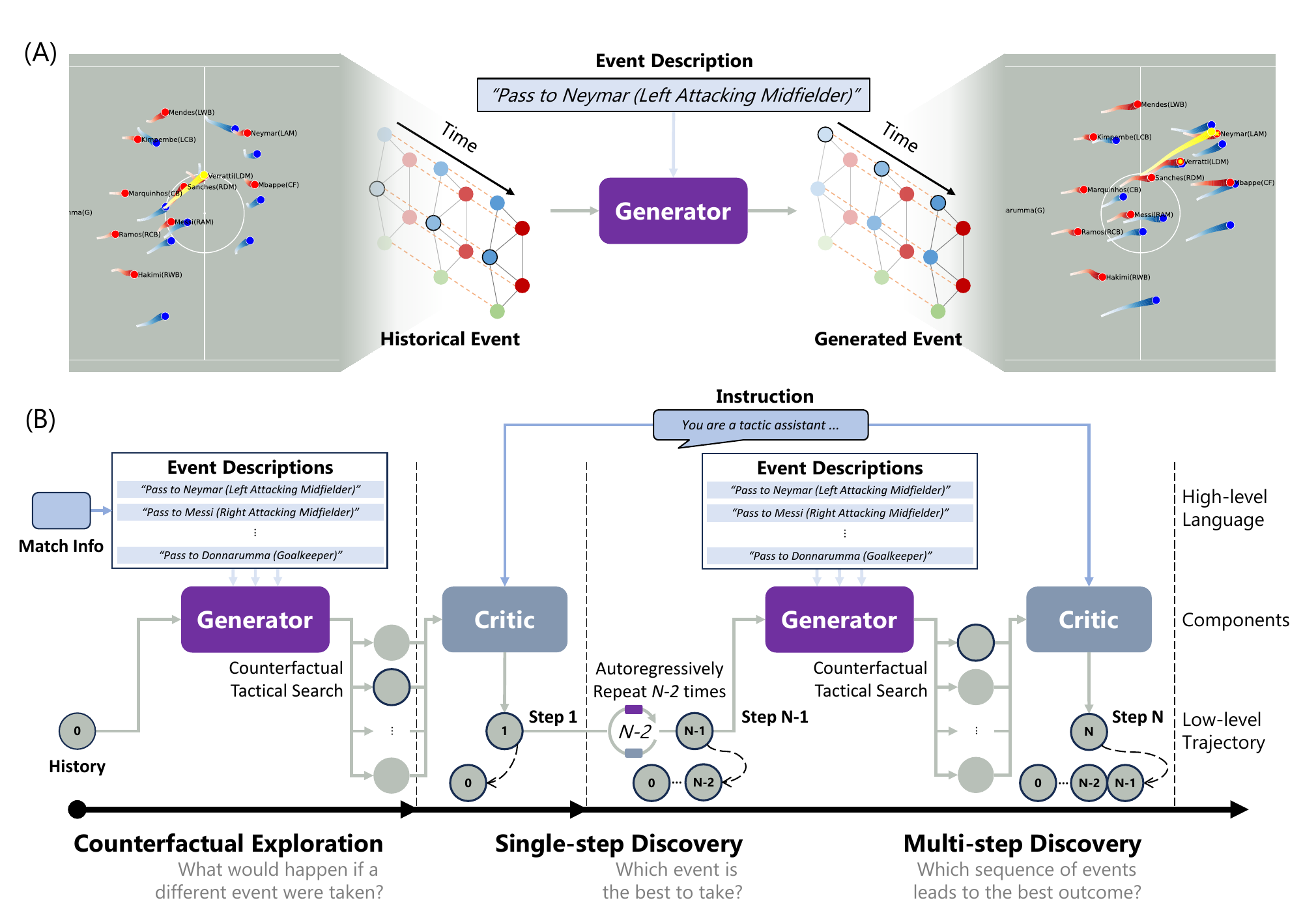}
    \caption{
        \textbf{An overview of TacEleven.} 
        \textbf{(A)}, 
        \textbf{An intuitive example of LTG.} The model takes an event description and a historical spatiotemporal graph as input and outputs a future spatiotemporal graph; both graphs are visualized as tactical sketches.
        In the tactical sketches, the attacking direction is from left to right, with red representing teammates, blue representing opponents, and yellow representing the ball. Trajectories are illustrated from light to dark, culminating at the circles. Players performing events are highlighted with yellow hexagons and their corresponding targets are highlighted with yellow crosses. In this example, the output of the generator shows that Verratti passes to Neymar. \textcolor{blue}{Note that minor abrupt bends in the trajectories are attributable to noise in the data.}
        \textbf{(B)}, 
        \textbf{\textcolor{blue}{The generator–critic framework} integrates three tasks with progressive tactical complexity: counterfactual exploration, single-step discovery and multi-step discovery.} In counterfactual exploration, the generator use counterfactual tactical search to produce tactical proposals aligned with high-level event descriptions. In single-step discovery, the critic selects the most effective proposals with a high-level instructions. Multi-step discovery selects proposals iteratively, with the generator producing successive steps in an autoregressive manner. Together, these tasks create a counterfactual tactical tree search, enabling long-sequence tactic generation.
        }
    \label{fig:overview}
\end{figure}

\textcolor{orange}{To address tactic discovery in open-play scenarios, we propose a framework called \textit{TacEleven}. Within this framework, a tactic is decomposed into a sequence of event-scale \textit{text-to-trajectory pairs}. Each pair comprises a event description (e.g., \textit{``Pass to Neymar (Left Attacking Midfielder)"}) and a corresponding low-level trajectory that specifies how the event is executed. Consequently, generating an open-play tactic involves constructing a coherent sequence, consisting of one or more text-to-trajectory pairs, \textcolor{blue}{informed by a historical context and aligned with a high-level \textit{tactical instruction}}. 
Based on this formulation, TacEleven is composed of two core components: \textcolor{blue}{a \textit{language-controlled tactical generator (generator, LTG)} and a \textit{multimodal large language model-based tactical critic (critic, MTC)}}, as illustrated in Fig.~\ref{fig:overview}.}

The generator is trained on a curated dataset of over one million text-to-trajectory pairs that we constructed by creatively aligning event data with tracking data. By leveraging a spatiotemporal graph model \cite{GMAN}, it learns the latent correspondence between varied textual descriptions and concrete full-field trajectories, thus enabling the generation of diverse tactical proposals, as shown in Fig.~\ref{fig:overview}(A). 
\textcolor{teal}{Subsequently, the critic evaluates the quality and feasibility of the tactical proposals using multimodal reasoning and selects the one that best aligns with the tactical instruction. This evaluation process enhances the explainability of the tactic discovery procedure.} To generate a long sequence consisting of multiple text-to-trajectory pairs, we adopt an autoregressive tactical tree search. Specifically, the trajectory from the selected tactical proposal serves as the historical context, and the generator subsequently produces the next tactical proposals, which are then re-evaluated by the critic. \textcolor{blue}{TacEleven, operating through a closed-loop interaction between the generator and the critic, constitutes a \textit{generator-critic} framework, as shown in Fig.~\ref{fig:overview}(B). The methodology section formalizes the spatiotemporal graph model and the generator-critic framework.
}

TacEleven executes across three tasks with progressive tactical complexity: \textit{counterfactual exploration}, \textit{single-step discovery}, and \textit{multi-step discovery}. 
In the counterfactual exploration task, we provide variant event descriptions as counterfactual language conditions to the generator, which simulates and produces corresponding alternative tactical proposals. In the single-step discovery task, the critic leverages its multimodal reasoning ability to select the optimal proposal from the generated candidates, conditioned on a high-level tactical instruction, allowing for immediate tactical adjustments based on real-time inputs. In the multi-step discovery task, the generator–critic loop is executed in an autoregressive manner, enabling TacEleven to construct coherent long-sequence tactics by decomposing tactics into sequential events. Overall, users only need to provide a high-level instruction and a historical trajectory, and TacEleven can automatically discover open-play tactics.



\textcolor{blue}{
In the next section, we demonstrate the effectiveness of TacEleven through improvements in performance metrics such as Expected Goals (xG) \cite{xG}, Expected Threat (xT) \cite{xT} and Pitch Control (PC) \cite{PC} for each task. Case studies further illustrate the explainability of TacEleven. Moreover, the realism and effectiveness of the discovered tactics are statistically validated through questionnaires administered to domain experts, with evaluation criteria explicitly mapped to the progressive tasks. Through these tasks, TacEleven can be used to discover long-sequence open-play football tactics through an intuitive, language-based interaction paradigm. Coaches and analysts can explore improved tactics at any point during a match simply by prompting TacEleven, and different preferences expressed in the prompt lead to stylistically distinct tactics.
}

\section{Result}

\textcolor{orange}{To evaluate the effectiveness and explainability of TacEleven, we design three tasks with progressive tactical complexity: counterfactual exploration (CF), single-step discovery (SS), and multi-step discovery (MS). Each task combines both qualitative and quantitative analyses to ensure reliable and interpretable evaluation.}

\textcolor{orange}{The counterfactual exploration task assesses the LTG by providing different counterfactual event descriptions to test whether it can generate corresponding trajectories. Scaling curves are plotted to illustrate how its performance scales with dataset size and model capacity (Sec.~\ref{subsec:cf_gen}). The single-step discovery task quantitatively measures the improvement of discovered single-step tactics, and visualizes representative scenarios to demonstrate explainability (Sec.~\ref{sec:single-step}). The multi-step discovery task evaluates TacEleven’s ability to produce long-sequence tactics through the same quantitative metrics and similar qualitative analysis (Sec.~\ref{sec:multi-step}). Each section briefly describes TacEleven for clarity. We refer readers to Sec.~\ref{sec:method} for details.}

\textcolor{orange}{In addition, to verify the realism and effectiveness of the discovered tactics, we conducted a questionnaire involving 36 participants, including coaches and analysts from AJ Auxerre\footnote{\url{https://www.aja.fr}}. Statistical analysis of the responses validates both the realism and the effectiveness of the tactics discovered by TacEleven (Sec.~\ref{sec:questionnaire}).}

\subsection{Counterfactual exploration}
\label{subsec:cf_gen}
\begin{figure*}[htbp]
  \centering

  \begin{subfigure}[t]{.24\textwidth}\centering
    \includegraphics[width=\linewidth]{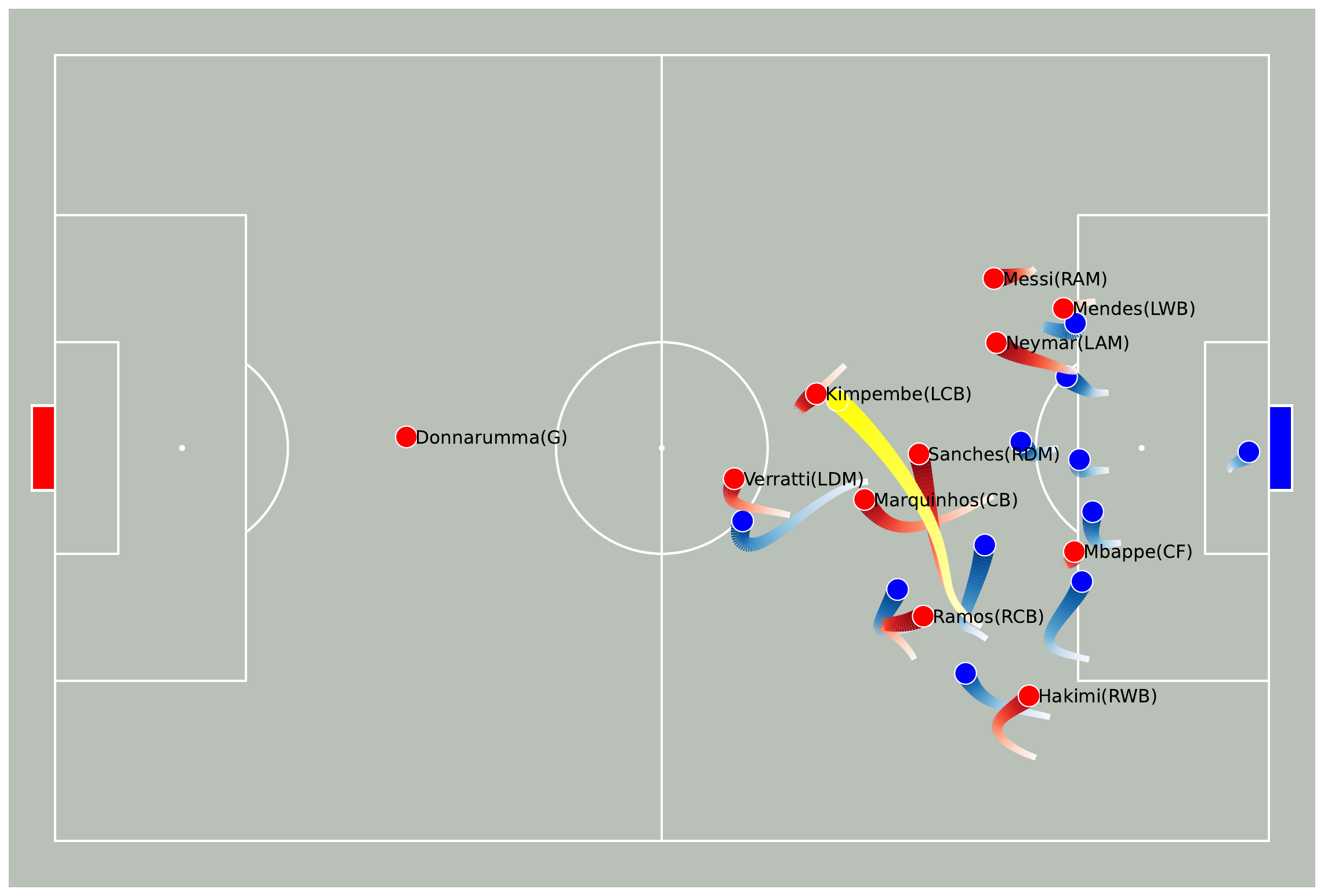}
    \subcaption{History}
    \label{subfig:CF:a}
  \end{subfigure}\hfill
  \begin{subfigure}[t]{.24\textwidth}\centering
    \includegraphics[width=\linewidth]{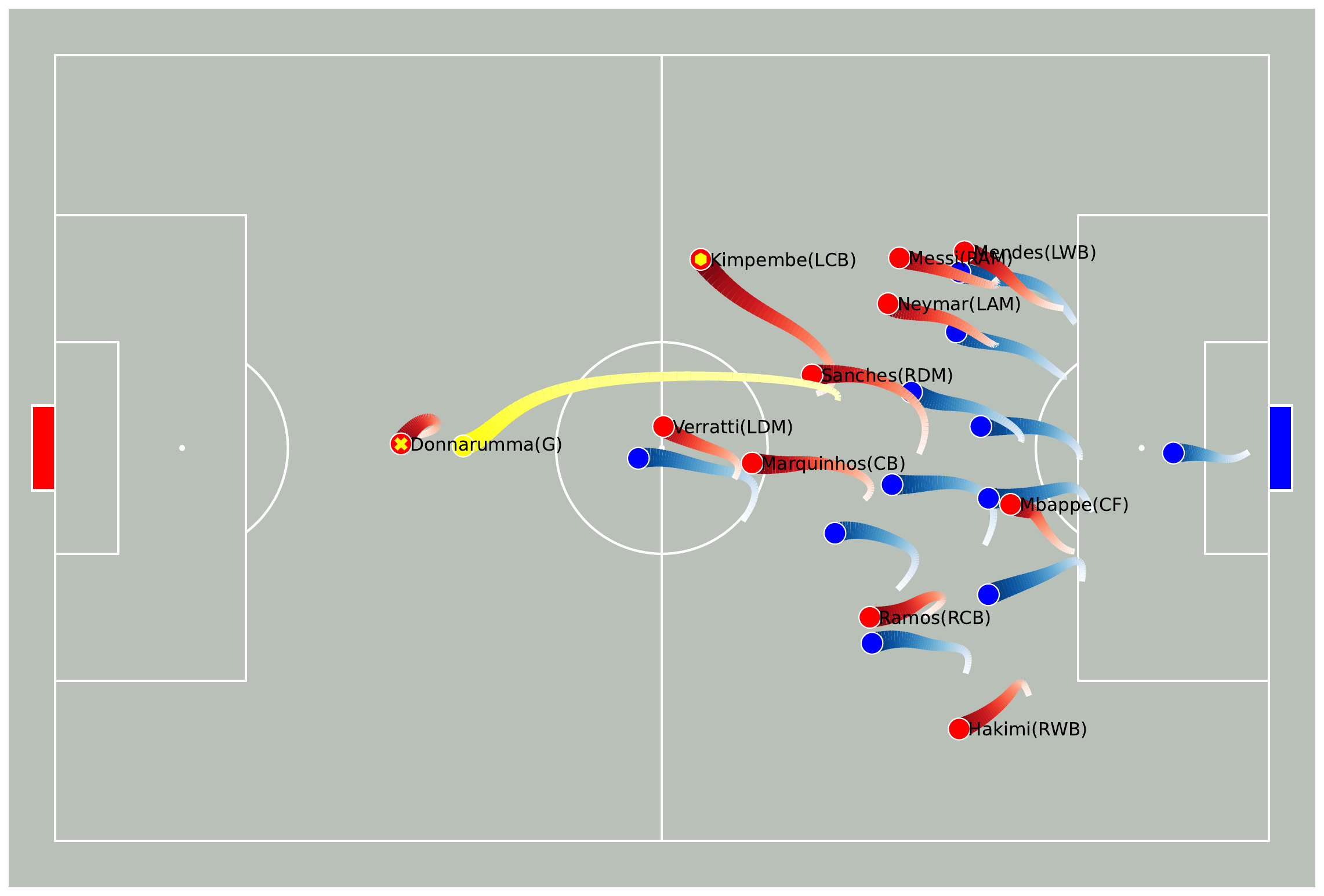}
    \subcaption{Goalkeeper}
  \end{subfigure}\hfill
  \begin{subfigure}[t]{.24\textwidth}\centering
    \includegraphics[width=\linewidth]{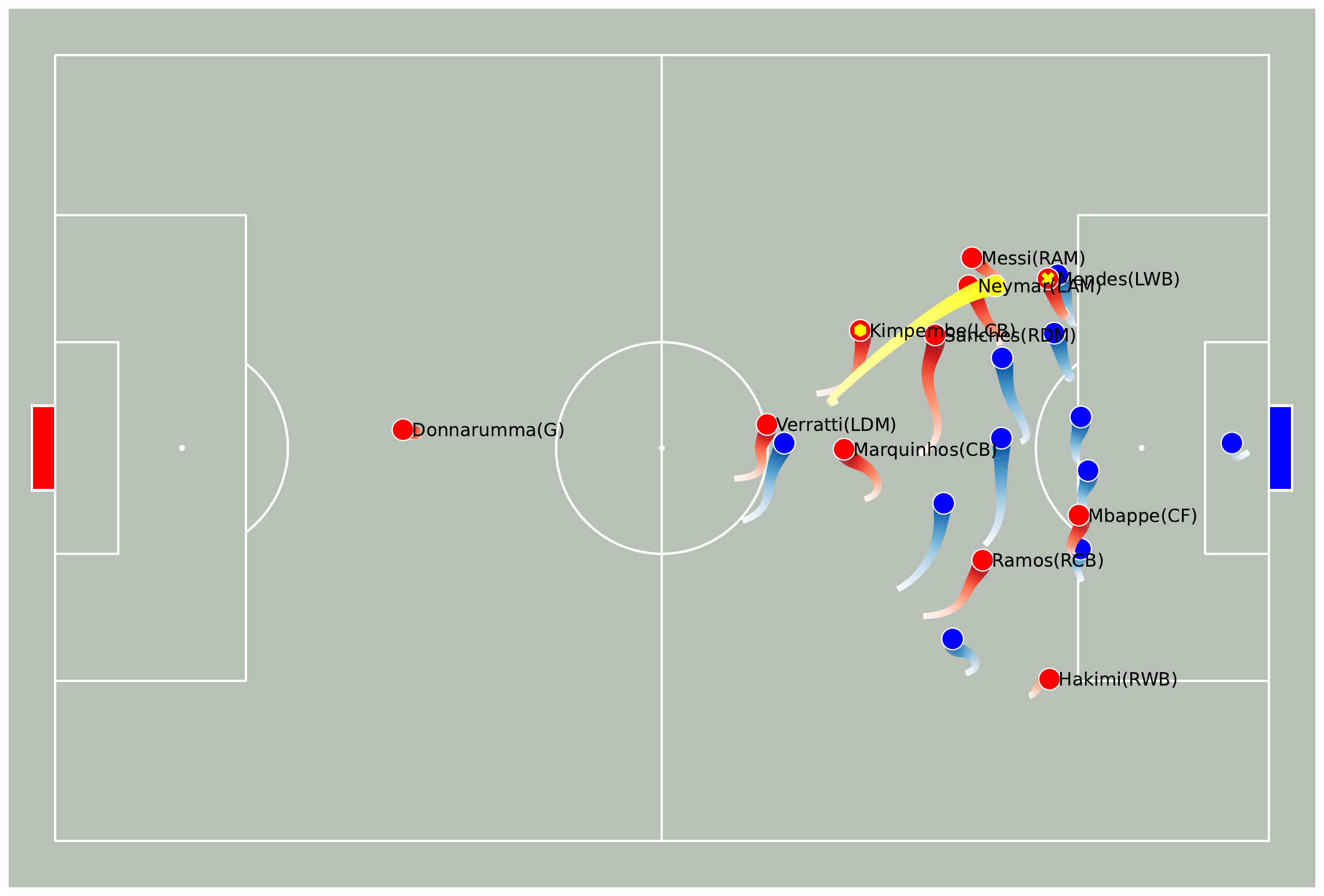}
    \subcaption{Left Wing Back}
  \end{subfigure}\hfill
  \begin{subfigure}[t]{.24\textwidth}\centering
    \includegraphics[width=\linewidth]{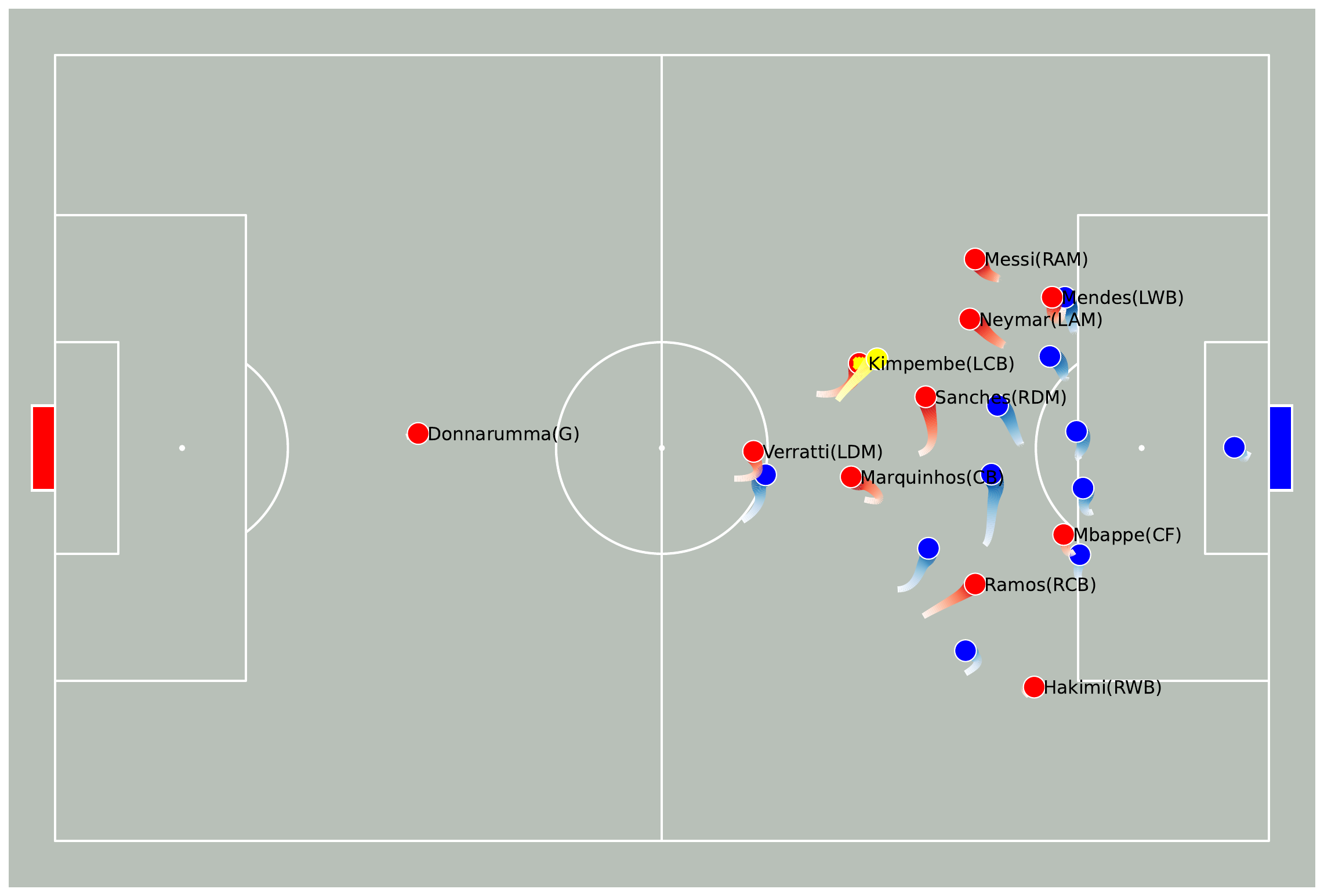}
    \subcaption{Left Center Back (Carrying)}
  \end{subfigure}

  \vspace{6pt}
  \begin{subfigure}[t]{.24\textwidth}\centering
    \includegraphics[width=\linewidth]{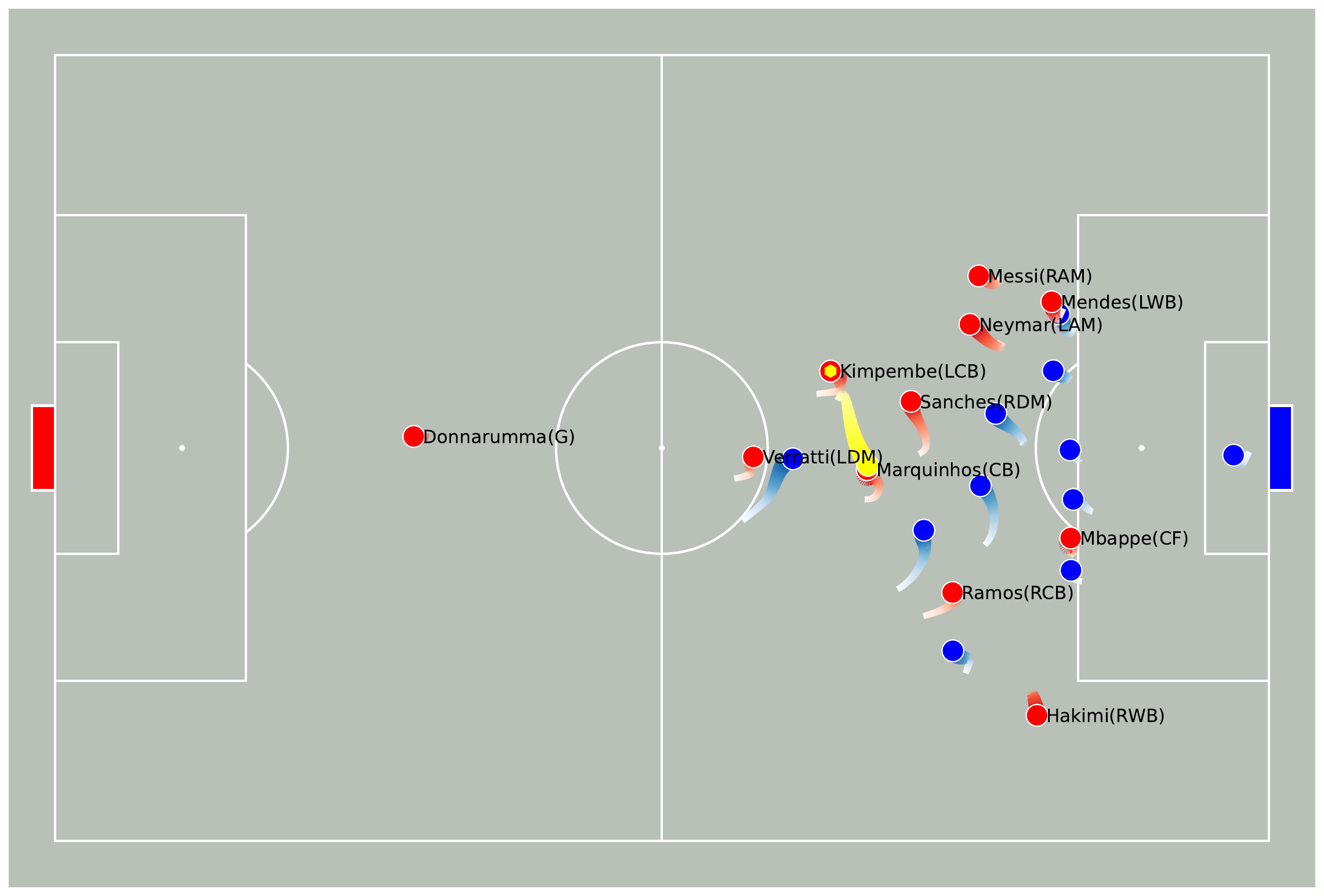}
    \subcaption{Center Back}
  \end{subfigure}
  \begin{subfigure}[t]{.24\textwidth}\centering
    \includegraphics[width=\linewidth]{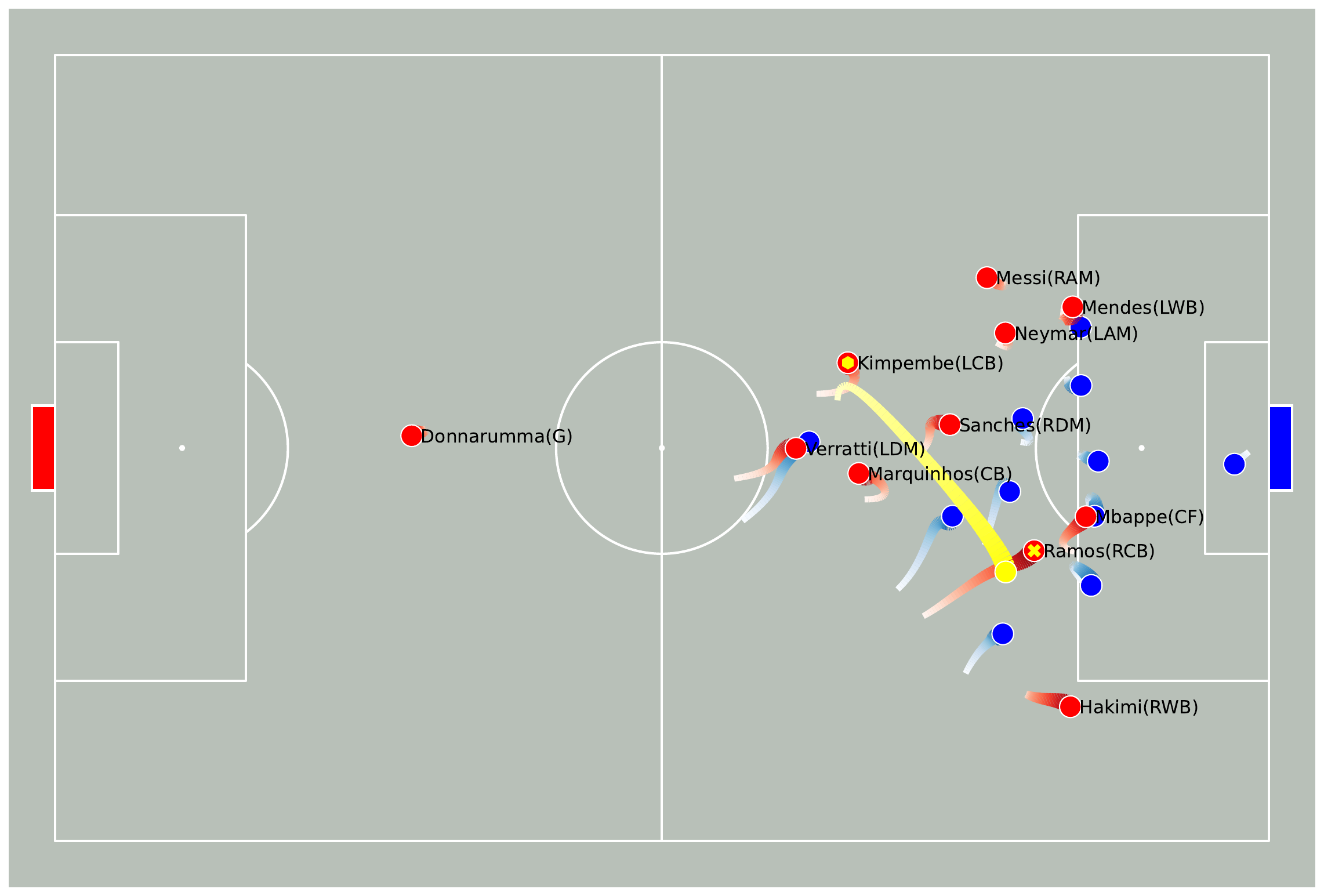}
    \subcaption{Right Center Back}
  \end{subfigure}\hfill
  \begin{subfigure}[t]{.24\textwidth}\centering
    \includegraphics[width=\linewidth]{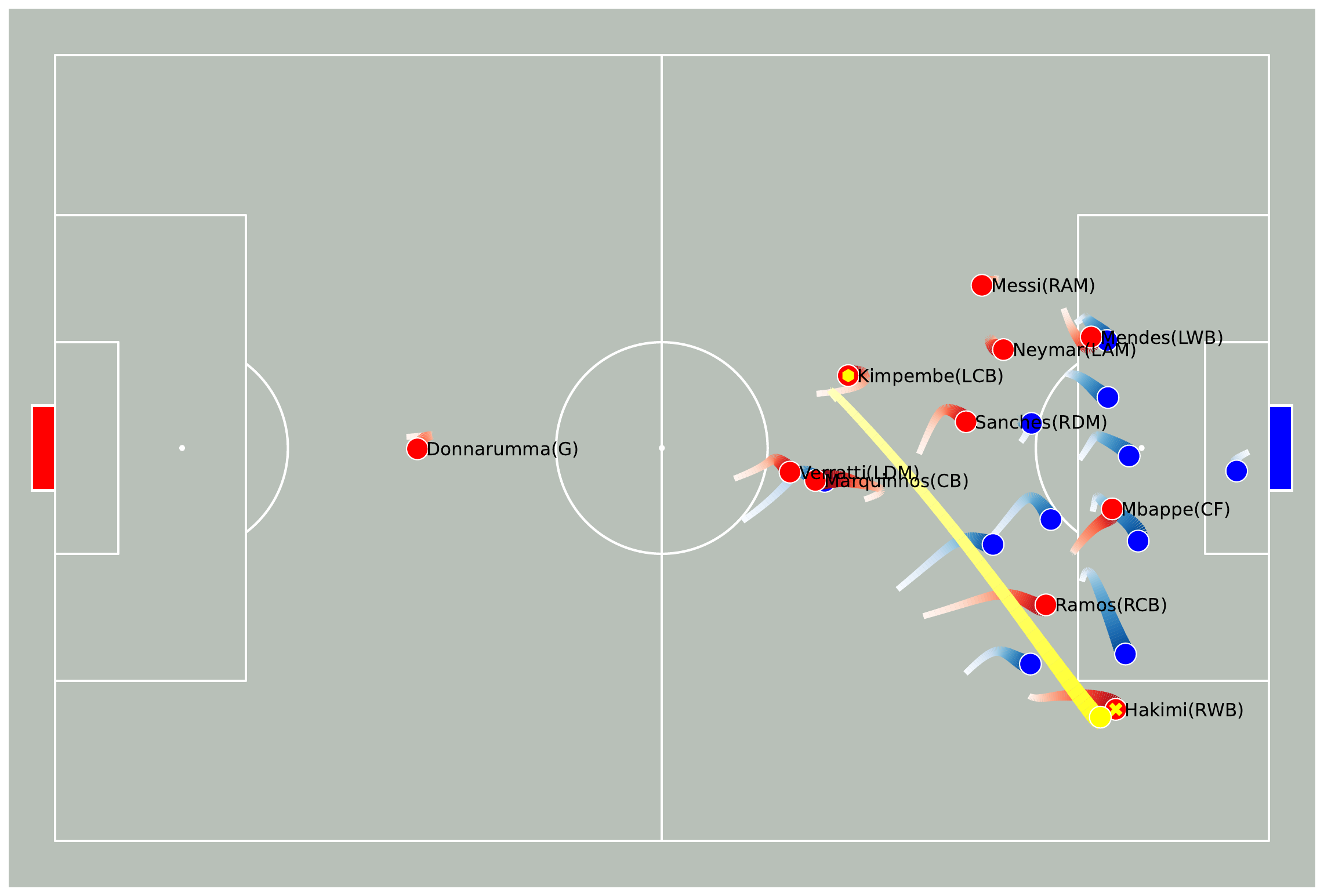}
    \subcaption{Right Wing Back}
  \end{subfigure}\hfill
  \begin{subfigure}[t]{.24\textwidth}\centering
    \includegraphics[width=\linewidth]{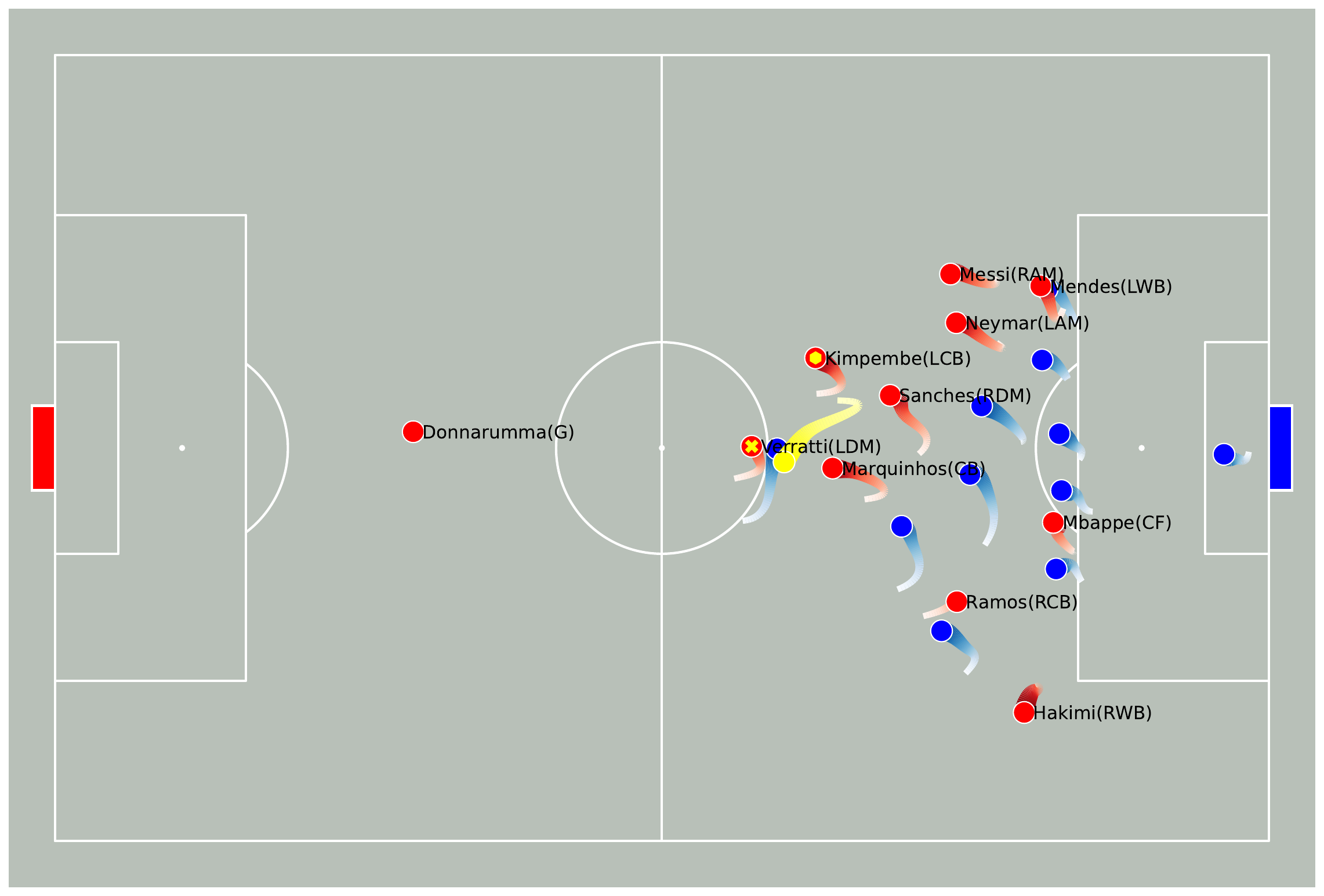}
    \subcaption{Left Defensive Midfielder}
    \label{subfig:unsuc}
  \end{subfigure}

  \vspace{6pt}
  \begin{subfigure}[t]{.24\textwidth}\centering
    \includegraphics[width=\linewidth]{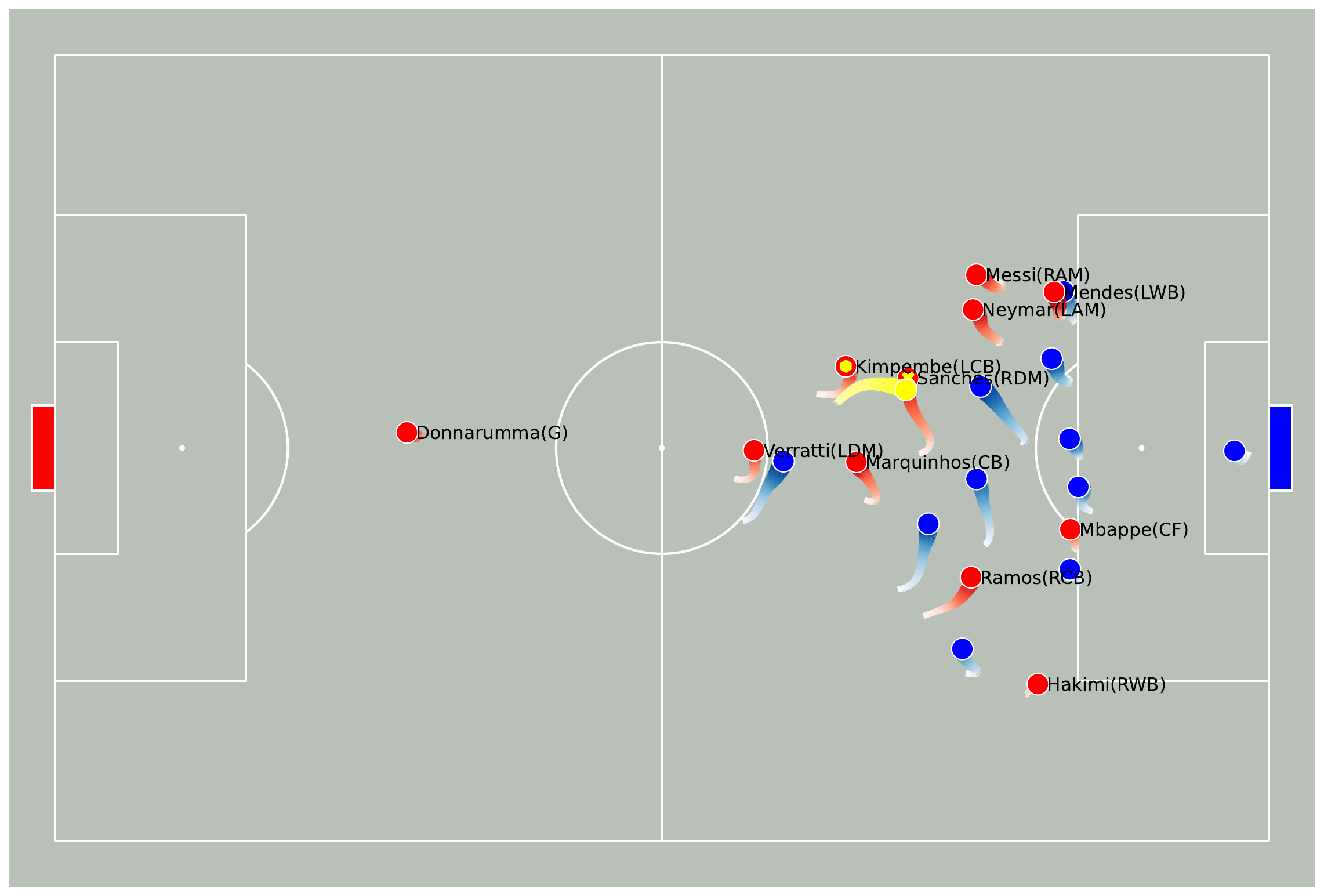}
    \subcaption{Right Defensive Midfielder}
  \end{subfigure}\hfill
  \begin{subfigure}[t]{.24\textwidth}\centering
    \includegraphics[width=\linewidth]{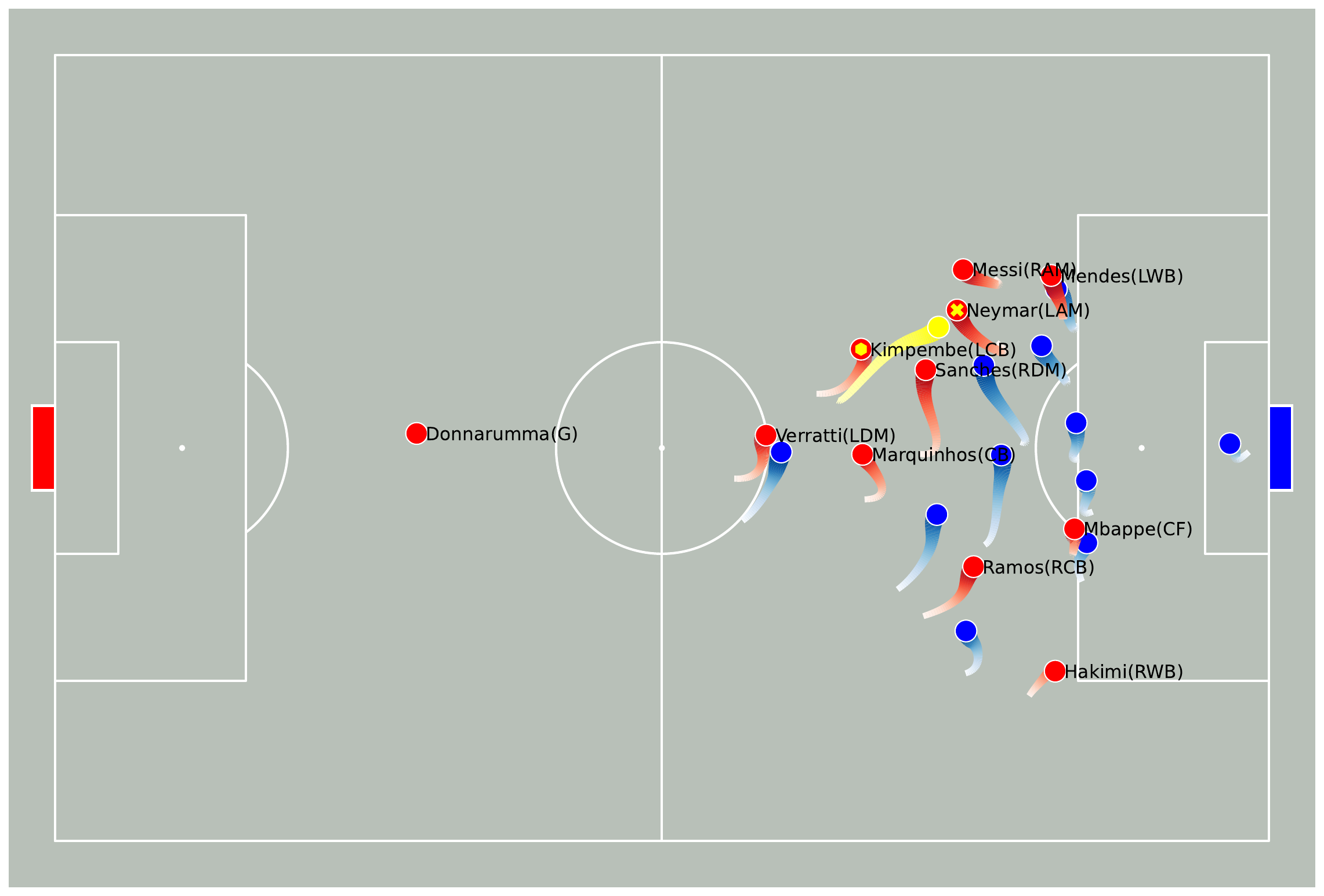}
    \subcaption{Left Attacking Midfielder}
  \end{subfigure}\hfill
  \begin{subfigure}[t]{.24\textwidth}\centering
    \includegraphics[width=\linewidth]{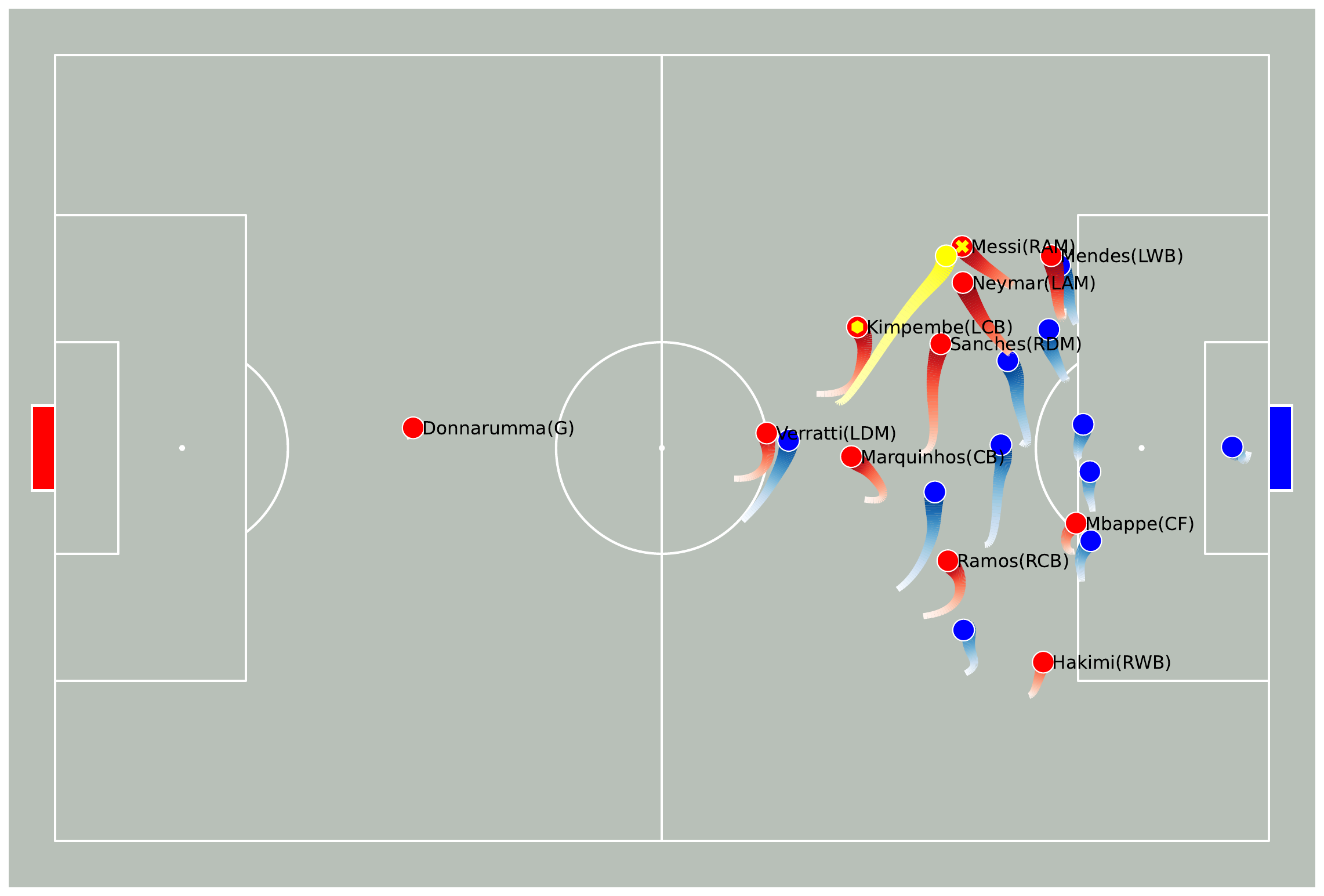}
    \subcaption{Right Attacking Midfield}
  \end{subfigure}\hfill
   \begin{subfigure}[t]{.24\textwidth}\centering
    \includegraphics[width=\linewidth]{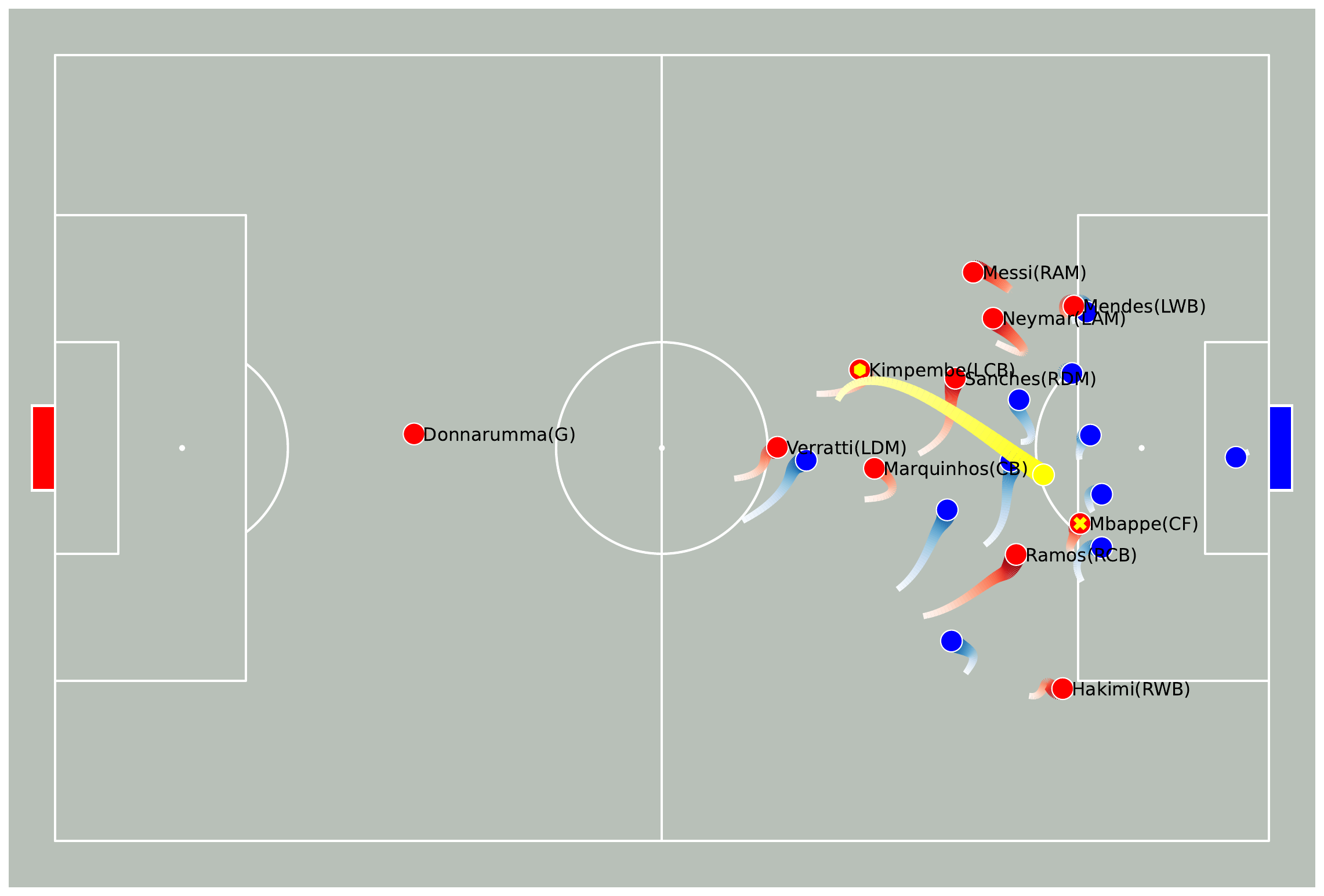}
    \subcaption{Center Forward}
    \label{subfig:CF}
  \end{subfigure}
  

  \caption{\textbf{Different counterfactual full-filed trajectories generated by LTG.} (a) shows the historical trajectory used as part of the LTG input. (b)–(l) present the generated tactical proposals given different event descriptions under the same historical trajectory. The subcaptions of (b) to (l) indicate either the intended pass target or the player carrying the ball for each event.}
  \label{fig:roles}
\end{figure*}

\textbf{Qualitative analysis.} The LTG supports generating counterfactual football event trajectories following event descriptions. Fig.~\ref{fig:roles} illustrates the trajectories generated by LTG under eleven different descriptions within the same historical context (Fig.~\ref{subfig:CF:a}). The enumerated event descriptions include ten variations of passes to different teammates and one description corresponding to a carry event. This demonstrates the ability of the LTG to generate compelling trajectories conditioned on event descriptions, enabling counterfactual tactic exploration simply by modifying the event description. In addition, typical movement patterns can be observed, such as the center-forward’s hook-shaped run to avoid being offside (Fig.~\ref{subfig:CF}). Some passes that failed due to interception by opponents were simultaneously predicted, as shown in Figure \ref{subfig:unsuc}.

\textbf{Quantitative analysis.} \textcolor{orange}{The LTG is a spatiotemporal graph-based generative model trained on a curated dataset. To evaluate how its performance scales with dataset size and model size, we test the LTG on a test dataset include 322,877 text-to-trajectory pairs, and design two tailored metrics: \textit{counterfactual alignment error}, and \textit{factual trajectory error}, which measure counterfactual consistency and factual accuracy of the generation respectively.
\textcolor{blue}{The former quantifies the endpoint alignment between the ball trajectories and the corresponding receiver trajectories derived from counterfactual event descriptions, with all trajectories produced in parallel by the LTG.} 
The latter computes the mean squared error between the generated trajectory and the ground-truth trajectory when the ground-truth event description is provided to the LTG. The metrics are detailed in Appendix \ref{appendix:eval}. In summary, the former reflects the generalization ability to counterfactual event descriptions, while the latter captures its precision in trajectory generation.}

\begin{figure}[tbp]
    \centering
    \includegraphics[width=0.7\linewidth]{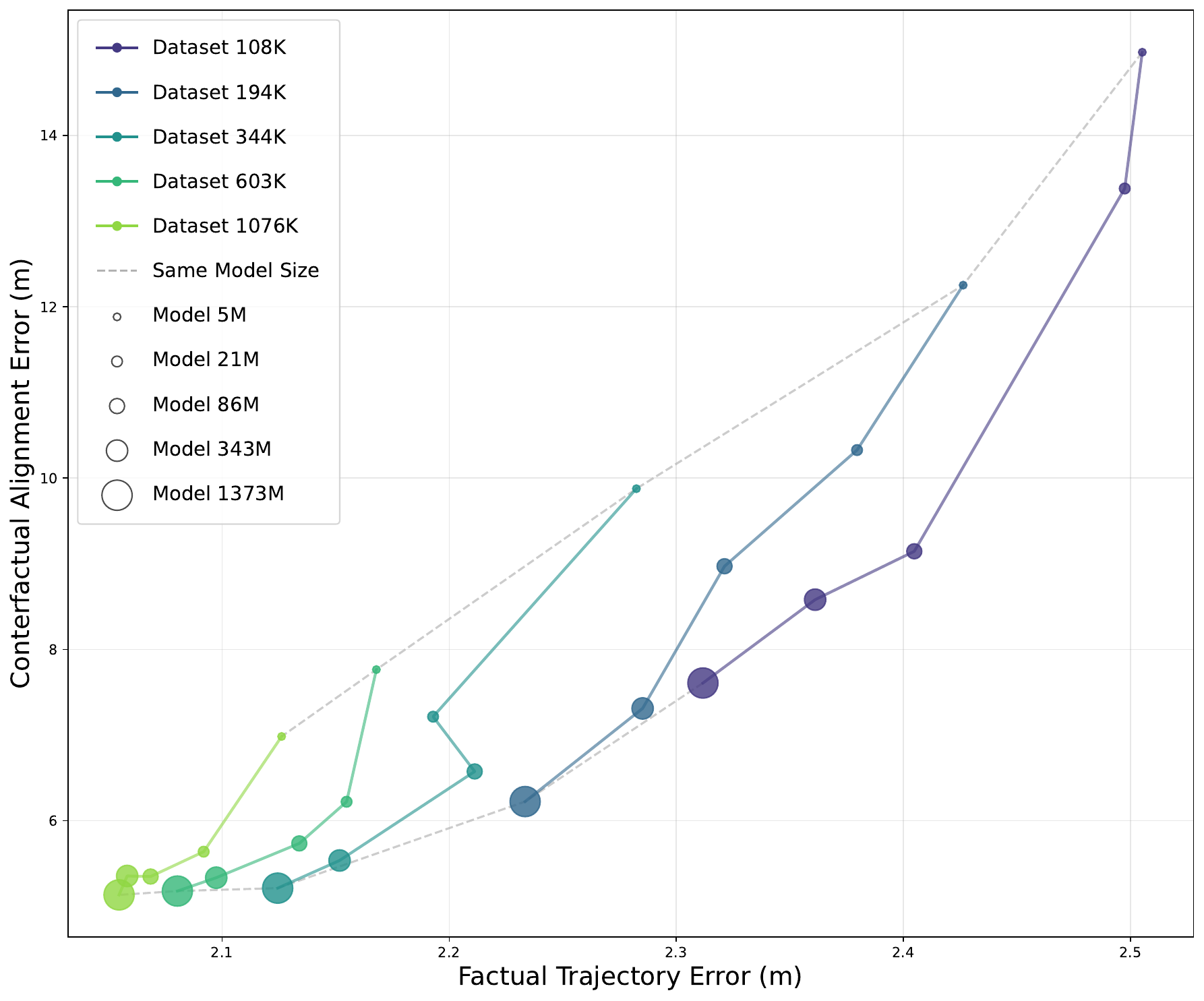}
    \caption{
        \textbf{Generalizability of different scaling models and datasets.} 
        \textcolor{blue}{Factual Trajectory Error (m) and Counterfactual Alignment Error (m) are shown on the horizontal and vertical axes, respectively. Solid lines connect models trained on datasets of different sizes, color-coded from blue (108K) to green (1076K). Marker size indicates model scale, ranging from 5M to 1,373M parameters. Dashed grey lines connect models of identical size, highlighting the influence of data scale under fixed model capacity.}
    }
    \label{fig:scaling}
\end{figure}

We scales dataset size from 108K to 1,076K, and scales model size with parameters from 5M to 1,373M to see how the two metrics change accordingly (Fig.~\ref{fig:scaling}). Overall, both the counterfactual alignment error and the factual trajectory error decrease rapidly as the dataset and model size increase, \textcolor{blue}{eventually converging to highly precise final errors of 5 m and 2 m, respectively.} Larger datasets also substantially narrow the performance gap between small (5M) and large (1,373M) models, suggesting that competitive performance can be achieved with a much smaller model when sufficient data are available. Interestingly, for the 1,373M model, the counterfactual alignment error decreases with increasing dataset size and then stabilizes. This behavior occurs mainly because pass events often involve lead passes, where the ball is intentionally passed ahead of the receiver to allow him to reach it in stride, and the model learns to capture this pattern effectively. In conclusion, the LTG demonstrates promising scaling capabilities, maintaining robust generalization performance with increased model size and dataset size, even under more diverse and generalized description settings in the future research. We find that at our dataset scale, the 1,373M model achieves decent performance while maintaining reasonable inference speed, making it a favorable choice. All results in following experiments are conducted using the LTG with 1,373M parameters trained on the 1,076K dataset.

In Table~\ref{tab:eval_xG_xT_etc}, we report the quantitative evaluation of the improvement of proposed tactics using four position-based metrics: consistency ($C$), xG, xT, and PC (detailed in Appendix~\ref{appendix:eval}). These metrics correspond to the robustness of the language condition, the on-ball value, and the off-ball value, respectively. The consistency metric $C$ indicates whether the player specified in the event description belongs to the top-k player–ball endpoint distance set in the counterfactual scenario, where the distance is calculated as the counterfactual alignment error. Each $\Delta$ value represents the change from the starting to the terminal position, where positive increments in $xT(A)$ or $PC(A)$ indicate higher attacking potential and spatial dominance, while negative values in $xT(D)$ or $PC(D)$ reflect stronger defensive suppression of the opponent’s threat. 
\textcolor{orange}{In the CF task, we use the LTG to generate tactical proposals on the test dataset by randomly selecting event descriptions, denoted as Random, and compare its metrics with the ground-truth results, denoted as Factual. 
The Factual achieves a $C$ of 96\%, while the Random reaches 84\% with out-of-distribution event descriptions, demonstrating relatively strong generalization ability. For the other metrics, the differences between the Factual and the Random are minor, indicating that LTG, having been trained on expert data, is capable of generating high-quality tactics even under random event conditions.
}

\begin{table}[tbp]
\centering

\caption{\textbf{Evaluation results on different methods.} 
In the “Task” column, “CF”, “SS”, and “MS” denote the counterfactual exploration, single-step discovery, and multi-step discovery tasks, respectively. The “Configuration” column lists the method type, positional categories, and instruction style.
“Random” denotes a method that randomly selects from generated proposals, whereas “Factual” denotes selecting the ground-truth results in the “CF” task. “Entire”, “Back”, “Middle”, and “Front” respectively denote evaluations conducted on the entire team, back-field, middle-field, and front-field players in the “SS” task. “Aggressive”, “Conservative”, and “Neutral” denote the instruction styles adopted by the MTC in the “MS” task.
$C$ measures the consistency between counterfactual outcomes and language descriptions, $xG$ estimates the goal probability, $xT$ quantifies the threat, and $PC$ reflects the pitch control. $\Delta$ denotes the change from the initial to the terminal position. $A$ and $D$ denotes the attacking and the defending team.
}
\label{tab:eval_xG_xT_etc}
    \begin{tabular}{clrrrrrr}
    \toprule
     Task & Configuration & $C\uparrow$  &  $\Delta xG\uparrow$  &  $\Delta xT(A)\uparrow$  &  $\Delta xT(D)\downarrow$  &  $\Delta PC(A)\uparrow$  &  $\Delta PC(D)\downarrow$ \\
    \midrule
    \multirow{2}{*}{CF}    & Factual       & $\mathbf{96.0 \%}$ &  $ +0.000$ &  $+0.055$ & $-0.054$ &  $+0.002$ & $-0.002$ \\
                           & Random      & $84.0 \%$ &  $-0.008$ &  $+0.066$ & $-0.047$ &  $-0.004$ & $+0.004$ \\
    \midrule
    \multirow{4}{*}{SS} & Entire & $92.0 \%$ &  $\mathbf{+0.016}$ &  $\mathbf{+0.131}$ & $\mathbf{-0.099}$ &  $\mathbf{+0.014}$ & $\mathbf{-0.014}$ \\
                        & Back   & $\mathbf{93.3 \%}$ & $+0.009$ & $+0.131$ & $-0.008$ &  $+0.010$ & $-0.010$ \\
                        & Middle & $93.3 \%$ &  $\mathbf{+0.033}$ &  $+0.101$ & $-0.131$ &  $+0.016$ & $-0.016$ \\
                        & Front  & $87.5 \%$ &  $+0.010$ & $\mathbf{+0.319}$ & $\mathbf{-0.133}$ &  $\mathbf{+0.033}$ & $\mathbf{-0.033}$ \\
    \midrule
    \multirow{3}{*}{MS} & Aggressive   & $89.2 \%$ & $\mathbf{+0.017}$ & $+0.219$ & $-0.118$ &  $\mathbf{+0.024}$ & $\mathbf{-0.024}$ \\
                        & Conservative   & $88.7 \%$ & $-0.003$ & $-0.004$ & $-0.040$ &  $+0.001$ & $-0.001$ \\
                        & Neutral   & $\mathbf{90.6 \%}$ & $+0.014$ & $\mathbf{+0.237}$ & $\mathbf{-0.119}$ &  $+0.022$ & $-0.022$ \\
    \bottomrule
    \end{tabular}
\end{table}

\subsection{Single-step discovery}
\label{sec:single-step}
\newcolumntype{Y}{>{\centering\arraybackslash}X}  
\newcolumntype{Z}{>{\raggedright\arraybackslash}X} 

\begin{figure*}[htbp!]
    \centering
    \begin{tabularx}{\textwidth}{cYYY} 
    \toprule
     & \textbf{Historical} & \textbf{Factual} & \textbf{Discovered} \\
    \midrule
    
    \multirow{2}{*}{{\textbf{1}}} &
    \includegraphics[width=\linewidth]{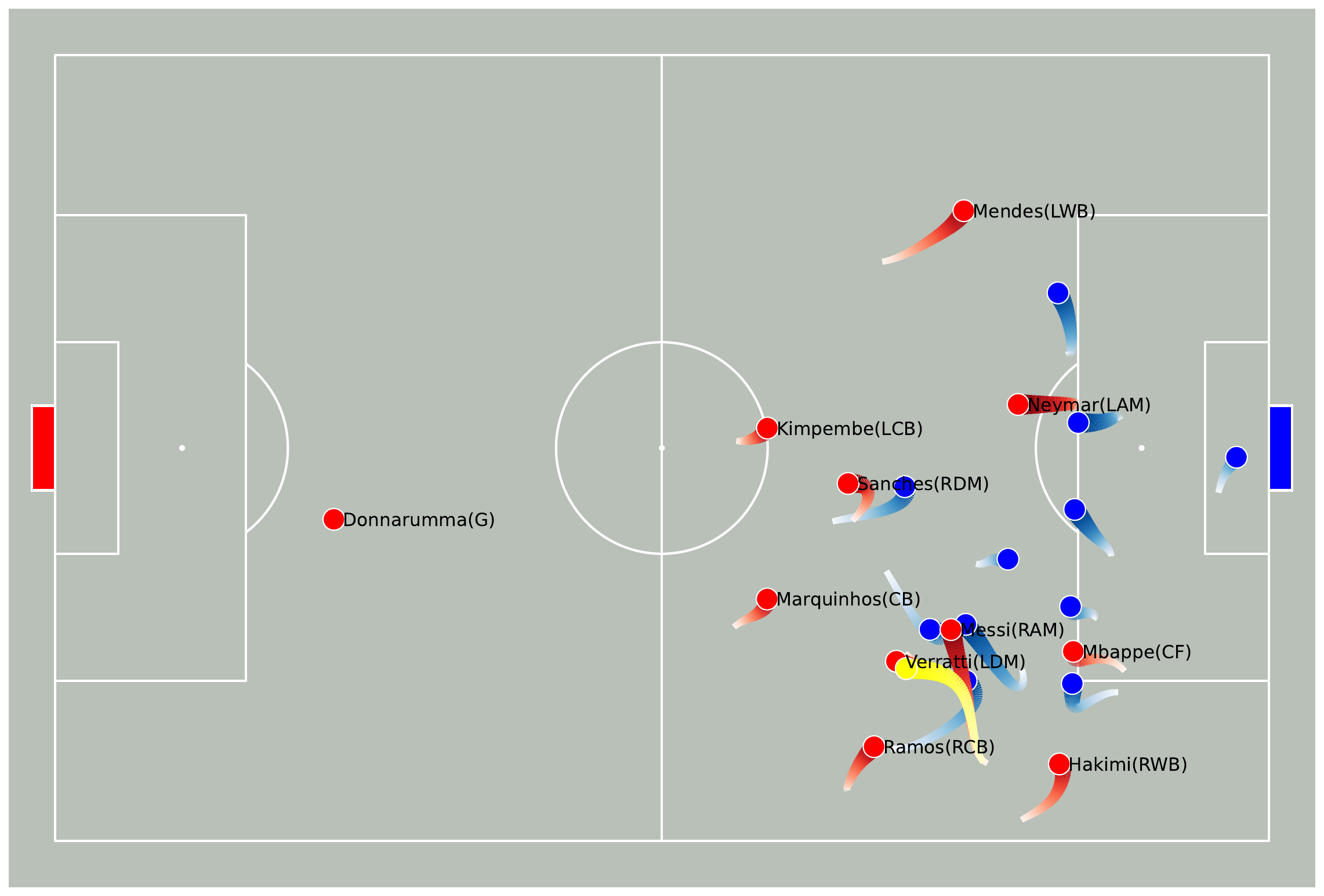} &
    \includegraphics[width=\linewidth]{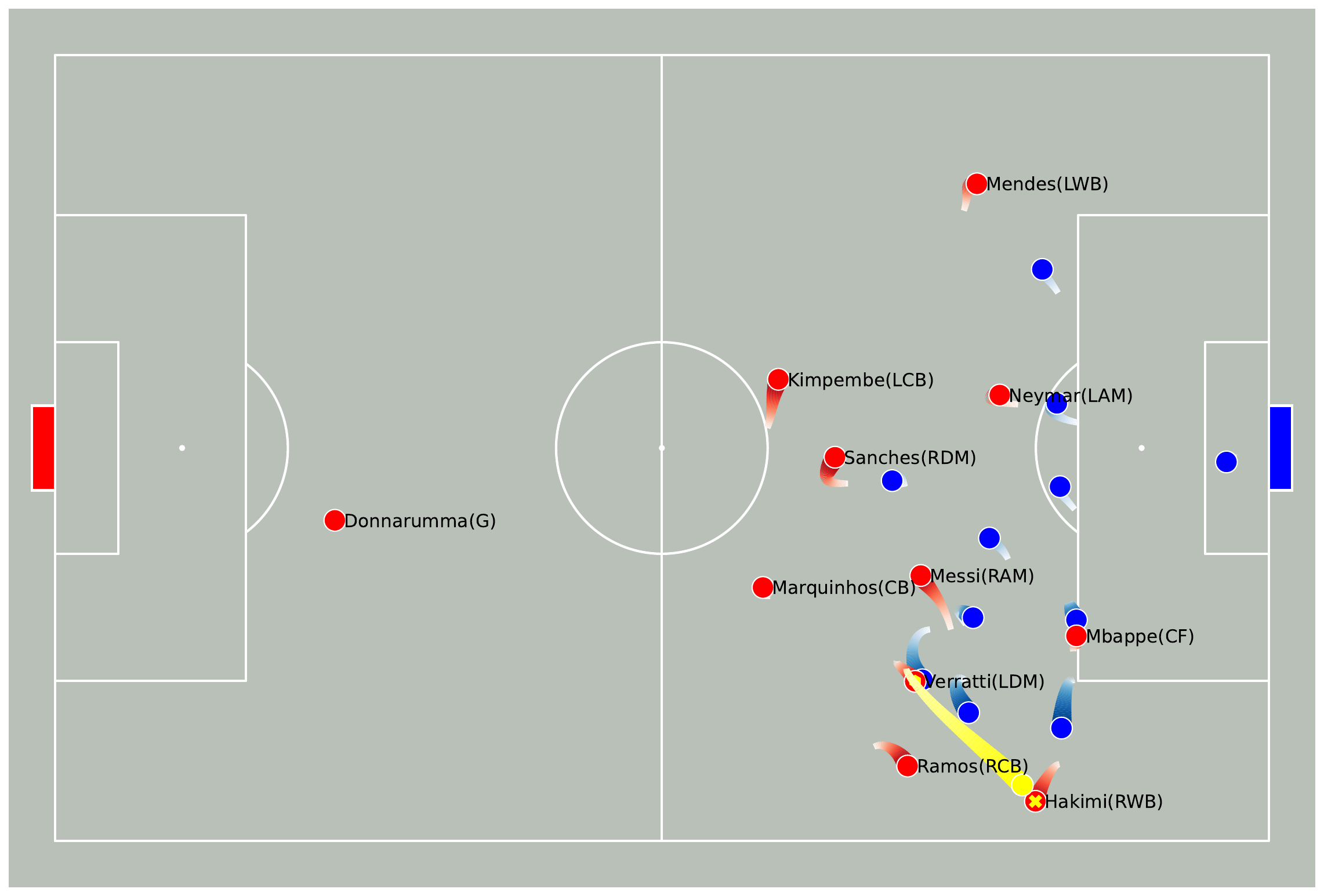} &
    \includegraphics[width=\linewidth]{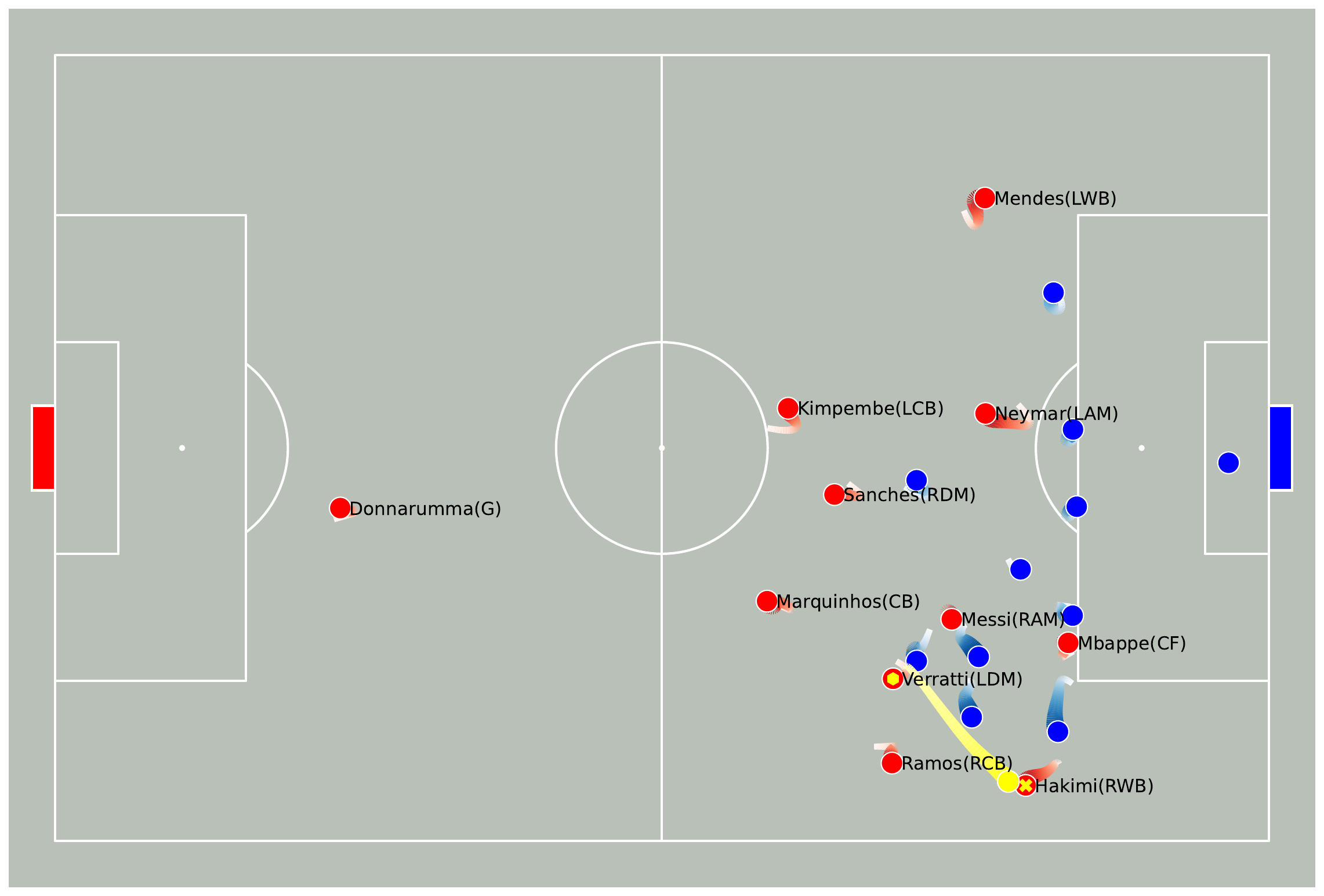} \\
    & \multicolumn{3}{c}{%
        \begin{tcolorbox}[colback=gray!10, colframe=gray!40, boxrule=0.3pt, width=0.9\linewidth]
        \tiny \textbf{MLLM reasoning}: 
        The pass to Achraf A. Hakimi is optimal. It balances risk and reward by leveraging Hakimi’s pace and positioning to exploit Monaco’s defensive width. The moderate risk is mitigated by Hakimi’s reliability, while the high scoring advantage aligns with PSG’s aggressive style. Other options either lack creativity (center-back passes) or carry excessive risk (direct passes to forwards). Hakimi’s role as a wing-back makes him the most logical recipient for a progressive pass, ensuring tactical continuity and maximizing spatial exploitation.
        \end{tcolorbox}
    } \\[2pt]

    \midrule
    
    \multirow{2}{*}{{\textbf{2}}} &
    \includegraphics[width=\linewidth]{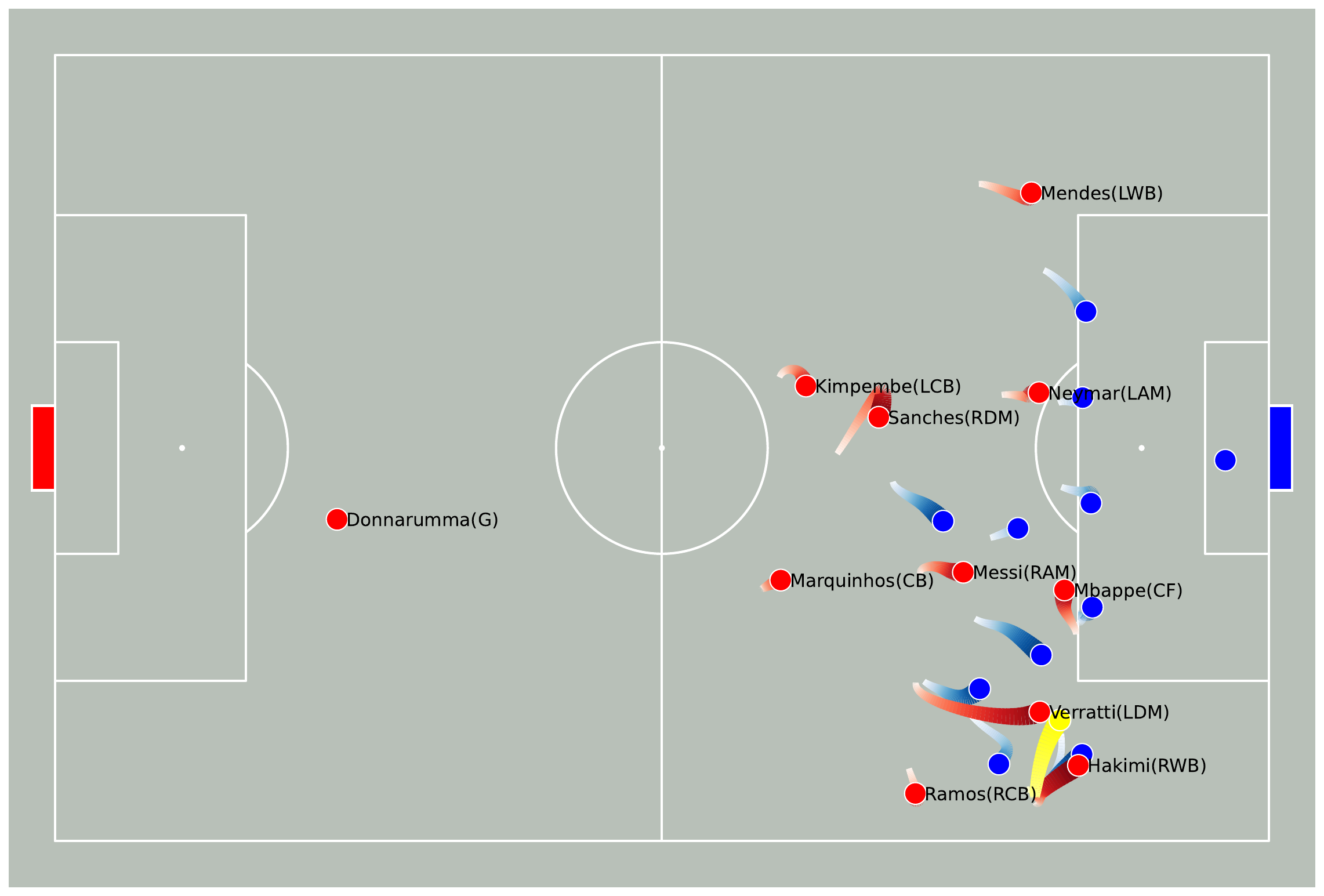} &
    \includegraphics[width=\linewidth]{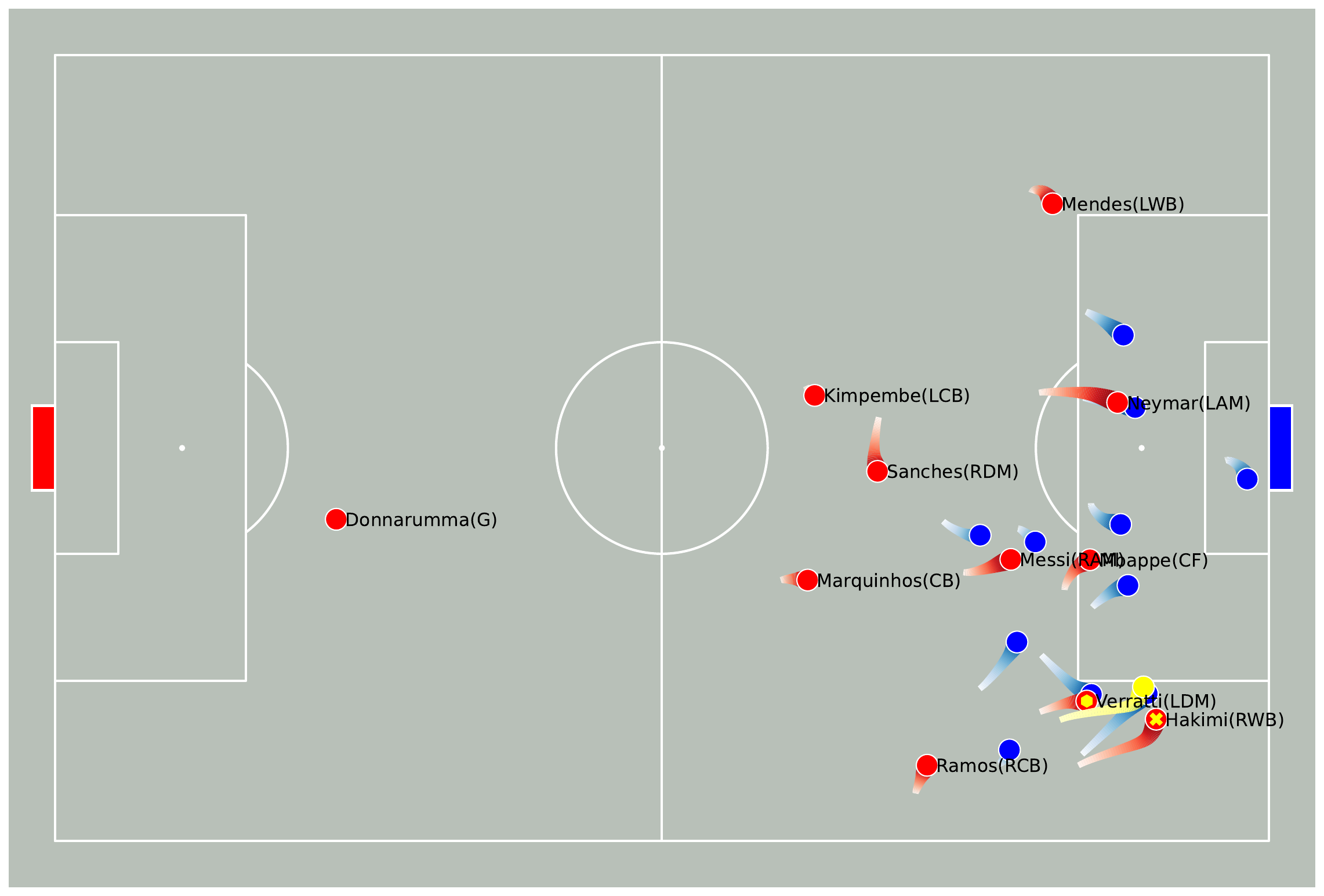} &
    \includegraphics[width=\linewidth]{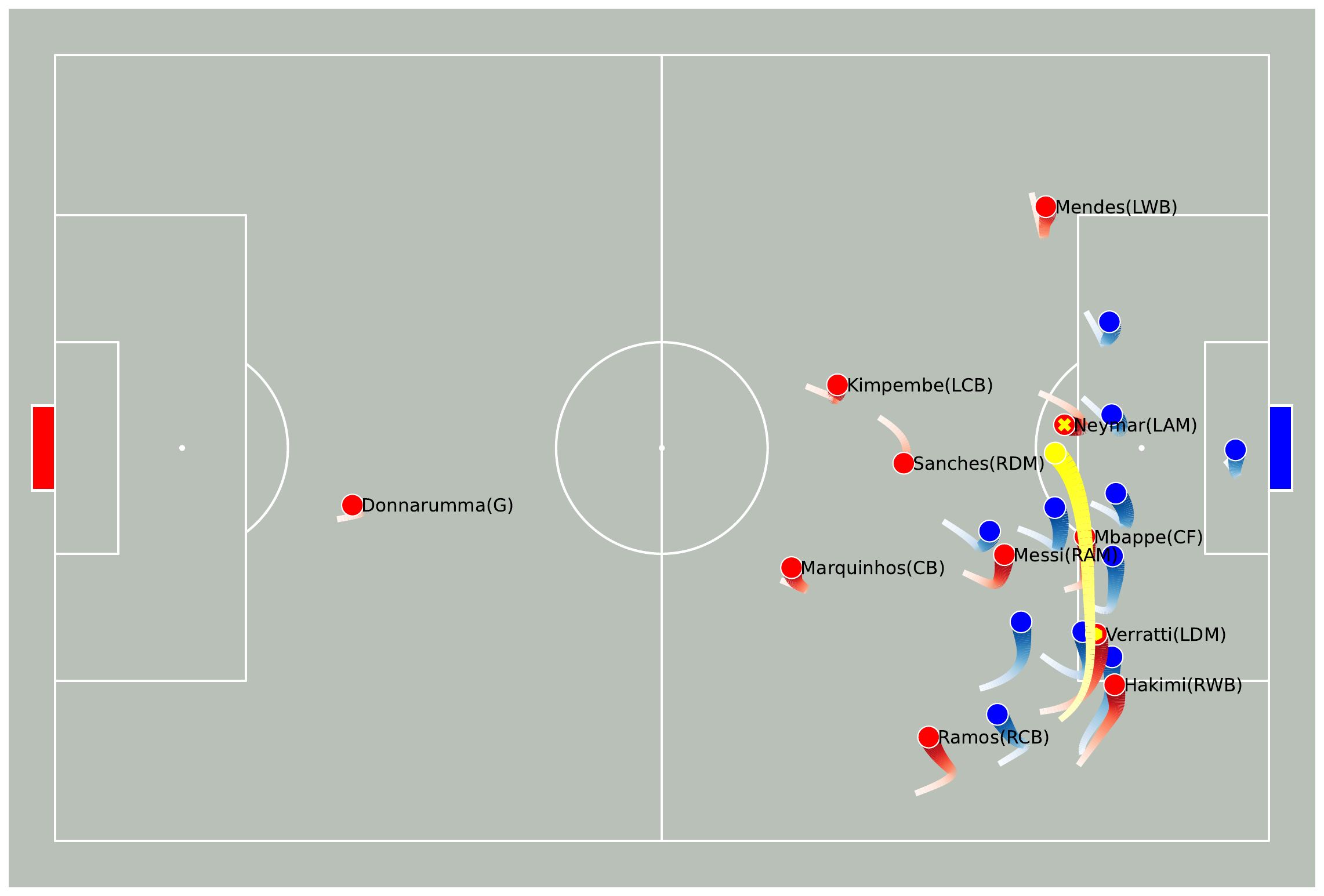} \\
    & \multicolumn{3}{c}{%
        \begin{tcolorbox}[colback=gray!10, colframe=gray!40, boxrule=0.3pt, width=0.9\linewidth]
        \tiny \textbf{MLLM reasoning}: 
        Neymar emerges as the best option. His proximity to Verratti minimizes interception risk, while his technical ability ensures effective continuation. The pass aligns with PSG’s fluid attacking style, leveraging Neymar’s creativity to unlock Monaco’s compact defense. Other high-reward options (e.g., Mbappe) carry excessive risk, whereas lower-risk passes (e.g., Kimpembe) lack strategic impact. Neymar’s suitability and execution likelihood justify prioritizing this choice.  
        \end{tcolorbox}
    } \\[2pt]
    
    \midrule
    
    \multirow{2}{*}{{\textbf{3}}} &
    \includegraphics[width=\linewidth]{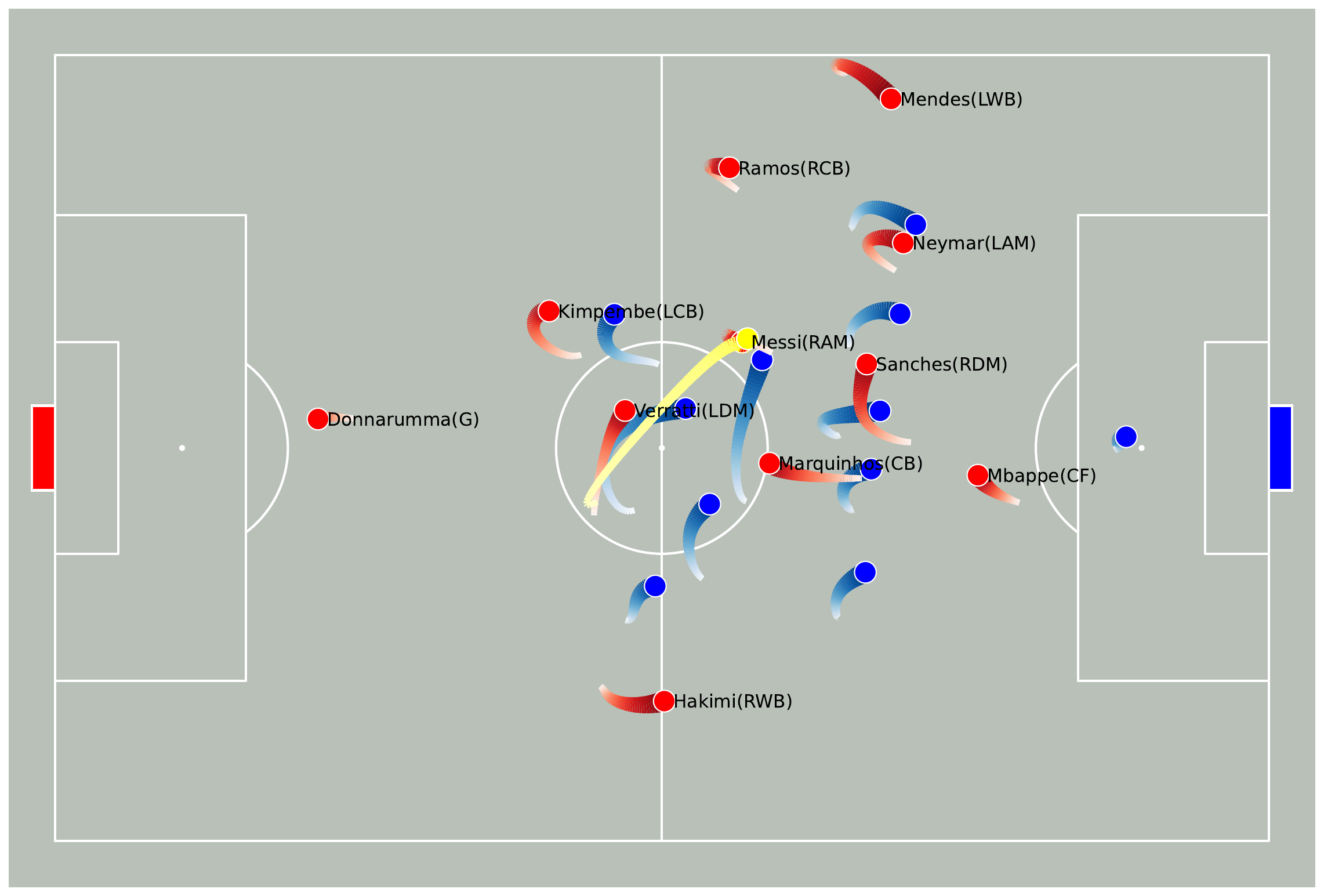} &
    \includegraphics[width=\linewidth]{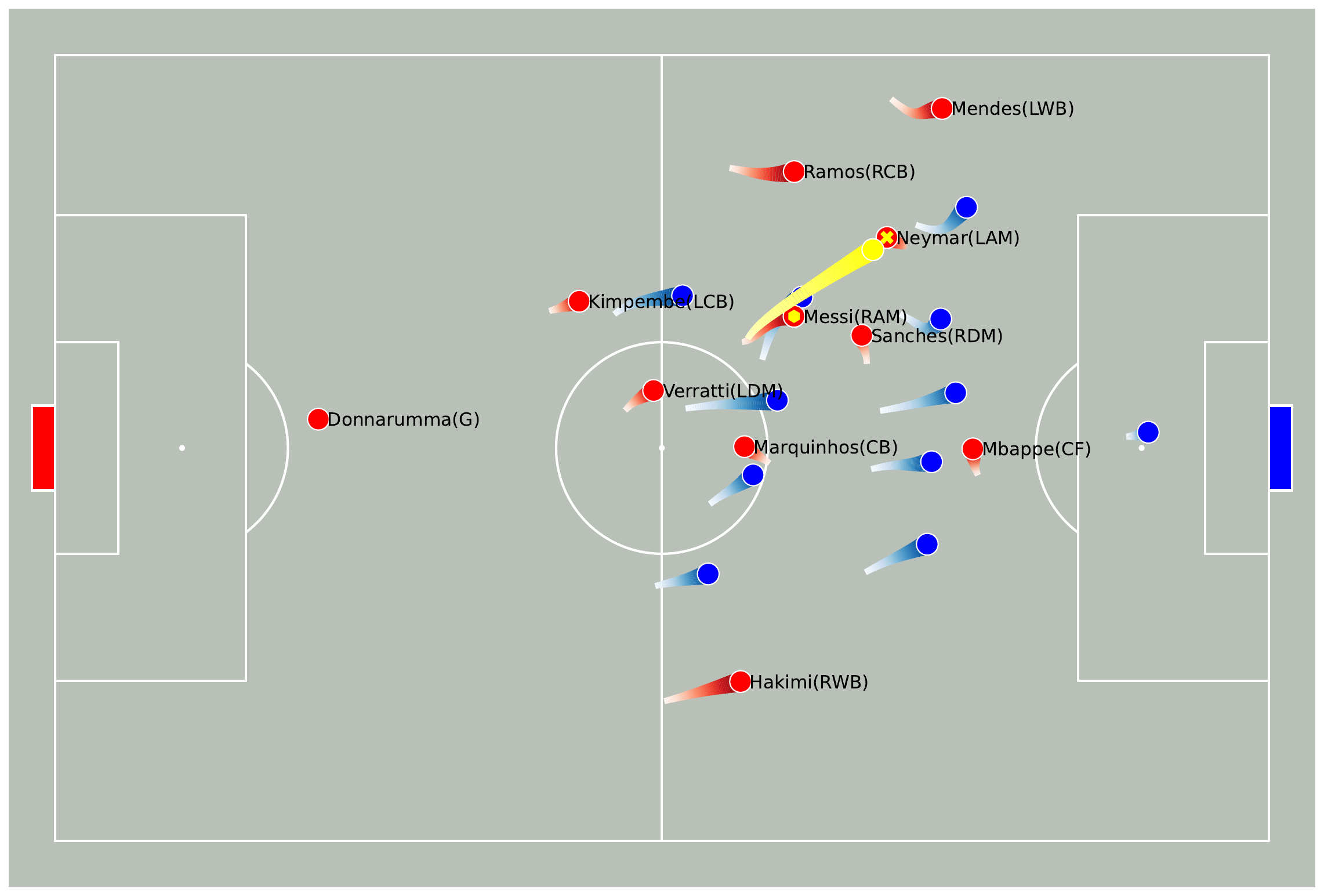} &
    \includegraphics[width=\linewidth]{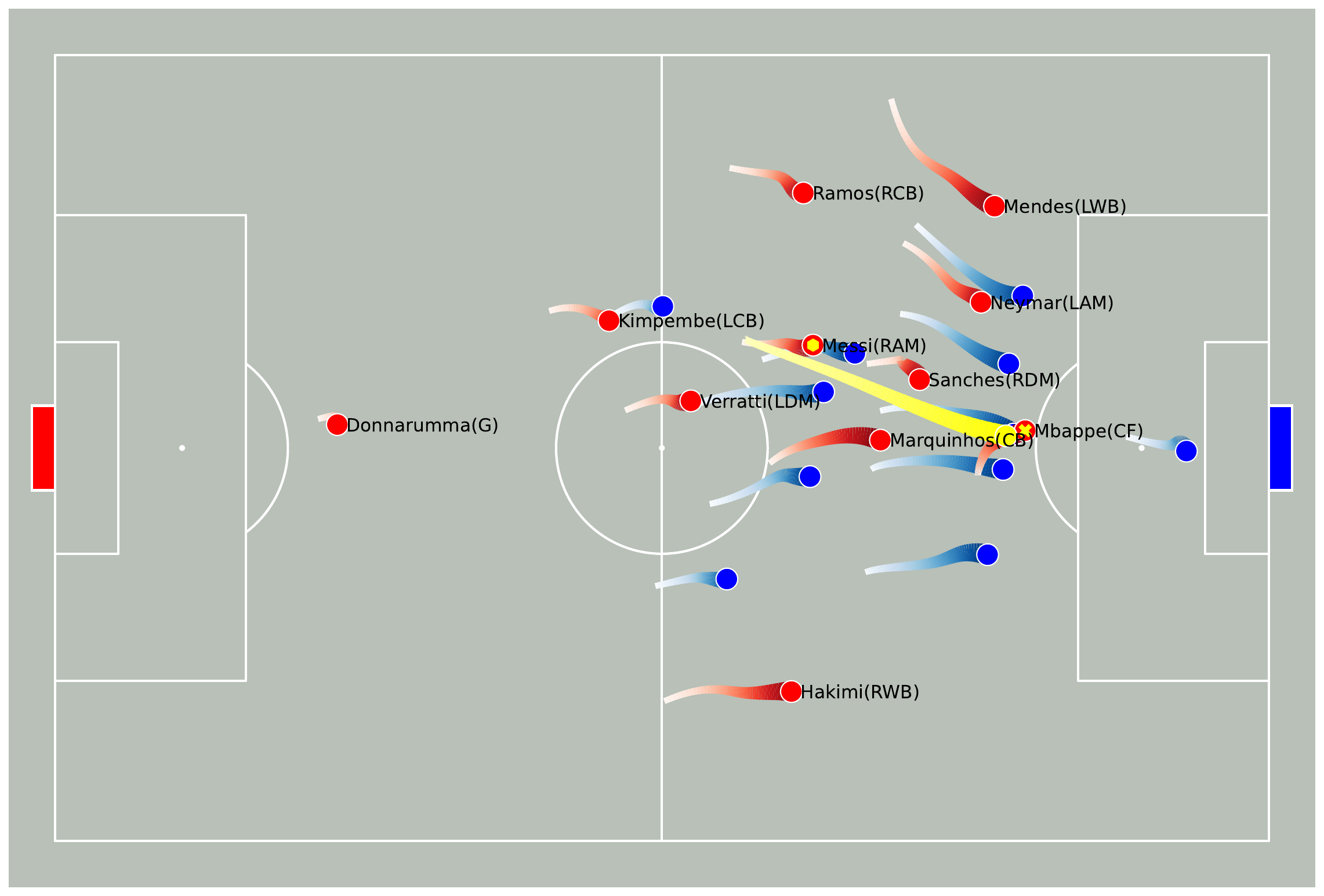} \\
    & \multicolumn{3}{c}{%
        \begin{tcolorbox}[colback=gray!10, colframe=gray!40, boxrule=0.3pt, width=0.9\linewidth]
        \tiny \textbf{MLLM reasoning}: 
        The pass to Kylian Mbappe is optimal due to its direct alignment with scoring objectives, leveraging Mbappe’s speed and positioning. The sketch confirms a clear path, minimizing interception risk while maximizing penetration. Other options either lack urgency (e.g., Verratti) or introduce unnecessary complexity (e.g., cross-field passes). Mbappe’s role as a center-forward ensures tactical coherence, making this the highest-impact decision.
        \end{tcolorbox}
    } \\
    \bottomrule
    \end{tabularx}
    \caption{\textbf{Historical / Factual / Discovered trajectories with explanations below each scenario.}
    In Scenario 1, the result illustrates how MTC identifies a pass from Verratti to Hakimi as the optimal choice, balancing the tactical continuity and spatial exploitation. In Scenario 2, MTC selects a pass to Neymar considering his creative style. In Scenario 3, within the middle-third build-up, MTC favors a pass to Mbappe considering his speed and positioning.}
    \label{fig:single-step}
\end{figure*}

\textbf{Qualitative analysis.} 
\textcolor{blue}{Fig.~\ref{fig:single-step} presents three representative scenarios demonstrating the result of the single-step tactic discovery. Each case illustrates the historical context, the factual event, the optimal proposal selected by MTC, and the corresponding reasoning generated by multimodal large language model (MLLM)} \textcolor{teal}{based on LTG proposals. In our experiments, we use QVQ-Max, a vision-reasoning model by the Qwen team\footnote{\url{https://qwen.ai/research}}, as the MLLM.}

\textcolor{blue}{
The chosen canonical scenarios exemplify how our framework successfully addresses practical implementations.
In Scenario 1, the discovered tactical trajectory closely aligns with the factual tactical trajectory, exemplifying the execution of stereotypical actions in players' cognition during practical open-play situations.
In Scenario 2, confronted with a failed real-world scenario, the MTC identifies a more optimal proposal that successfully executes the action and gains an advantage, demonstrating the manifestation of creative decision-making in practical open play.
In Scenario 3, despite the factual scenario being successful, the MTC opts for a fundamentally different course of action, potentially attributable to broader field of player vision. This illustrates the diversity of tactical approaches or individual player styles observable in practical open-play contexts.}
\textcolor{orange}{Moreover, the reasoning process demonstrates that the MLLM can effectively reason over spatial structure, player characteristics to generate novel and advanced tactical alternatives, providing additional explainability for the internal decision logic of MTC. Overall, TacEleven exhibits strong performance in the single-step tactic discovery task, enabling the discovery of diverse and compelling tactical solutions.}

\textbf{Quantitative analysis.} \textcolor{orange}{We quantitatively evaluate the improvement of the discovered single-step tactics in the SS task of Table~\ref{tab:eval_xG_xT_etc}. 
According to player roles, we separately report the results for players in the front line (Front), midfield (Middle), back line (Back), and for the entire team (Entire). Overall, the single-step tactics discovered by TacEleven outperform the Factual in all metrics except consistency, indicating that the model successfully discovers superior single-step tactics. Compared with the Random, the consistency improves by 8\%, showing that MTC not only identifies better tactics in a transparent and explainable manner but also filters out unreasonable ones. Across different field zones, the performance metrics exhibit distinct patterns. The Front shows the most significant improvement, suggesting that in the attacking third, TacEleven can more effectively mine tactics that create offensive advantages.}

\subsection{Multi-step discovery}
\label{sec:multi-step}

\textbf{Qualitative analysis.} 
\textcolor{teal}{Fig.~\ref{fig:multi-step1} presents the results of multi-step tactic discovery, selected from a 2022 Ligue 1 match between Paris Saint-Germain and Monaco.
It illustrates a midfield penetration sequence that features a possession-oriented central buildup and emphasizes controlled advancement, structural compactness, and midfield overloads. }In step 1, after assessing the situation without stopping the ball, Sanches (RDM) immediately delivers a direct pass to Verratti (LDM), effectively bypassing three defending players.
In step 2, upon receiving the ball, Verratti (LDM) evades the defender and subsequently passes the ball to Mbappé (CF), who has dropped back to provide support. Notably, Mbappé (CF) performs a feint run, deceiving the opponent's central defender and drawing their attention. This coordinated movement successfully lures the opposing center-back into stepping forward, creating space behind the defensive line.
In step 3, observing the gap left by the advancing center-back, Neymar (LAM) makes an immediate forward run into the vacated space. In response, Mbappé (CF) delivers a precise through ball to Neymar (LAM)'s path. Simultaneously, Mbappé (CF) continues his forward movement, positioning himself for potential follow-up actions such as receiving a return pass or preparing for a rebound shot.
Collectively, compared to the ground truth (``Factual") shown in Fig.~\ref{fig:multi-step1}, the sequence predicted by the model exemplifies risk-controlled, central superiority, where compact passing and progressive stability facilitate a smooth transition into attack while generating significant offensive threat.

\newcolumntype{C}[1]{>{\centering\arraybackslash}p{#1}}
\begin{figure*}[tbp]
    \centering
    \setlength{\tabcolsep}{3pt}
    \begin{tabular}{C{0.24\textwidth}C{0.24\textwidth}C{0.24\textwidth}C{0.24\textwidth}}
        \toprule
        & \multicolumn{3}{c}{\textbf{Discovered}} \\[-0.3em]
        \cmidrule(lr){2-4}
        \textbf{Factual} & \textbf{Step 1} & \textbf{Step 2} & \textbf{Step 3} \\
        \midrule

        \includegraphics[width=\linewidth]{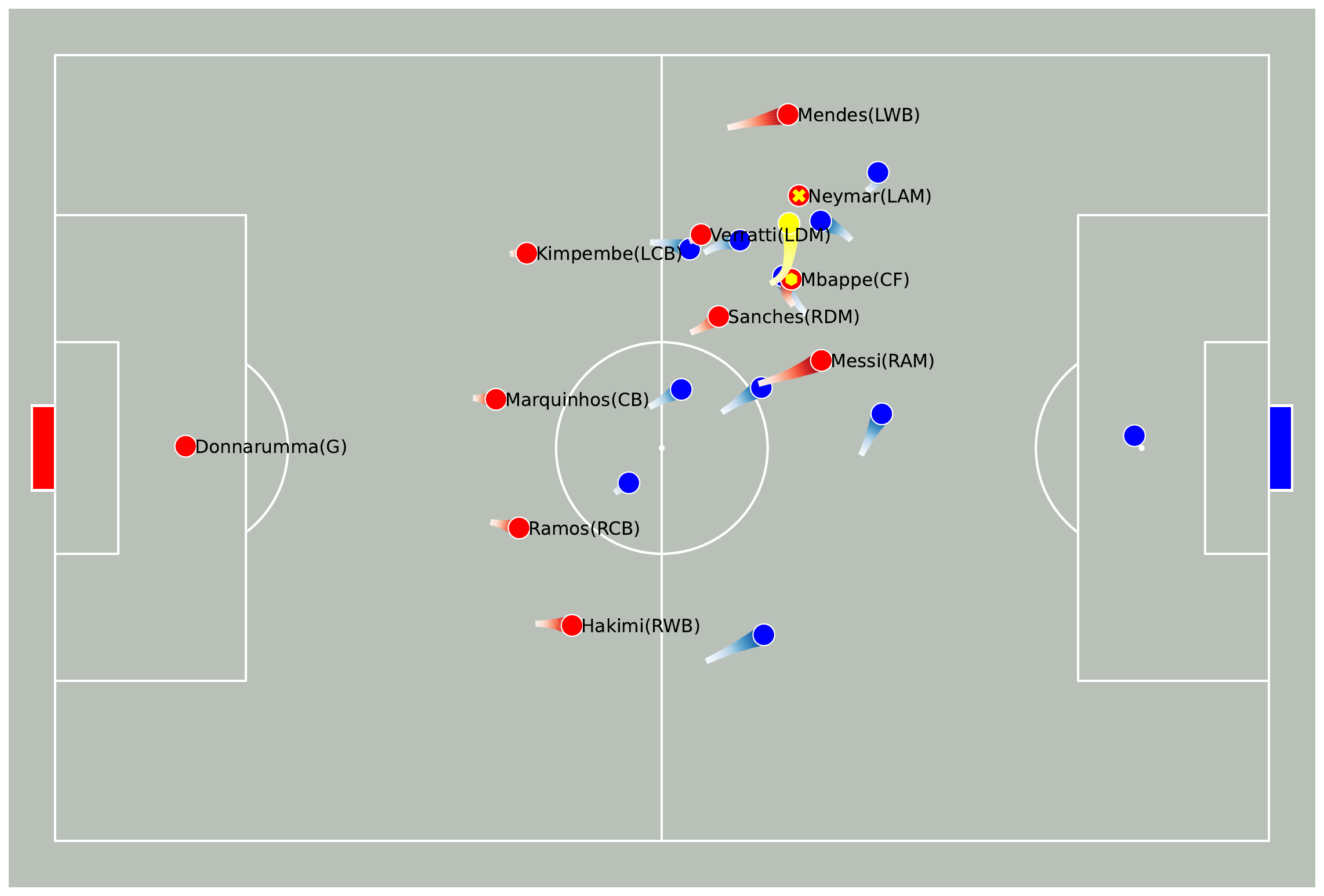} &
        \includegraphics[width=\linewidth]{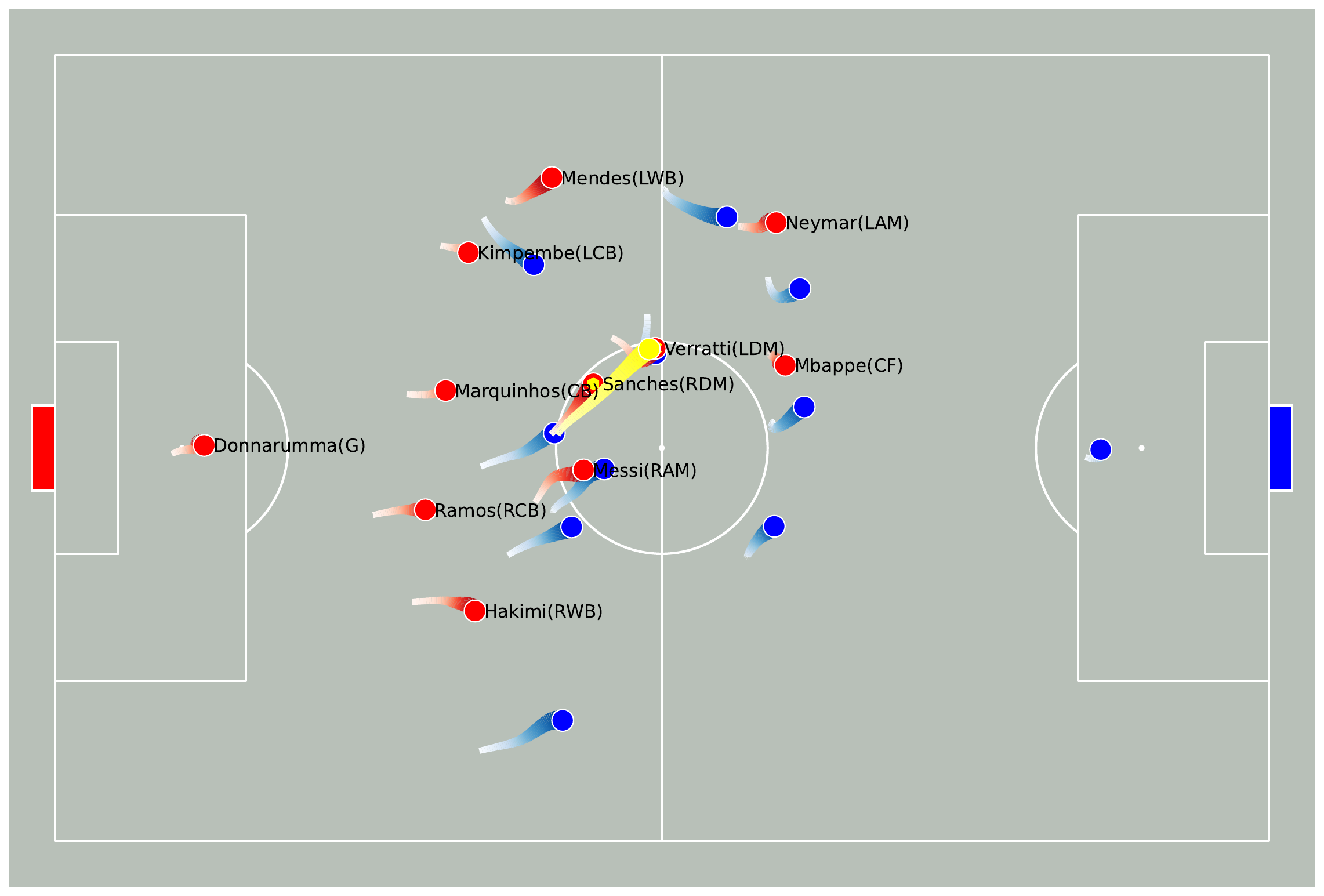} &
        \includegraphics[width=\linewidth]{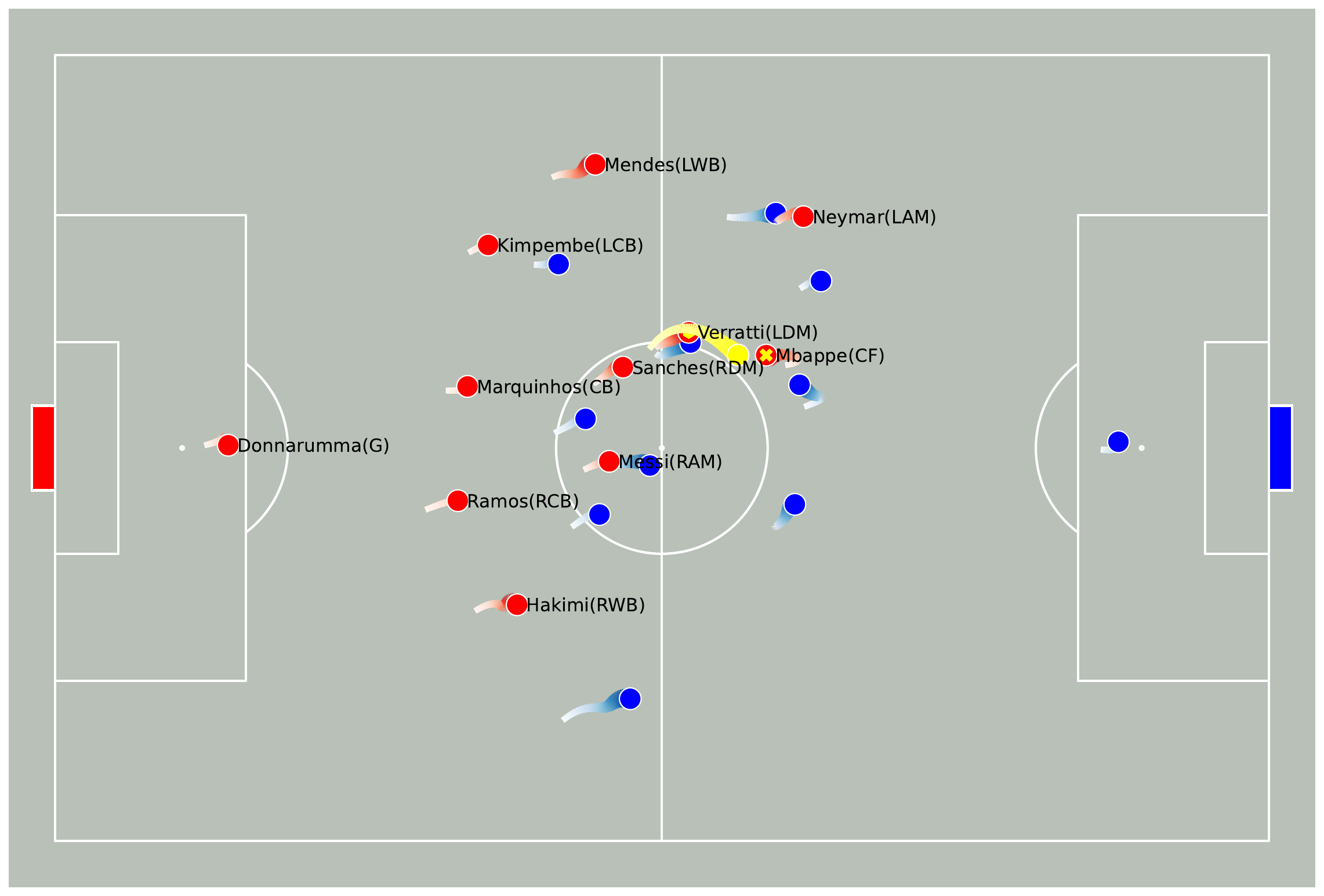} &
        \includegraphics[width=\linewidth]{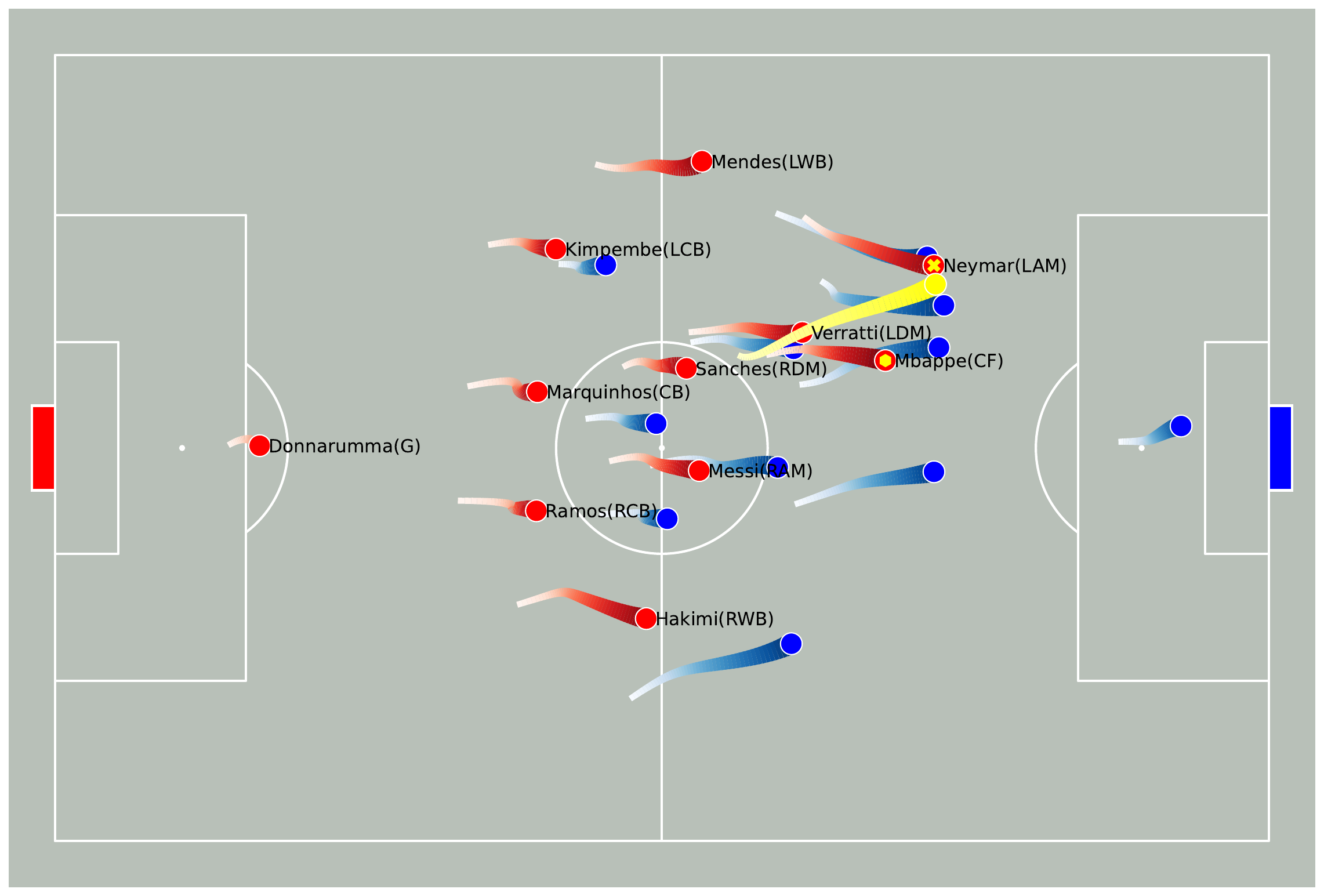} \\
        \bottomrule
    \end{tabular}

    \caption{Multi-step counterfactual exploration with an instruction: \textit{``Design a football tactic that focuses on collaboration among midfield players to execute localized passing sequences, break through the opposing defense, gain offensive advantages, and create potential breakthroughs and attacks. The objective is to create numerical superiority in central areas, enabling the team to outplay opponents in tight spaces and progress the ball forward."}}
    \label{fig:multi-step1}
\end{figure*}

\textbf{Quantitative analysis.}
\textcolor{teal}{In the MS task reported in Table~\ref{tab:eval_xG_xT_etc}, TacEleven achieves notable quantitative improvements. By varying the tactical instructions, we further generate three distinct styles of multi-step tactics: Aggressive, Conservative, and Neutral on the test dataset. In many cases, possession is lost within one or two steps, resulting in a scarcity of direct ground-truth trajectories for comparison in multi-step offensive discovery.} Therefore, we mainly evaluate the incremental advantages provided by the discovered multi-step tactics and to compare the differences among the three styles. We observe that multi-step tactics discovered under the aggressive and neutral styles tend to produce incremental advantages, whereas the conservative style is more effective at maintaining existing advantages. \textcolor{orange}{This demonstrates that TacEleven can flexibly produce stylistically distinct tactics by simply adjusting tactical instructions.}

\subsection{Questionnaire study with domain experts}
\label{sec:questionnaire}

\begin{figure}[htbp]
    \centering
        \begin{subfigure}[b]{0.45\textwidth}
            \includegraphics[width=\textwidth]{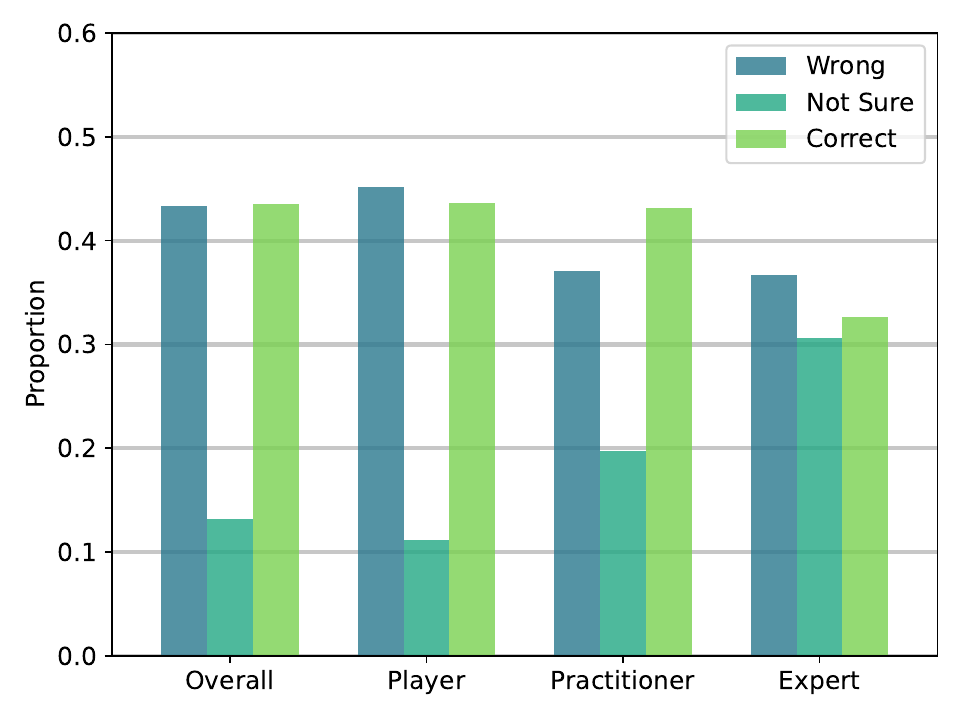} 
            \caption{single-step realism}
            \label{fig:bar-q1}
        \end{subfigure}
        \hfill 
        \begin{subfigure}[b]{0.45\textwidth}
            \includegraphics[width=\textwidth]{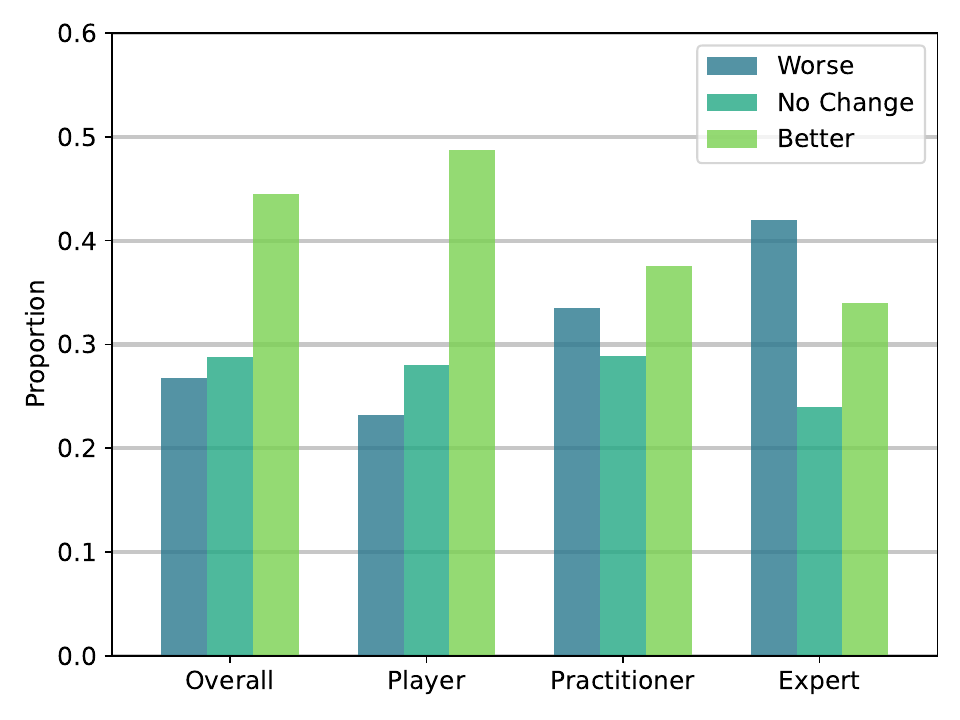} 
            \caption{single-step effectiveness}
            \label{fig:bar-q2}
        \end{subfigure}
        
        \begin{subfigure}[b]{0.45\textwidth}
            \includegraphics[width=\textwidth]{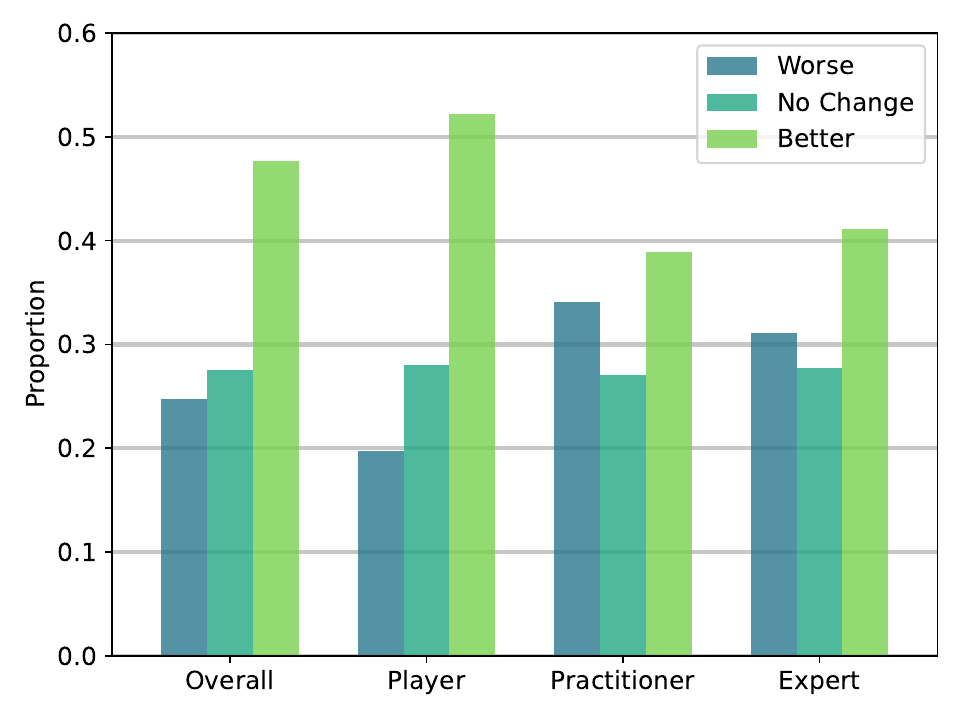} 
            \caption{multi-step effectiveness}
            \label{fig:bar-q3}
        \end{subfigure}
        \hfill 
        \begin{subfigure}[b]{0.45\textwidth}
            \includegraphics[width=\textwidth]{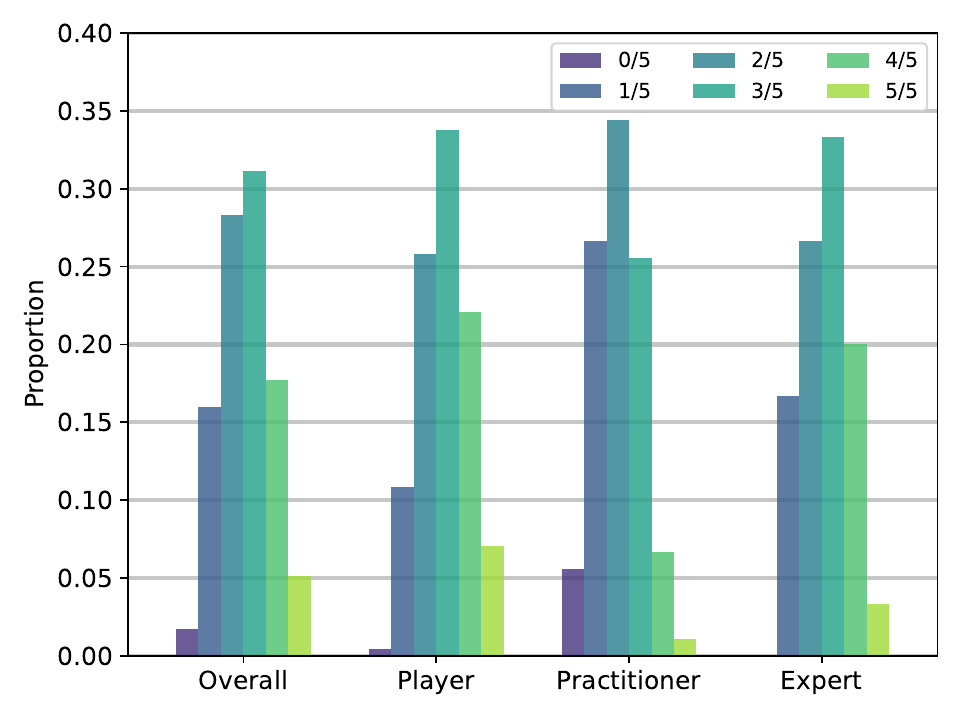} 
            \caption{multi-step 5-shot adoption rate}
            \label{fig:bar-q4}
        \end{subfigure}
    \caption{Bar plots of questionnaire results across four participant groups. Each bar shows the percentage of participants selecting each option in a given questionnaire.}
    \label{fig:bar}
\end{figure}

\begin{figure}[htbp]
    \centering
        \begin{subfigure}[b]{0.45\textwidth}
            \includegraphics[width=\textwidth]{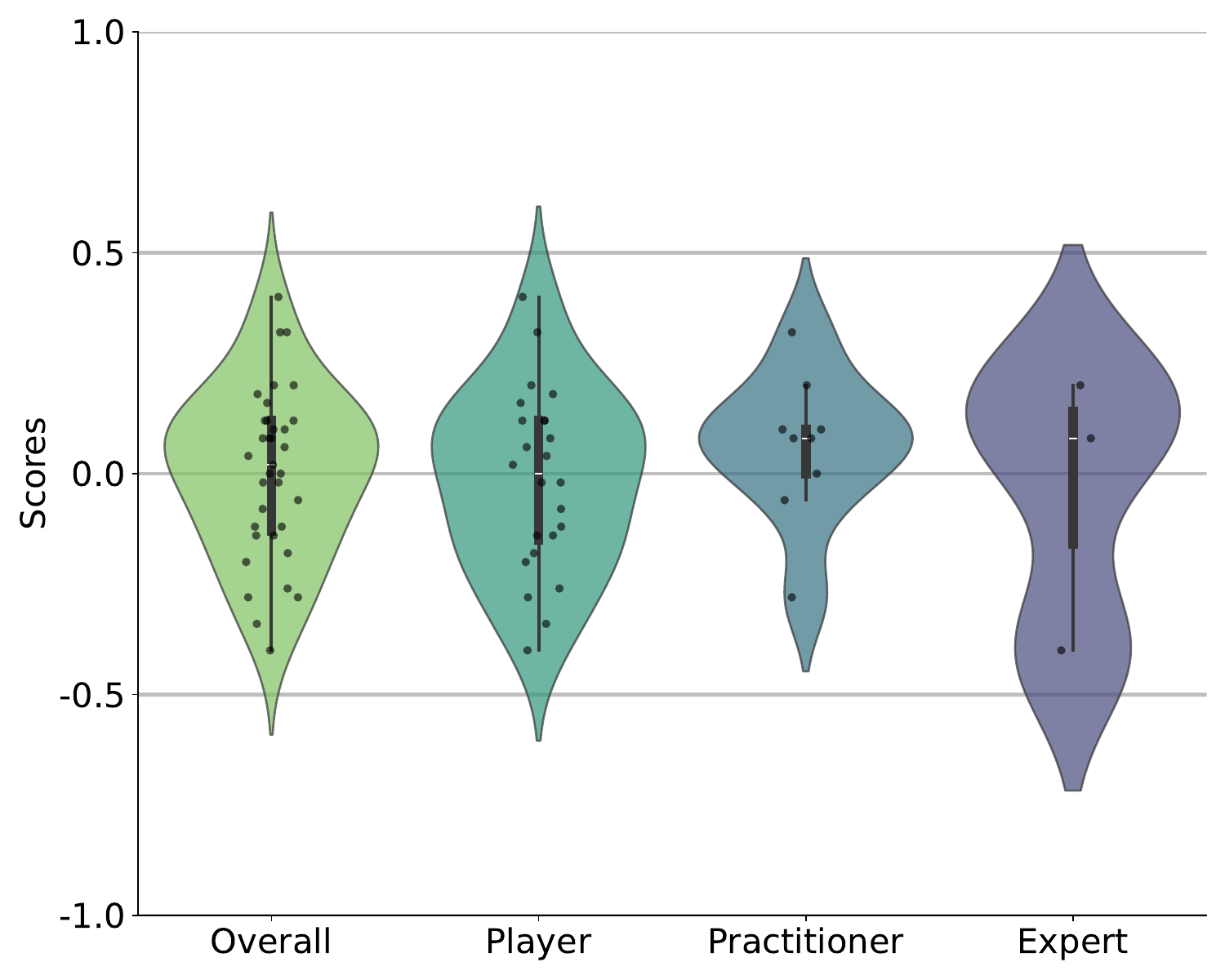} 
            \caption{single-step realism}
            \label{fig:violin-q1}
        \end{subfigure}
        \hfill 
        \begin{subfigure}[b]{0.45\textwidth}
            \includegraphics[width=\textwidth]{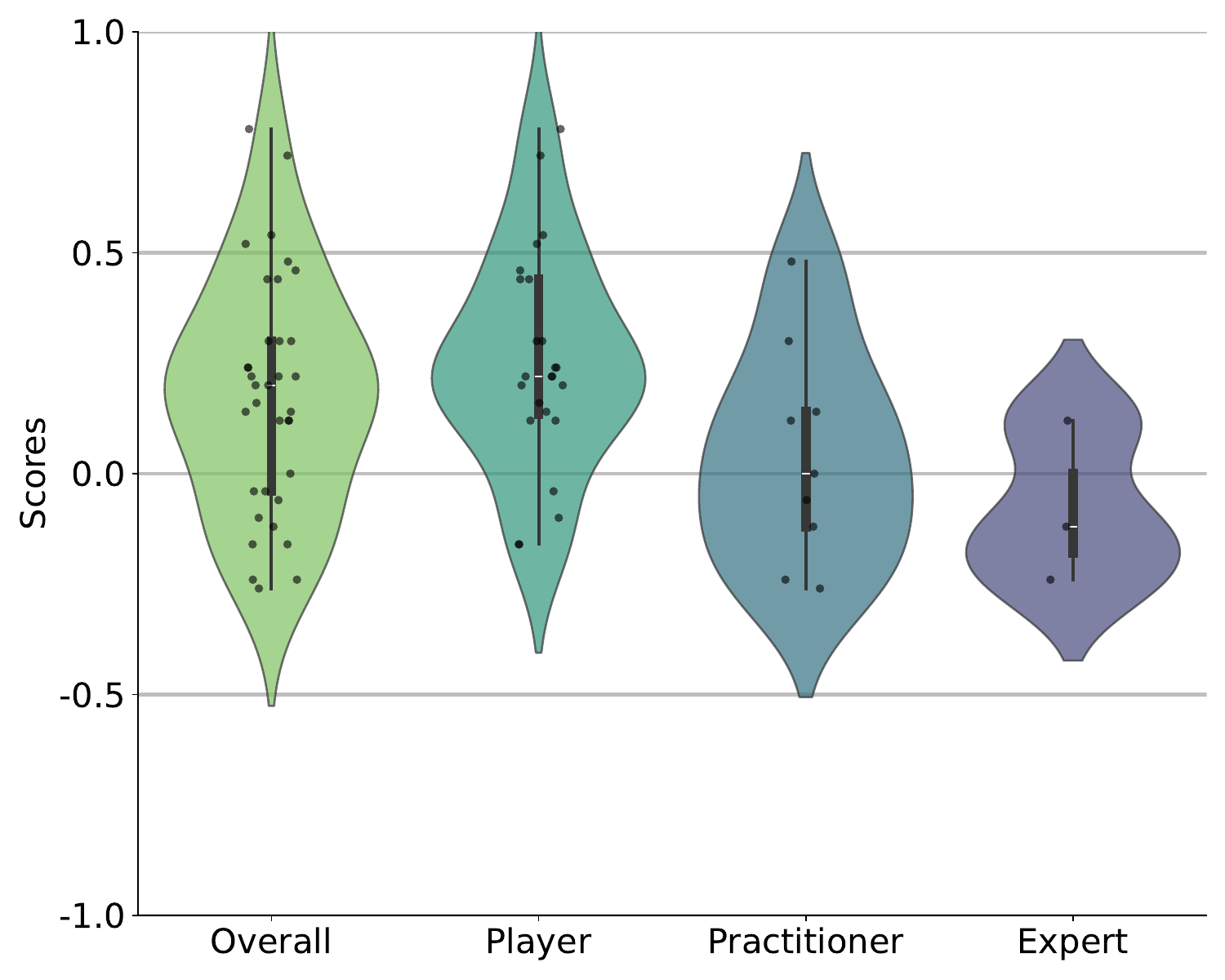} 
            \caption{single-step effectiveness}
            \label{fig:violin-q2}
        \end{subfigure}
        
        \begin{subfigure}[b]{0.45\textwidth}
            \includegraphics[width=\textwidth]{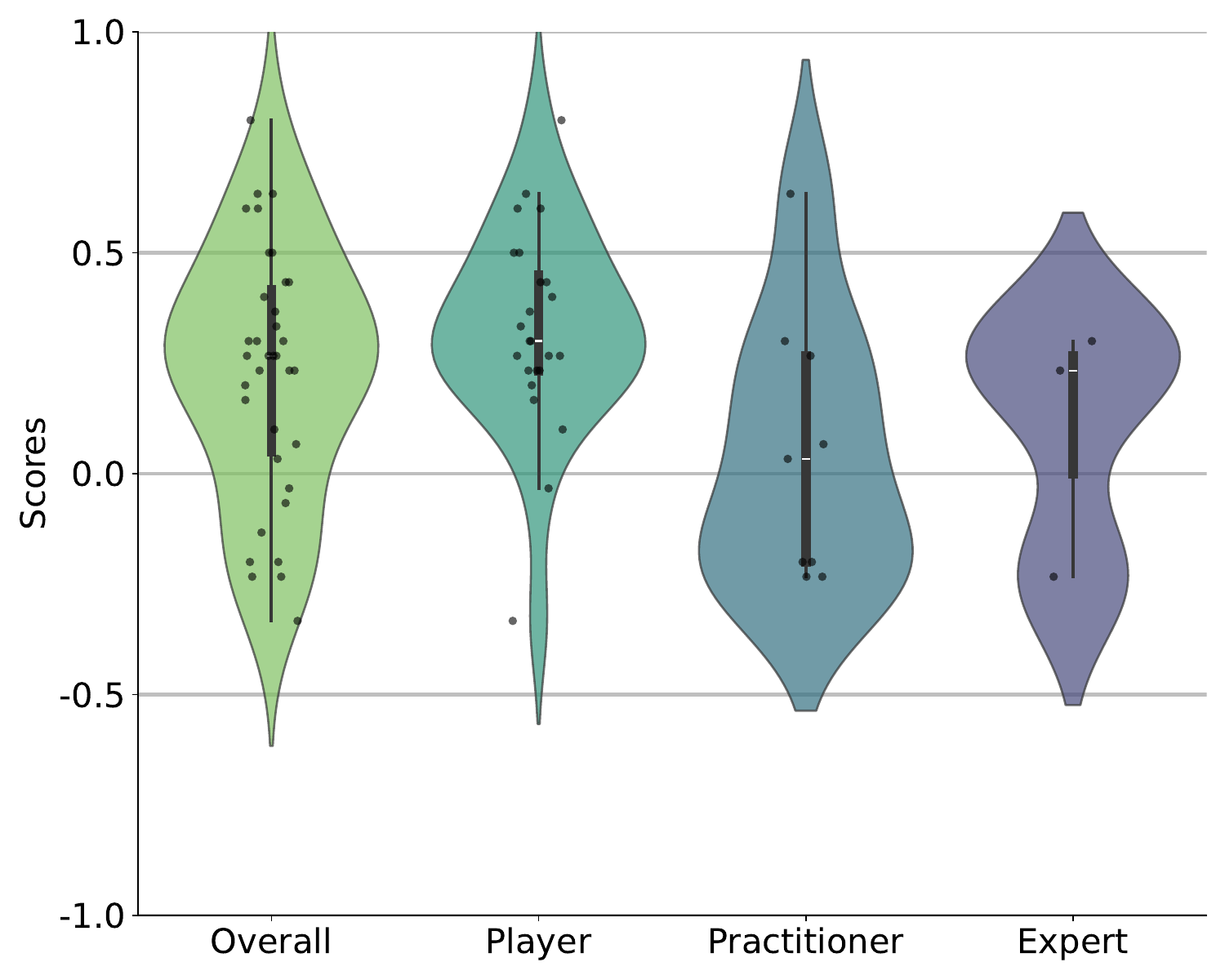} 
            \caption{multi-step effectiveness}
            \label{fig:violin-q3}
        \end{subfigure}
        \hfill 
        \begin{subfigure}[b]{0.45\textwidth}
            \includegraphics[width=\textwidth]{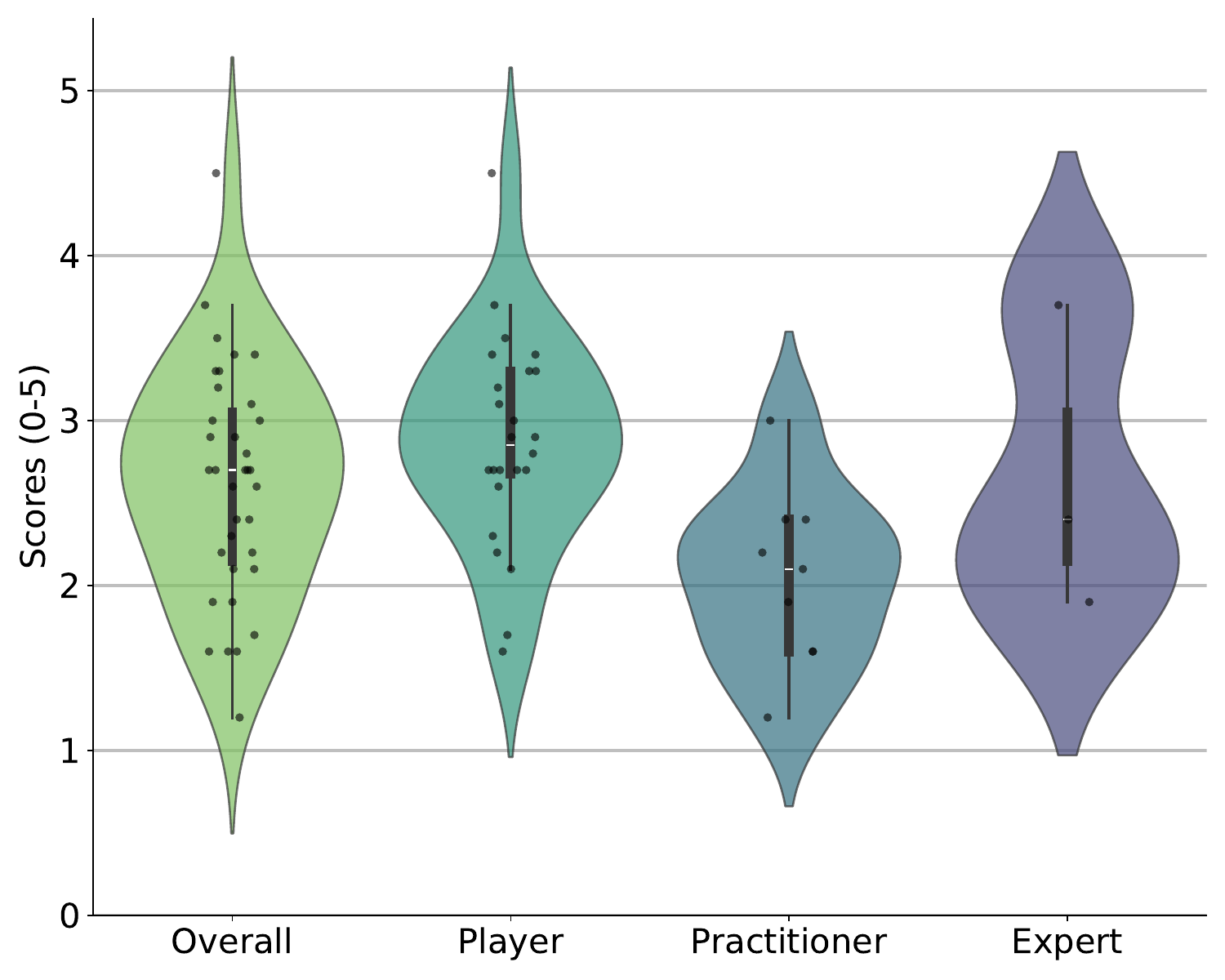} 
            \caption{multi-step 5-shot adoption rate}
            \label{fig:violin-q4}
        \end{subfigure}
    \caption{Violin plots of questionnaire results across four participant groups. Each black dot represents a participant’s mean score across all samples in a given questionnaire. The plots illustrate the distribution of judgments within and across groups, showing overall agreement that TacEleven produces realistic and effective tactical alternatives.
}
    \label{fig:violin}
\end{figure}





Through well-defined metrics and the visualization of TacEleven's discovery process and results, we benchmarked the model's capabilities and demonstrated its explainability. In practical applications, when TacEleven is used to assist coaches and analysts in tactical decision-making, two core questions arise: whether the discovered tactics are realistic and whether they are effective in improving play quality. These two dimensions, realism and effectiveness, rely on human subjective judgment. To objectively evaluate them, we collaborated with experts from AJ Auxerre (3 participants), practitioners such as coaches majoring in football (9 participants), and football players (24 participants), to conduct a comprehensive questionnaire study. Specifically, we designed four questionnaires to assess (1) the realism of discovered single-step tactic, (2) the effectiveness of discovered single-step tactic, (3) the effectiveness of discovered multi-step tactic, and (4) the adoption rate of five-shot tactic discovery in typical scenarios. The questionnaire results are presented in Fig.~\ref{fig:bar} and Fig.~\ref{fig:violin}, with detailed questionnaire content provided in Appendix~\ref{appendix:QA}.

For the first questionnaire, we randomly select 50 scenarios from the test dataset and discover one single-step tactic for each scenario. The factual and discovered tactics are visualized as two images in the questionnaire, with their positions randomized. Participants are asked to identify which image represents the factual tactic, choosing among three options: A, Left is factual; B, Right is factual; C, Cannot distinguish. The results show that participants achieve an accuracy of 43.49\% when identifying factual tactics (Fig.~\ref{fig:bar-q1}), indicating that the single-step tactics discovered by TacEleven are nearly indistinguishable from factual ones. We assign 1 for correct judgments, -1 for incorrect judgments, and 0 for indistinguishable samples, allowing us to analyze the distribution of responses across participant groups (Fig.~\ref{fig:violin-q1}). The mean scores for all three groups remain close to zero (players 0.00, practitioners 0.08, experts -0.05), suggesting that even experienced experts find it difficult to differentiate discovered tactics from factual ones. This result demonstrates the realism of the discovered single-step tactics and supports the realism of multi-step tactics, which are generated through an autoregressive process.

For the second questionnaire, we use the same 50 scenarios from the first questionnaire, presenting both TacEleven-discovered and factual single-step tactics, with each clearly labeled as generated by TacEleven or factual. Participants are asked to judge whether the discovered tactics are better than the factual tactics, choosing among three options: A, Better; B, Worse; C, No Change. On average, nearly half of the discovered tactics, 44.46 \% are considered better than the factual ones (27 \% are worse and 28 \% show no change). Among experts, the proportion judged as better decreases to 34.00 \% (41 \% are worse and 24 \% show no change) (Fig.~\ref{fig:bar-q2}).
It is encouraging that even the strictest experts regard 34 \% of the discovered tactics as improvements over the factual ones. Since TacEleven can rapidly and inexpensively generate large numbers of tactical alternatives, this proportion highlights its potential to uncover tactics that exceed real-world strategies in quality. 
Similar to the first questionnaire, we assign 1 for option A, -1 for option B, and 0 for indistinguishable samples, allowing us to analyze the distribution of responses across participant groups (Fig.~\ref{fig:violin-q2}). The results show that from players to practitioners to experts, judgments become increasingly strict as domain knowledge increases. 
For the third questionnaire, we use 30 sampled cases and ask participants to judge whether the multi-step tactics discovered by TacEleven are better than the factual tactics, choosing among three options: A, Better; B, Worse; C, No Change. In multi-step scenarios, the ratings from the expert group increase significantly, with 41.11 \% of the discovered tactics considered better than the factual ones (31 \% are worse and 28 \% show no change) (Fig.~\ref{fig:bar-q3}). We assign 1 for option A, -1 for option B, and 0 for indistinguishable samples. The overall rating distribution shifts upward compared with single-step scenarios, with experts showing a clear positive shift (Fig.~\ref{fig:violin-q3}). This pattern suggests that proposing multi-step tactics is substantially more difficult for human than proposing single-step tactics. Hence, in this setting, the multi-step tactics discovered by TacEleven offer greater analytical value, often revealing possibilities that may be overlooked by humans. Experts, with deeper domain knowledge of Ligue 1 tactics, can better recognize the superiority of the tactics discovered by TacEleven.

For the fourth questionnaire, we simulate real-world usage scenarios. We select ten notable failure cases from the Paris Saint-Germain FC vs. Monaco match in the 2022–23 Ligue 1 season that attracted public discussion, and discover five different multi-step tactics for each case. Participants are asked to indicate how many of these five multi-step tactics are adoptable. The overall average number of adoptable samples is 2.63, corresponding to an adoption rate of 52.50 \% (Fig.~\ref{fig:bar-q4}, Fig.~\ref{fig:violin-q4}). This finding aligns with the proportion of multi-step tactics considered better in the third questionnaire, and the two results reinforce each other.

In summary, the tactics discovered by TacEleven demonstrate both realism and effectiveness, showing strong potential to assist coaches and analysts in tactical decision-making within real-world scenarios.

\section{Discussion}

Our study introduces TacEleven, a generative framework for football open-play scenarios, enabling three progressive tactical tasks: counterfactual exploration, single-step discovery, and multi-step discovery. Through counterfactual exploration, we simulate alternative tactical proposals with counterfactual language conditions; single-step discovery allows for immediate tactical adjustments based on real-time inputs; and multi-step discovery facilitates long-horizon planning via an autoregressive approach that decomposes tactics into sequential events. These capabilities collectively empower TacEleven to dynamically analyze and generate realistic football tactics, as validated by both qualitative and quantitative analyses, including expert evaluations that confirm its realism and effectiveness.

The key innovations of our work address critical limitations in existing football analytics. While prior studies predominantly focus on static scenarios (e.g., TacticAI\cite{TacticAI}) or are constrained to short-horizon dynamic tasks with limited condition flexibility, TacEleven introduces a decoupled event-sequence approach that enables dynamic, long-sequence tactical prediction under diverse language conditions.
Moreover, by incorporating language-based conditions through MLLMs, we enable human-computer interaction and automate top-level tactical guidance. This multimodal integration not only enhances flexibility but also bridges the gap between data-driven analysis and practical coaching needs.

In detail, we introduce several technical contributions that fundamentally advance football analytics. First, the autoregressive tactical decomposition reformulates tactical discovery as a sequential generation problem by decomposing complex tactics into discrete event steps. Furthermore, to address language generalization challenges, we develop a hierarchical language control architecture featuring strategic-level parsing where an MLLM decomposes instructions into executable descriptions, tactical-level instantiation that maps descriptions to trajectories with spatiotemporal attention mechanisms, and dynamic adjustment via real-time generator-critic feedback loops. Finally, our physics-informed temporal alignment method resolves data misalignment through event-driven interpolation leveraging football physics, achieves multi-source fusion of tracking data and tactical annotations.

Validated through quantitative metrics and a questionnaire-based qualitative assessment, TacEleven demonstrates superior performance in generating realistic, effective tactics achieving capabilities previously unattainable in football analytics. Looking ahead, future research will focus on enhancing the robustness of language-conditioned control by generating diverse tactical descriptions, enabling universal language-driven decision-making. This development will not only refine TacEleven into a more nuanced and adaptive tactical system, but also extend its validated framework to broader domains such as autonomous driving and unmanned swarm systems. Together, these directions underscore TacEleven's role as a scalable and transferable paradigm for complex decision-making scenarios.

\section{Methodology}
\label{sec:method}

We first provide a clear definition of football tactics in open play, represent them using spatiotemporal graph sequences, and formally define the counterfactual tactical exploration task, as well as the single-step and multi-step tactical discovery tasks (Sec.~\ref{sec4.1:problem_formulation}). Next, we describe the preprocessing procedures and the construction of the final dataset (Sec.~\ref{sec4.2:data}). Finally, we outline the architecture and training configurations used in LTG (Sec.\ref{sec4.3:LTG}), along with the technical details of the tactical tree search employed in MTC (Sec.~\ref{sec4.4:TTS}).

\subsection{Problem Formulation}
\label{sec4.1:problem_formulation}

\textbf{Tactic definition.} To model tactics in open play, we first define a set of \textit{meta-actions} $\mathbbm{A}$, the most fundamental and atomic actions in football. These meta-actions are both minimal and complete, capturing the essence of individual events (e.g., pass, carry, shot, etc.) while forming the building blocks for more complex tactical patterns.
By combining different meta-actions $a_1, a_2, \ldots \in \mathbbm{A}$, a tactic $\mathbf{T}$ can be formed as $\langle a_1, a_2, \ldots, a_N \rangle$, where $N$ denotes the step number of the meta-actions, introducing abundant fine-grained features. Compared to relying solely on semantic descriptions, the incorporation of these fine-grained features offers more precise and detailed low-level representations, enabling a deeper understanding of tactical structures.

\textbf{Spatiotemporal graph representing.} \textcolor{blue}{The spatiotemporal graph (STG) offers a powerful framework for modeling these relationships, combining spatial and temporal features with semantic descriptions. This representation not only facilitates the integration of multi-modal data but also enhances the model's ability to capture the complexities of football tactics, making it a cornerstone of our approach.}
A STG $\mathcal{G}$ can comprehensively represent $\mathbbm{A}$. The STG $\mathcal{G} = (\mathcal{V}, \mathcal{E}, \mathbf{X}, A, T, s)$ is assumed to be static, with nodes $\mathcal{V}$, edges $\mathcal{E} \in \mathcal{V} \times \mathcal{V}$, node feature sequence $\mathbf{X}$, a stable adjacency matrix $A$, the temporal index $T$ of $\mathbf{X}$, and an global language description $s$. Within the context of football games, $\mathcal{V}$ is defined as the set of 23 entities (22 players and 1 ball) present on the pitch, and $\mathcal{E}$ is constructed as the set of all possible pairwise connections among them.
Furthermore, the STG is assumed to encode relevant features. In particular, a node feature sequence $\mathbf{X} \in \mathbbm{R}^{|T| \times |\mathcal{V}| \times d}$ with feature dimension $d$, a adjacency matrix $A \in \mathbbm{R}^{|\mathcal{V}|\times |\mathcal{V}|}$ with binary elements, and a language description $s$ are provided, where $|T|$ denotes the length of the temporal sequence, and $|V|$ denotes the number of nodes.
The relevant entries of these objects serve as input features for each node and edge. 
The $\mathbf{X}_v \in \mathbbm{R}^{|T| \times d}$ represents the attributes of an individual player or the ball $v \in \mathcal{V}$, (e.g., position, velocity, and role).
The edge $A_{uv} \in \{0,1\}$ represents the attribute of a particular pair of entities $(u,v)\in \mathcal{E}$, (e.g., whether two players are teammates). 
The language description $s$ is a text segment that characterizes the corresponding tactic instance (e.g., event type, participating players, etc.).

\textbf{Counterfactual tactical exploration task.} Counterfactual tactical exploration aims to achieve cross-modal alignment, translating the global language description $s_n$ with the history $a_{n-1}$ into the future $a_n$, where $n$ denotes the meta-action step. 
In each step, a historical instance $\mathcal{G}_{n-1} = (\mathcal{V}, \mathcal{E}, \mathbf{X}_{n-1}, A, T_{n-1}, s_{n-1})$ as the input, and a predicted instance $\mathcal{G}_{n} = (\mathcal{V}, \mathcal{E}, \mathbf{X}_{n}, A, T_{n}, s_{n})$ as the output, we calculate $\mathcal{G}_{n} = f_{\theta}(\mathcal{G}_{n-1} \mid s_{n})$ capable of inferring trajectories in response to $s_{n}$, where $\theta$ denotes the parameters of the neural network $f_{\theta}(\cdot)$. To achieve this, we utilize the LTG as the underlying model.
From this perspective, it aims to predict how the game would proceed if paused at a given moment, informed by historical trajectories and possible tactical intent as shown in Fig.~\ref{fig:overview}-A.
Future research could explore additional tasks with our tactic definition (e.g., tactic summary, tactic completion, etc.). 
The task is formulated as a regression problem and trained using stochastic gradient descent with the loss function of mean squared error. 
While similar to pedestrian or vehicle trajectory prediction, this task incorporates alignment with the language modality, allowing for more intuitive and human-interpretable control through natural language.

\textbf{Tactical discovery task.} Before the MTC in each step, the meta-actions are predicted by the trained LTG $f^{*}_{\theta}(\cdot)$.
Given the historical meta-action $s_{n-1}$, a counterfactual description set can be obtained as $\mathcal{D}_n = c(s_{n-1})$, where $c(\cdot)$ serves as a rule specifically tailored to generate a finite set of possible choices.
Using various action description $s_{n} \in \mathcal{D}_n$, diverse counterfactual proposals $\mathcal{G}_{n} = f^{*}_{\theta}(\mathcal{G}_{n-1} \mid s_{n})$ can be generated. The tactical discovery task is formulated as an optimization search in the counterfactual proposals for each step with a top-level instruction $P$.
We calculate $s_n$ using a reasoning function $f_{\psi}(\cdot)$, which takes as input $(s_1, \dots, s_{n-1}, \mathcal{G}_{0}, \dots \mathcal{G}_{n-1})$ and the instruction $P$ to mine the best descriptions ($\mathcal{G}_{0}$ denotes the historical context). This entire mechanism forms an iterative process, which is depicted in Fig.~\ref{fig:overview}-B.

\subsection{Data Preprocessing}
\label{sec4.2:data}

Football data is multi-modal and multi-layered containing 4 main parts \cite{sarmento2014match,de2018player,pu2024orientation}: (1) tracking trajectory data, which tracks all on-pitch players and the ball at 10 frames per second, (2) event data, which annotates key actions—such as passes, shots, and carries—with corresponding timestamps, outcomes, involved players, and additional contextual information; (3) player data, which records the players' profiles, including their heights, weights, ages, and roles; (4) match data, which provides unstructured match information, including coaching staff, team rosters, and the physical size of the pitch.

The misalignment of timestamps across multiple datasets often introduces noise that can severely degrade model performance, which has also been reported in other concurrent works \cite{biermann2023synchronization, zhang2025navigating}.
Hence, our first step is to align the event data's timestamps to the tracking trajectory's temporal framework.
In the result, timestamp alignment significantly improves our task metrics, reducing temporal misalignment from an average of 0.9 seconds to under 0.1 seconds.
The alignment pipeline identifies two key temporal anchors: (i) local extrema in the ball’s acceleration and (ii) the minimum ball-to-player distance. Specifically, we detect the time points at which the ball’s acceleration reaches a local extremum, treated as a common signature of the onset and termination of all on-ball events. 
Among each extremum, we finally choose the moment when the corresponding player is closest to the ball. 
This pipeline yields refined timestamps that more faithfully align with the corresponding moments in the tracking trajectory data. The designed method details in Appendix \ref{appendix:align}. 

Furthermore, to construct the prepared $\mathcal{G}$, the raw data are filtered after the timestamp alignment, due to incomplete information or identifier inconsistencies. After filtering, 1,076,258 valid meta-actions remained for training and evaluation. 
Subsequently, the node feature sequence $\mathbf{X}$ is extracted from the tracking trajectory data within $|T|=5$ by uniform temporal sampling.
the adjacency matrix $A$ encodes as an all-ones matrix representing a fully connected graph.
Finally, rich tactical description annotations, integrated from the event data, are appended to these instances.

\subsection{Language-controlled Tactical Generator Details}
\label{sec4.3:LTG}

\textbf{Embedding across time and language.} We propose an TL-embedding that jointly encodes temporal dynamic and semantic information. 
Specifically, we augmenting each time step's representation with a fixed sinusoidal positional encoding \cite{attention}, to obtain a temporal embedding in $\mathbbm{R}^{D}$, where $D$ denotes the latent space dimension. Given $p$ historical and $q$ future time steps, the temporal embedding at each time step $t \in \{ t_1, t_2, \ldots, t_{p+q} \}$ is denoted as $e_{t}^{\mathrm{time}} \in \mathbbm{R}^{D}$. 
\textcolor{blue}{
In practice, $p = q = |T|$. To generate trajectories corresponding to events of variable lengths, we predict only $|T|$ uniformly sampled points along each trajectory, including the starting point. It is worth noting that for different trajectories, the durations may vary.
}
Subsequently, a pretrained language model (e.g., BERT\cite{bert}) is employed to transform each token in the language description $s$ to a word-level embedding $e^{\mathrm{language}}_{s_k} \in \mathbbm{R}^{D}$, where $s_k$ denotes the $k$-th token in $s$. These embeddings are then aggregated by an attention layer to produce a sentence-level representation $e^{\mathrm{language}}_{s} \in \mathbbm{R}^{D}$.
Finally, the temporal and language embeddings are integrated to form the TL-embedding, capturing rich representations. In specific, for any vertex $v$ at time step $t$ under description $s$, the TL-embedding is defined as: $e^{\mathrm{TL}}_{t,s} = \mathrm{Concat}(e_{t}^{\mathrm{time}}, e^{\mathrm{language}}_{s}) \in \mathbbm{R}^{2D}$, where the operation $\mathrm{Concat}(\cdot)$ denotes the concatenation of vectors along the last dimension. Accordingly, the TL-embedding across the temporal index $T$ under instruction $s$ is represented as $E^{\mathrm{TL}}_{T, s} \in \mathbbm{R}^{(p+q) \times 3D}$. 
We utilize the TL-embedding in each layer of the network architecture, which is introduced in the following part.

\textbf{Spatiotemporal attention.} In our model, the attention mechanism \cite{attention} is incorporated to effectively capture spatiotemporal dependencies and align multi-modal inputs, thereby strengthening the model's predictive performance.
An attention mechanism is determined by a similarity function between the query matrix $Q$ and the key matrix $K$, and the resulting weighted combination is applied to the value matrix $V$.
In practice, these matrices are projected into designated dimensions with multiple axes, typically structured as $Q, K, V\in \mathbbm{R}^{N \times |T| \times |\mathcal{V}| \times D}$, where $N$ is the batch size. 
We define the spatiotemporal attention with as $\mathrm{Attention}^{\mathrm{ST}}(X^Q, X^K, X^V)$, with three input matrices $X^Q, X^K, X^V \in \mathbbm{R}^{N \times |T| \times |\mathcal{V}| \times d_{X^{*}}}$, where $d_{X^{*}}$ respectively denotes the feature dimension of them. 
The non-linear projections separately transform the input matrices: 
\begin{equation}
    \begin{aligned}
        Q &= \mathrm{ReLU}(X^Q \cdot W^Q + b^Q) \\
        K &= \mathrm{ReLU}(X^K \cdot W^K + b^K) \\
        V &= \mathrm{ReLU}(X^V \cdot W^V + b^V)
    \end{aligned}
\end{equation}
where $W^Q$, $W^K$, $W^V$, $b^Q$, $b^K$, $b^V$ are learnable parameters, the symbol $\cdot$ denotes batched matrix multiplication along the last two dimensions, and $\mathrm{ReLU}$ is the activation function. 
The $\mathrm{Attention}^{\mathrm{ST}}$  is instantiated in two forms: spatial attention and temporal attention, each designed to capture interactions across nodes and time steps, respectively:
\begin{equation}\label{eq:att_s}
    \mathrm{Attention}^{\mathrm{ST}}_{\mathrm{spatial}}(X^Q, X^K, X^V) = \mathrm{softmax}(\frac{Q \cdot K^{\top_{-1,-2}}}{\sqrt{D}}) \cdot V
\end{equation}
\begin{equation}\label{eq:att_t}
    \mathrm{Attention}^{\mathrm{ST}}_{\mathrm{temporal}}(X^Q, X^K, X^V) = \big( \mathrm{softmax}(\frac{Q^{\top_\mathrm{st}} \cdot K^{\top_\mathrm{st} \top_{-1,-2}}}{\sqrt{D}}) \cdot V^{\top_\mathrm{st}} \big)^{\top_{\mathrm{st}}}
\end{equation}
where $\mathrm{softmax}(\cdot)$ is an activation function that normalizes a vector into a probability distribution. The operation $\top_{\mathrm{st}}$ denotes the transposition of axes corresponding to the spatial and temporal dimensions, whereas $\top_{-1, -2}$ denotes the transposition of the last two axes.

It is worth noting that we adopt a multi-head attention mechanism \cite{cordonnier2021multiheadattentioncollaborateinstead} to enhance the model's capacity to capture complex dependencies. When only one input is provided, it is assumed to be shared across the $X^Q$, $X^K$, and $X^V$.

\textbf{Network architecture.} We adopt a variational layered autoencoder \cite{VAE} structure to effectively model the uncertainty and diversity in football tactics. We apply several layered attention blocks to capture the complex representation, which are categorized into two types: the \textit{spatiotemporal attention block} and the \textit{cross-time attention block}.
We denote the input to the $l^{\mathrm{th}}$ block as $H^{(l-1)} \in \mathbbm{R}^{N \times |T| \times |\mathcal{V}| \times D}$.
The historical node feature sequence $\mathbf{X}_{\mathrm{hist}}$ is initially transformed into the hidden representation $H^0$.
We utilize the \textit{spatiotemporal attention block}, which consists of spatial attention module, temporal attention module, and gate fusion module:
\begin{equation}
    \begin{aligned}
        H^{(l)}_S &= \mathrm{Attention}^{\mathrm{ST}}_{\mathrm{spatial}}\big(\mathrm{Concat}(H^{(l-1)}, E^{\mathrm{TL}}_{T, s})\big) \\
        H^{(l)}_T &= \mathrm{Attention}^{\mathrm{ST}}_{\mathrm{temporal}}\big(\mathrm{Concat}(H^{(l-1)}, E^{\mathrm{TL}}_{T, s})\big) \\
        H^{(l)}   &= g \odot H^{(l)}_S + (1-g) \odot H^{(l)}_T
    \end{aligned}
\end{equation}
with 
\begin{equation}
    g = \sigma(H^{(l)}_S \cdot W_{g,1} + H^{(l)}_T \cdot W_{g,2} + b_g)
\end{equation}
where $W_{g,1} \in \mathbbm{R}^{D\times D}$, $W_{g,2} \in \mathbbm{R}^{D\times D}$, and $b_g\in \mathbbm{R}^D$ are learnable parameters, $\odot$ denotes the element-wise product, $\sigma$ denotes the sigmoid activation function, and $g$ is the gate. The gated fusion mechanism adaptively regulates spatial and temporal dependencies at each vertex and time step.
The \textit{spatiotemporal attention block} serves as the fundamental building unit for both the encoder and decoder, each comprising $L$ stacked layers. The employed TL-embeddings are $E^{\mathrm{TL}}_{\mathcal{V}, T_{n-1}, s}$ for the encoder and $E^{\mathrm{TL}}_{\mathcal{V}, T_{n}, s}$ for the decoder, respectively.

The \textit{cross-time attention block}, placed between the encoder and decoder, enables temporal alignment by projecting the encoded historical information into the temporal space of the prediction target with the temporal attention:
\begin{equation}
    H^{(0^{\prime})} = \mathrm{Attention}^{\mathrm{ST}}_{\mathrm{temporal}}( E^{\mathrm{TL}}_{T_{n}, s_n}, E^{\mathrm{TL}}_{T_{n-1}, s_n}, H^{(L)})
\end{equation}
where $H^{(0^{\prime})}$ denotes the initial input to the decoder, and $H^{(L)}$ serves as the output of the final encoder layer.

To enhance the model expressiveness we integrate a variational layer into the architecture.
Inspired by the Variational Autoencoder (VAE) framework\cite{VAE}, this layer captures latent stochastic variables that condition the deterministic decoder. 
Specifically, a neural network parameterized by $\phi$ encodes the hidden feature $H^{(L)}$ into a latent distribution $q_\phi(z|\cdot)$, from which a latent variable $z$ is sampled. The sampled $z$ is then fused with the deterministic representation before decoding, allowing the model to generate diverse yet coherent outputs.
During training, we optimize the variational objective using the evidence lower bound (ELBO), which combines the reconstruction loss and the Kullback–Leibler (KL) divergence between the approximate posterior $q_\phi(z|\cdot)$ and the prior $p(z)$:
\begin{equation}
    \mathcal{L}(\theta, \phi) =
    \mathbbm{E}_{q_\phi(z \mid \cdot)} (  \log p_\theta(\cdot \mid z) )
    - \beta \mathbbm{KL} (q_\phi(z \mid \cdot) \mid\mid p(z) )
\end{equation}
where $p_{\theta}(\cdot \mid z)$ is the likelihood function parameterized by the model weights $\theta$, $q_\phi(z \mid \cdot)$ is the approximate posterior distribution parameterized by $\phi$, $p(z)$ is a prior distribution over the latent variables, assumed to be a standard normal $\mathcal{N}(0, I)$ in our case, $\mathbbm{KL}(\cdot \mid \mid \cdot)$ denotes the KL divergence; and $\beta$ is a weighting coefficient that controls the strength of the regularization term.
We make an ablation study shown in Appendix \ref{appendix:ablation} to show that the variational formulation is particularly beneficial for tactic generation. 

\textbf{Training details.} The model was trained on the processed dataset partitioned into $70 \%$ training and $30 \%$ test sets using stratified sampling (random seed=42), optimized with Adam\cite{Kingma2014AdamAM} (initial learning rate=1e-4, $\beta_1$=0.9, $\beta_2$=0.999, L2 regularization strength=0.01, batch size=256) for 50 epochs under a cosine decay schedule with linear warm-up. We set the 1B model with $L=6$ and $D=512$. Experiments ran on a Linux server with 8× NVIDIA RTX A6000 GPUs (48GB VRAM).

\begin{figure}[ht]
    \centering
    \includegraphics[width=.8\textwidth]{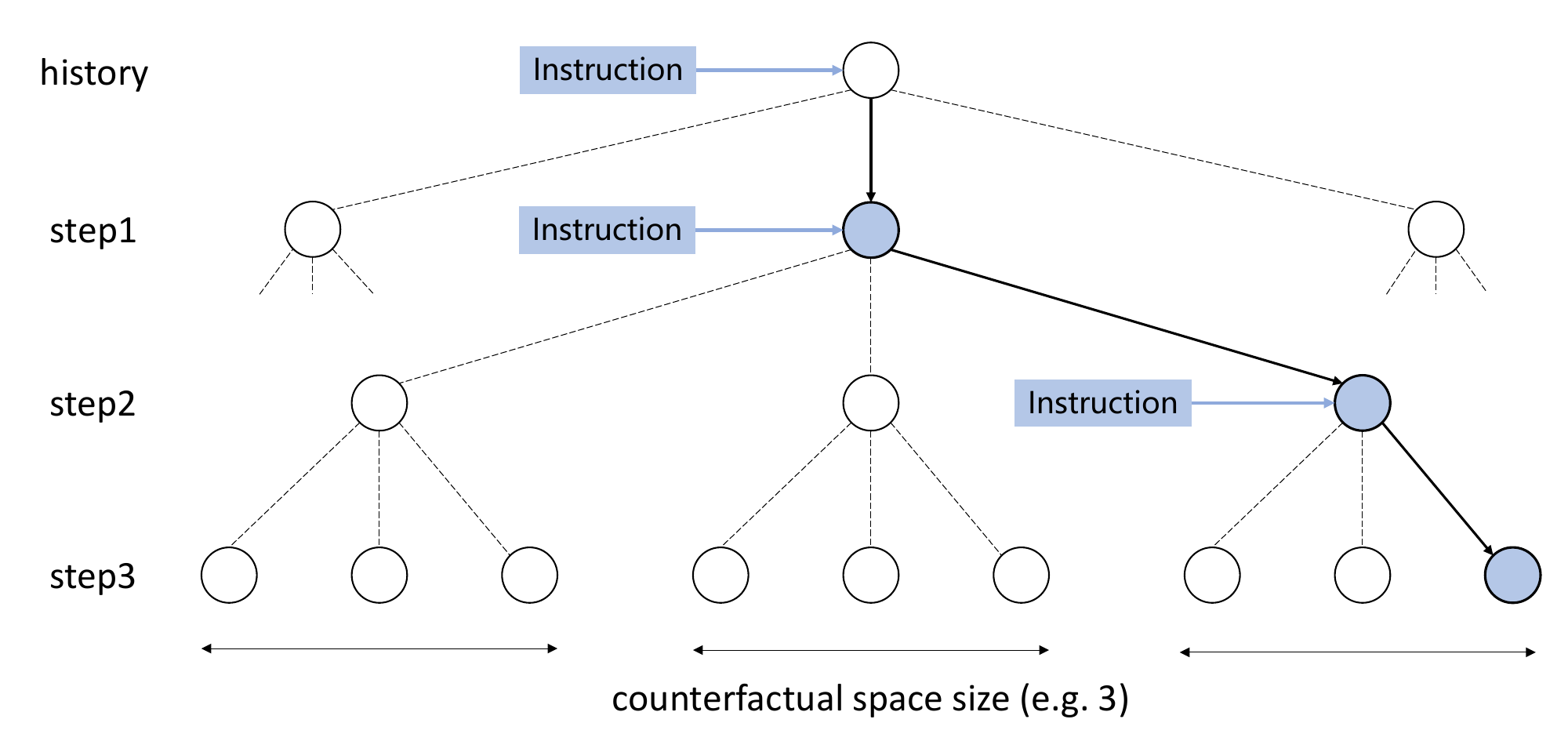}
    \caption{Tactic Discovery with Tree Search}
    \label{fig:tree}
\end{figure}

\subsection{Tactical Tree Search}
\label{sec4.4:TTS}

\textcolor{blue}{The tactical tree search is motivated by the need for a systematic approach to tactical discovery in football games. Tree search methods are widely adopted in AI-powered scientific discovery for efficiently exploring large combinatorial spaces \cite{segler2018planning,Dieb_Ju_Shiomi_Tsuda_2019,wang2023scientific,yamada2025aiscientistv2workshoplevelautomated}. By modeling the process as a tree search, we can effectively explore the counterfactual space and identify optimal tactical sequences that align with high-level instructions.}
The MTC leverages an MLLM $\mathcal{M}$ (QVQ-Max in practice) to accomplish the tactic discovery task.
Whether for single-step or multi-step tactical discovery, we uniformly model the process as a tactical tree search. As illustrated in Fig.~\ref{fig:tree}, the root node represents the historical context $\mathcal{G}_{0}$, the degree of each node is the size of the counterfactual space $|\mathcal{D}_n|$, and the depth of the tree corresponds to the predefined steps $N$. In the process of tactic discovery, $\mathcal{M}$ starts from the root node and searches for a path such that the formed meta-action sequence best aligns with the instruction $P$. Leveraging the planning and reasoning capabilities of MLLMs, $\mathcal{M}$ as $f_{\psi}(\cdot)$ is capable of making optimal search decisions at each step. We meticulously designed prompts to encompass the entire tactic discovery pipeline along with a tactical style pool, as represented in the Appendix \ref{appendix:prompt}.

\newpage
\printbibliography

\newpage
\section{Appendix}
\appendix
\section{Football Tactical Evaluation Calculation} \label{appendix:eval}
Here, we present the calculation formulas for all the metrics mentioned in the Result section.

The \textbf{Factual Trajectory Error} (FTE) and the \textbf{Counterfactual Alignment Error} (CAE) are two metrics to measure factual accuracy and counterfactual consistency of the LTG respectively.

The FTE is calculated as the mean squared error (MSE) between the output trajectory under the factual description and the ground truth:
\begin{equation}
    FTE = \frac{1}{|T|P} \sum_{t=1}^{|T|} \sum_{i=1}^{P} d_{i}^{t},
\end{equation}
where $d_i^t = \text{distance}(p_i^t, b^t)$ denotes the distance between the player $p_i^t$ and the ball $b^t$ at time $t$, $P$ denotes the number of the teammates, $|T|$ denotes the number of steps.

Given the historical meta-action $s_{n-1}$, a counterfactual description set can be obtained as $\mathcal{D}_n = c(s_{n-1})$, where $c(\cdot)$ serves as a rule specifically tailored to generate a finite set of possible choices.
The CAE is calculated as the endpoint alignment between the ball trajectories and the corresponding receiver trajectories derived from $\mathcal{D}_n$:
\begin{equation}
    CAE = \frac{1}{|\mathcal{D}_n|}\sum_{i=1}^{|\mathcal{D}_n|} d_{i_{\text{recipient}}}^{-1},
\end{equation}
where $t=0$ and $t=-1$ denotes the start and end of the event, $i_{\text{recipient}}$ denotes the index of the event recipient.

The \textbf{Consistency} between the trajectory and the language instruction is evaluated by ranking the carrier's distance to the ball among all teammates:
\begin{equation}
    C = \Big( d_{i_{\text{carrier}}}^0 \in \text{Top}_k(D^0) \Big) \land \Big( d_{i_{\text{recipient}}}^{-1} \in \text{Top}_k(D^{-1}) \Big)
\end{equation}
where $i_{\text{carrier}}$ denotes the index of the event carrier, $D^t = \{ d_1^t, d_2^t, \cdots, d_P^t \}$ denotes the set of distances among all $P$ teammates at time $t$, the symbol $\land$ denotes the logical operation ``and", the $\text{Top}_k$ denotes the top-ranked candidates considered ($k=3$ in practice)

\begin{figure}[htbp]
    \centering
        \begin{subfigure}[b]{0.45\textwidth}
            \includegraphics[width=\textwidth]{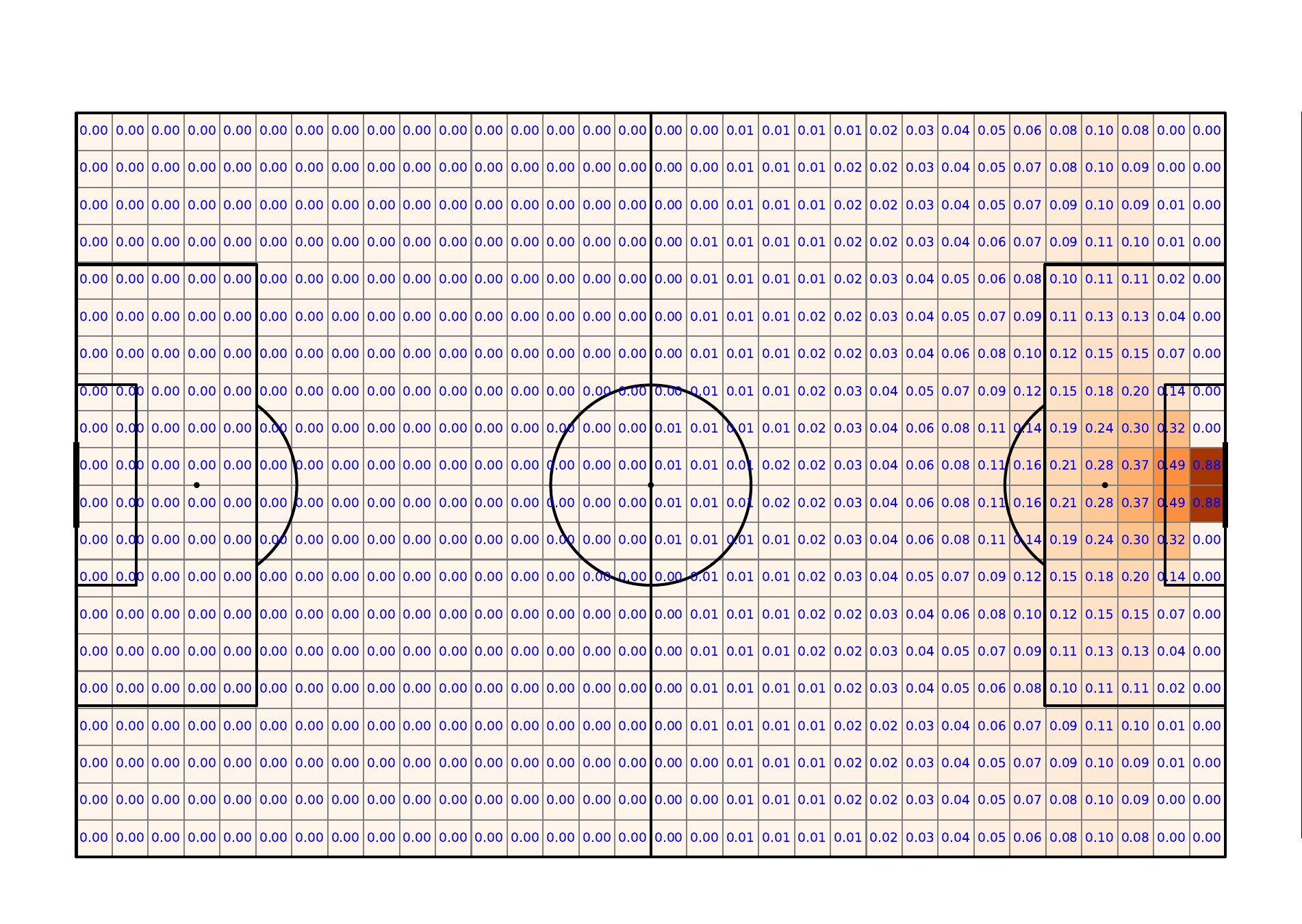} 
            \caption{expected goals}
            \label{fig:xg}
        \end{subfigure}
        \hfill 
        \begin{subfigure}[b]{0.45\textwidth}
            \includegraphics[width=\textwidth]{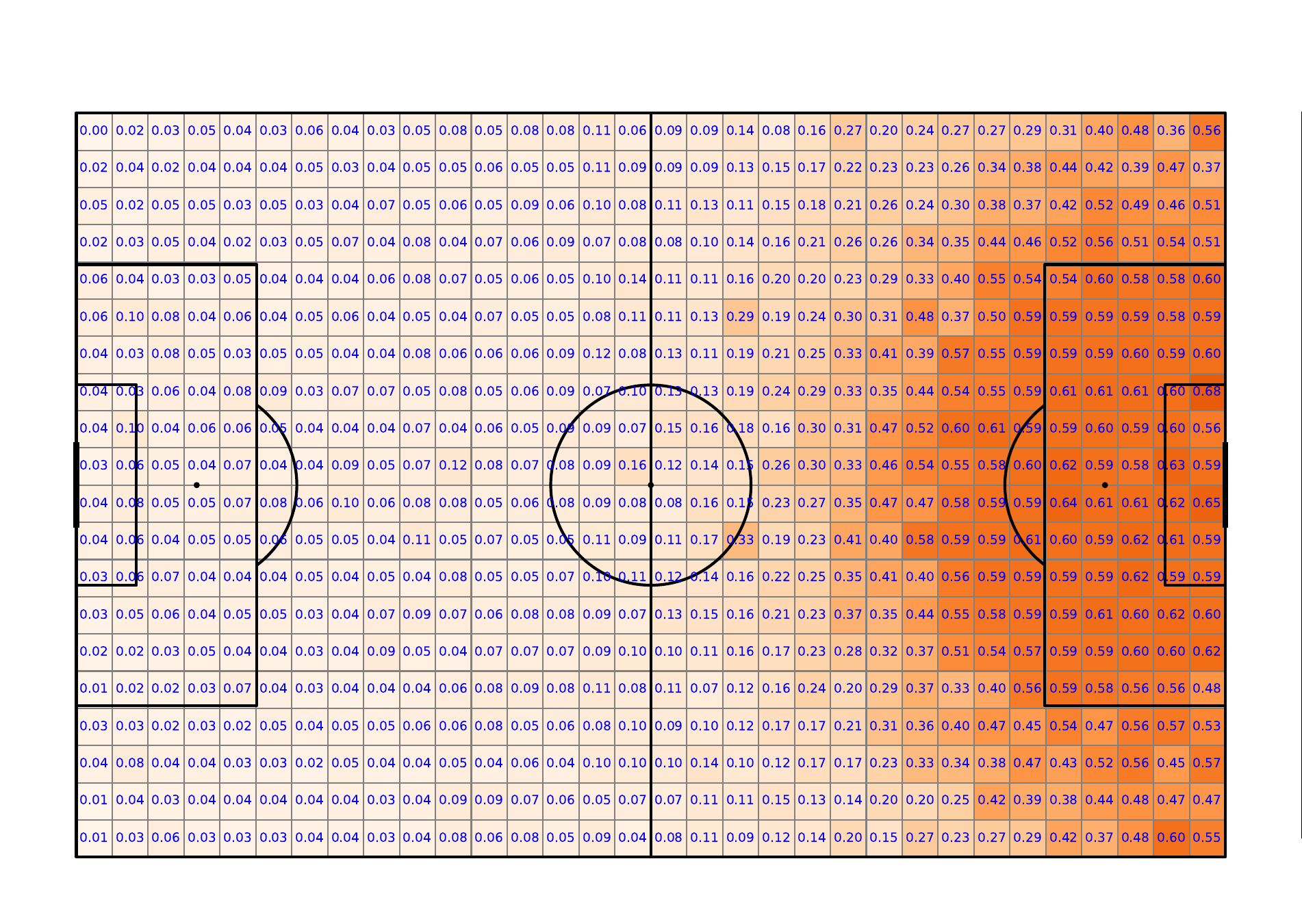} 
            \caption{expected threat}
            \label{fig:xt}
        \end{subfigure}
    \caption{xG and xT in grid pitch}
    \label{fig:eval}
\end{figure}

The \textbf{Expected Goals} (xG) and \textbf{Expected Threat} (xT) are calculated in grid pitch shown in figure \ref{fig:eval}. 
We utilized an xG model pretrained, which has been directly employed in numerous football studies. \footnote{CaleyXGCalculator: \url{https://github.com/krivonogov/xg}} We collect all event data from the seven matches comprising the quarterfinals, semifinals, and final of the 2024-2025 UEFA Champions League. These matches is popular, authoritative, and advanced. Based on this dataset, we train our xT model.

The \textbf{Pitch Control} is calculated using the physics-based method referenced in the main text.

The $\mathrm{xT(A)}$ or $\mathrm{PC(A)}$ indicate the \textit{attacking} potential and spatial dominance, while the $\mathrm{xT(D)}$ or $\mathrm{PC(D)}$ reflect the \textit{defensive} potential of the opponent’s threat and control. They are obtained by sampling the data of players over time and spatial scales and averaging it on a grid pitch.

\section{Questionnaire Design} \label{appendix:QA}

Through well-defined metrics and the demonstration of TacEleven's mining process and results, we benchmarked the model's capabilities and demonstrated its interpretability. When actually used to assist coaches and analysts in tactical decision-making, the core questions are whether the tactics mined by TacEleven are authentic and credible, and whether they can discover tactics superior to real situations. These two metrics—authenticity and effectiveness—depend on human subjective judgment. Therefore, we collaborated with experts from Auxerre (5 people), football enthusiasts (20 people), and youth team members (20 people) to conduct a questionnaire survey to objectively evaluate the authenticity and effectiveness of tactics mined by TacEleven.

We designed four comprehensive questionnaires to evaluate different aspects of TacEleven's tactical mining capabilities:

\subsubsection*{Questionnaire 1: Single-step Tactic Authenticity}
\begin{itemize}
    \item Objective: Evaluate the model's ability to generate realistic single-step tactical events.
    \item Format: Participants are presented with two events that share identical historical contexts—one real and one model-generated. They must identify which event actually occurred.
    \item Question: "Both events have the same historical context. One is real, the other is model-generated. Please identify which event actually occurred."
    \item Options: left is real / right is real / cannot distinguish.
    \item Scale: 50 questions, estimated completion time: 8 minutes
\end{itemize}

\subsubsection*{Questionnaire 2: Single-step Tactic Effectiveness}
\begin{itemize}
    \item Objective: Assess whether model-generated single-step tactics outperform real situations.
    \item Format: Participants compare real events with model-generated alternatives in identical historical contexts.
    \item Question: "Both events have the same historical context. The left event is real, the right is model-generated. Please evaluate if the model-generated result is better than the real outcome."
    \item Options: better / worse / no change
    \item Scale: 50 questions, estimated completion time: 8 minutes
\end{itemize}

\subsubsection*{Questionnaire 3: Multi-step Tactic Effectiveness}
\begin{itemize}
    \item Objective: Evaluate the quality of multi-step tactical sequences generated by the model.
    \item Format: Random selection of 30 scenarios with both real tactics and model-generated tactical sequences.
    \item Question: "Please evaluate whether the model-generated multi-step tactical sequence is better than the real tactical outcome."
    \item Options: better / worse / no change
    \item Scale: 30 questions, estimated completion time: 12 minutes
\end{itemize}

\subsubsection*{Questionnaire 4: 5-shot Tactical Mining Adoption}
\begin{itemize}
    \item Objective: Assess the practical utility of multiple tactical options in typical game scenarios.
    \item Format: Focused analysis of 10 typical attacking failure scenarios from Paris Saint-Germain vs Monaco matches, featuring top players (Messi, Neymar, Mbappé, and others).
    \item Question: "Given the real tactical situation and 5 model-generated potential outcomes, how many tactics are practically effective based on your professional analysis?"
    \item Options: 0 / 1 / 2 / 3 / 4 / 5
    \item Scale: 10 questions, estimated completion time: 12 minutes
\end{itemize}

The design of our questionnaire was meticulously guided by a comprehensive set of key considerations to ensure both scientific rigor and practical utility. We adopted a strategy of progressive complexity, beginning with single-step tactical evaluations and systematically advancing to multi-step tactical sequences, thereby enabling a granular assessment of the model's capabilities across different levels of tactical sophistication. Time efficiency was carefully calibrated through precisely calculated question counts and realistic completion time estimates, ensuring meaningful expert participation without inducing survey fatigue or compromising response quality. To capture a holistic perspective, we intentionally incorporated diverse expertise levels by including professional football experts, dedicated enthusiasts, and active youth team players, each bringing unique insights from their respective domains of football knowledge. The questionnaire maintained strong real-world relevance by focusing on authentic match scenarios derived from actual game situations, with particular emphasis on encounters involving elite players and top-tier teams to enhance ecological validity. Finally, we implemented a balanced assessment framework that strategically combined binary choice paradigms for evaluating tactical authenticity with graded evaluation scales for measuring effectiveness, thus providing a comprehensive analytical approach that captures both the qualitative and quantitative dimensions of tactical quality. This multi-faceted design philosophy ensured that our evaluation methodology was not only theoretically sound but also practically applicable in real-world football analytic contexts.

The questionnaires were administered electronically and the experts were provided with clear visualizations of tactical situations and adequate time for each evaluation. The results from these questionnaires are presented in Fig.~\ref{fig:violin} and Fig.~\ref{fig:bar} in the main text, providing quantitative measures of TacEleven's performance in generating authentic and effective football tactics.

\section{Football Data Alignment Methodology}\label{appendix:align}

The integration of multi-modal football data presents significant challenges due to inherent temporal misalignments between different data sources. While event data provides semantically rich annotations of key actions (passes, shots, carries, etc.), these timestamps often demonstrate systematic offsets when compared to the corresponding physical events observable in tracking trajectory data. This discrepancy arises from various factors including human annotation latency, differences in data collection systems, and varying definitions of event initiation and termination across data providers.
As shown on the left side of Figure \ref{fig:alignment}, the ongoing dribble event is visualized as trajectories from the start time to the end time marked in the event data. The yellow arrow in the figure indicates the direction of the dribble, while the two yellow circles represent positional errors caused by misalignment. Obviously, the ball has already been passed at the final moment, indicating that the actual end time of the dribble should have been earlier.

To address this critical issue, we developed a novel alignment pipeline that leverages kinematic constraints from tracking data to recalibrate event timestamps, shown with a schematic diagram on the right side of Figure \ref{fig:alignment}. Our method operates on the fundamental principle that certain physical signatures in player and ball motion serve as reliable temporal anchors that can bridge the semantic and physical domains of football data. The alignment process consists of 
two meticulously designed stages:

\begin{figure}[htbp]
    \centering
    \includegraphics[width=\textwidth]{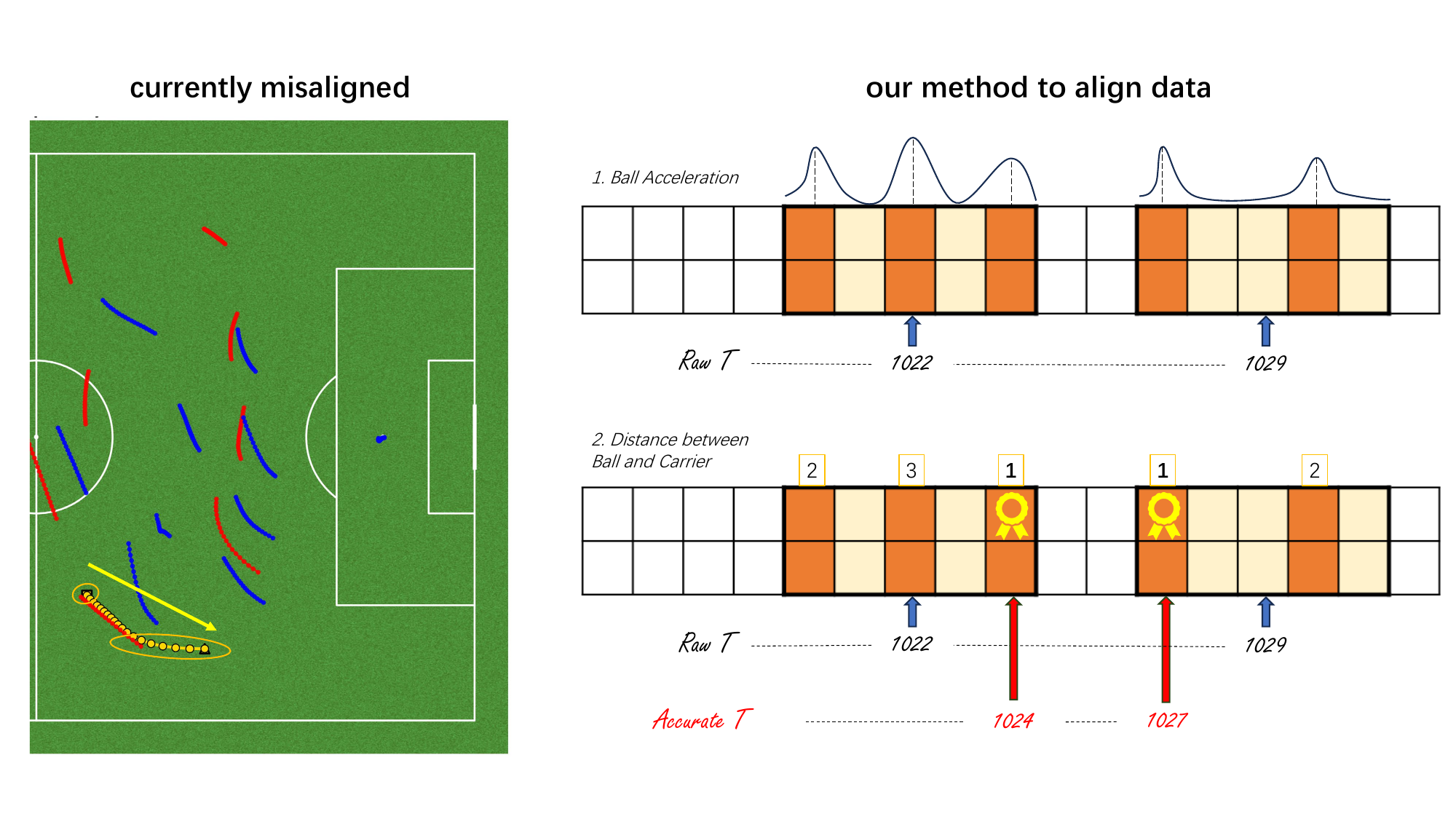}
    \caption{Our method to align football data}
    \label{fig:alignment}
\end{figure}

First, we identify candidate temporal anchors by detecting local extrema in the ball's acceleration signal. These peaks are hypothesized to correspond to moments of significant force application (e.g., kicks, passes, shots), which should align with the semantic events recorded in the event data. The instantaneous acceleration magnitude is computed directly from the raw velocity data using a backward difference method. Local maxima in this resulting signal are then identified, shown as the orange parts in the figure, using a peak detection algorithm that compares each point to its immediate neighbors.

Second, for each event annotation in the event data, we establish a search window centered on the originally reported timestamp. Within this temporal neighborhood, we identify all candidate anchors from the first stage and then apply a secondary refinement criterion: we select the moment where the distance between the ball and the primarily involved player reaches a local minimum. This dual-criterion approach ensures that we capture not only the kinetic signature of the event but also the spatial proximity context that characterizes player-ball interactions. Then, we implement a consensus mechanism that reconciles multiple candidate anchors when they appear within physiologically plausible time intervals, shown as the yellow medal in the figure. This mechanism employs a voting system that prioritizes anchors with stronger acceleration signals and closer player-ball distances, resulting in robust timestamp assignments that are consistent across both kinetic and spatial dimensions.

The output of this pipeline is a set of refined timestamps that demonstrate significantly improved alignment between event annotations and their corresponding physical manifestations in tracking data. Validation against manually annotated ground truth data shows that our method reduces temporal misalignment from an average of 0.9 seconds to under 0.1 seconds, representing a substantial improvement that enables more accurate fusion of multi-modal football data for tactical analysis and model training.

This alignment methodology not only facilitates the technical integration of diverse data sources but also enhances the physiological plausibility of the resulting fused dataset, enabling researchers to develop more accurate and robust models for football analytics that can simultaneously leverage the semantic richness of event data and the kinematic precision of tracking data.


\section{Ablation Study}\label{appendix:ablation}

A comprehensive ablation study was conducted to rigorously validate the effectiveness of individual components within our proposed framework and to verify the critical importance of our data alignment methodology. The experimental design systematically evaluated three key aspects of our approach: first, the impact of utilizing unaligned multi-modal data to establish the necessity of our temporal synchronization pipeline (w/o alignment); second, the contribution of the variational module by examining model performance when this probabilistic component was removed (w/o variation); and third, the value of the attention mechanism by replacing it with a functionally equivalent but structurally simpler multilayer perceptron (MLP) network (w/o attention). Each ablated model configuration was initialized from the same 1B-parameter foundation model and underwent identical training procedures, maintaining consistent hyperparameter settings, optimization strategies, and computational budgets across all experimental conditions to ensure a rigorous and fair comparative analysis.

The performance across all variants was quantitatively assessed using three standard evaluation metrics:  Factual Trajectory Error(FTE), Counterfactual Start-point Error(CSE) and Counterfactual Alignment Error(CAE), which collectively provide a comprehensive view of model capabilities. 
The results demonstrate that the complete model with all components intact achieves superior performance, thereby confirming that each element—the data alignment preprocessing, the variational module for probabilistic modeling, and the attention mechanism for feature refinement—makes a meaningful and non-redundant contribution to the overall system effectiveness. This systematic ablation provides empirical evidence supporting our architectural design choices and validates the utility of each technical innovation introduced in this work.

\begin{table}[thbp]
\centering
\caption{Ablation results}
\label{tab:ablation}
\begin{tabular}{cccc}
\toprule
Setting    & FTW $\downarrow$ (m) & CSE $\downarrow$ (m) & CAE $\downarrow$ (m) \\
\midrule
\textbf{our method} & \textbf{2.0546} & \textbf{2.2101} & \textbf{5.1344}  \\
w/o alignment       & 2.3781 & 9.7762 & 14.9311  \\
w/o variation       & 2.1006 & 6.5543 & 11.7736  \\
w/o attention       & 7.6545 & 13.2548 & 25.3484  \\

\bottomrule
\end{tabular}
\end{table}

\section{MLLM Prompt Design}\label{appendix:prompt}
In MTC, we employ an MLLM as an intelligent agent to execute a comprehensive planning, with the objective of selecting optimal tactics that align with high-level strategic instructions. The prompt design for this agent is meticulously structured to facilitate nuanced understanding and decision-making within the dynamic environment of a football match. The prompt explicitly defines the agent’s role as a tactical analyst, providing it with detailed contextual information including the current match state, relevant player profiles, historical tactical patterns, and specific strategic goals. It is then instructed to evaluate a range of potential tactical actions—such as passes, dribbles, or carries—by reasoning about their feasibility, effectiveness, and alignment with the stated game plan. The prompt further guides the model to incorporate multimodal inputs, such as spatial trajectories of players and temporal event sequences, into its reasoning process, ensuring that its recommendations are grounded in both visual and symbolic representations of the game. By integrating these elements into a coherent and context-rich instructional framework, the prompt enables the MLLM agent to function not merely as a passive evaluator but as an active participant in a closed-loop reasoning process, bridging the gap between abstract strategic intent and executable in-game actions. Below are the prompts for single-step and multi-step discovery for the MLLM.

\begin{promptblock}
Single-step Discovery:
You are a football analyst reviewer. Your task is to:
1. Review the given match situation and proposed events
2. Provide counterfactual suggestions if needed
3. Return your analysis in a structured format

Given the match situation:
{message.decision_task_dict["description"]}
The following is historical match situation visualization:
{"type": "image_url", "image_url": {"url": history_img}}
The following is predicted attacking tactic visualization under the factual instruction:
{"type": "image_url", "image_url": {"url": processed_images['prediction_image']}}
The followings are counterfactual optional predicted attacking tactic visualizations
Counterfactual Instruction: {cf_img_key} with following predicted attacking tactic visualization:
{"type": "image_url", "image_url": {"url": processed_images['cf_img_data']}}
Counterfactual Instruction: {cf_img_key} with following predicted attacking tactic visualization:
{"type": "image_url", "image_url": {"url": processed_images['cf_img_data']}}
...

In all visualizations: red = teammates, blue = opponents, yellow = ball. Trajectories progress from shallow to deep, ending at the scatter point.
Player nodes corresponding to each event are highlighted with yellow edges.
Attacking team(red) is on the left and defending team(blue) is on the right.
Beside the scatters, the attacking player names and role-initials are displayed.
In the predicted attacking tactic visualization, the transparent trajectories represent the historical visualizations for easy comparison. First and foremost, the sketch must reflect the instructions, with event relevant players close to the ball.

[reasoning]
In the [reasoning] part, it is necessary to include an analysis of each sketch.
The reasoning must address the following aspects in the given sequence:
1. check the consistency between the current tactic sketch and the language instructions, must refuse if the ball is not close to the expected recipient in the sketch,
2. assess the authenticity of the tactical sketch and evaluate its feasibility given the current match situation.
3. analyse the scoring advantage,
4. analyse the risk of losing ball,
5. player suitability considering player history attributes, based on your knowledge of the specific player,
6. tactical execution success rate.

You final answer must be strictly in following format:
[summary]
A horizontal comparison of the analyses for all sketches should be conducted in a list format, and based on this comparison, an optimal choice should be made.
[event]
The [event] must be chosen uniquely from {[json.loads(key) for key in processed_images['cf_images'].keys()]}
\end{promptblock}

\begin{promptblock}
Multi-step Discovery:
# Role: You are a football analyst reviewer. Your task is to:
            1. Review the given match situation and proposed events
            2. Provide counterfactual suggestions if needed
            3. Return your analysis in a structured format
            
## Profile
- **language**: English
- **description**: A highly specialized AI designed to evaluate and optimize tactical scenarios in sports, particularly focusing on visualized data analysis for strategic decision-making.
- **background**: Developed by a team of sports analysts and AI experts, this AI has been trained on vast datasets from professional sports, including soccer, basketball, and other team-based games. It excels in interpreting complex visualizations and translating them into actionable insights.
- **personality**: Analytical, detail-oriented, and objective. The AI provides clear, concise, and unbiased evaluations, ensuring that decisions are based on data-driven logic rather than intuition.
- **expertise**: Sports tactics, data visualization interpretation, strategic planning, and performance analysis.
- **target_audience**: Coaches, sports analysts, and team strategists who rely on visual data to make informed decisions during gameplay.

Given the match situation:
{message.decision_task_dict["description"]}
The following is historical match situation visualization
{"type": "image_url", "image_url": {"url": history_img}}
The following is predicted attacking tactic visualization
{"type": "image_url", "image_url": {"url": processed_images['prediction_image']}}
The followings are counterfactual predicted attacking tactic visualizations
Counterfactual Instruction: {cf_img_key} with following predicted attacking tactic visualization
{"type": "image_url", "image_url": {"url": cf_img_data}}
Counterfactual Instruction: {cf_img_key} with following predicted attacking tactic visualization
{"type": "image_url", "image_url": {"url": cf_img_data}}
...

This is the previous historical situation, including the decisions made by the attacking players before:{message.decision_task_dict["history"]}
## Initialization
You are evaluating a multi-stage tactical scenario with the MANDATORY STRATEGIC OBJECTIVE:
{scenario_requirements}

In all visualizations: red = teammates, blue = opponents, yellow = ball. Trajectories progress from shallow to deep, ending at the scatter point.
Player nodes corresponding to each event are highlighted with yellow edges.
Attacking team(red) is on the left and defending team(blue) is on the right.
Beside the scatters, the attacking player names and role-initials are displayed.
In the predicted attacking tactic visualization, the teammates with the ball is predicted, and the opponents are also predicted.
In the predicted attacking tactic visualization, the transparent trajectories represent the historical visualizations for easy comparison.
                              
You are REQUIRED to:
1. **Enumerate and analyze every possible passing option** that could realistically occur from the current carrier, with attention to lane openness, distance, angle, interception risk, pressure, receiver orientation, continuation options, and alignment with the scenario requirements. Additionally, consider tactical continuity and coherence - analyze how each option connects logically to previous moves and maintains strategic flow toward the objective.
2. **Individually evaluate each candidate event** from the list below. For each candidate, provide a compact but detailed micro-analysis (advantages, risks, execution likelihood, tactical fit).  
3. **Construct a horizontal comparison** of all candidate options against each other. Use structured reasoning to rank them based on consistency with scenario requirements, scoring advantage, risk, formation organization, player suitability, and tactical execution success rate.  
It is necessary to include an analysis of each sketch and  ALL {len(candidate)} candidate options, covering aspects including consistency between the current tactic and the language instructions(must), scoring advantage, risk of lossing ball, formation organization, player suitability, and tactical execution success rate. Finally, a horizontal comparison of the analyses for all sketches should be conducted in a list format, and based on this comparison, an optimal choice should be made.
4. **Select exactly one candidate** as the next optimal tactical move. 
5. **Check the consistency between the current tactic sketch and the language instructions, must refuse if the ball is not close to the expected recipient in the sketch,as well as the player suitability considering player history attributes, based on your knowledge of the specific player,

The candidate events are:
{[json.loads(key) for key in processed_images['cf_images'].keys()]}
## Output Obligation
- In the reasoning section, first present the **full per-pass analysis**, then the **candidate-focused comparison table**, and finally a concise justification for the best choice.  
- In the event section, copy the selected candidate **EXACTLY as it appears** (JSON only, no extra text). Output the event object like: {"event_type": "...", "carrier_name": "...", "carrier_role": "...", "recipient_name": "...", "recipient_role": "..."}
- Only one JSON object is allowed in the `###event` block. 
\end{promptblock}

\end{document}